\documentclass[aps,prx,reprint,superscriptaddress,nofootinbib,longbibliography]{revtex4-2}
\usepackage{amsmath,amssymb,booktabs,graphicx,xcolor}
\graphicspath{{../}{./}} 
\usepackage[colorlinks=true,linkcolor=blue,citecolor=blue,urlcolor=blue]{hyperref}

\newcommand{\gL}{\gamma_L^*}
\newcommand{\gT}{\gamma_T^*}
\newcommand{\VL}{V_L}
\newcommand{\VT}{V_T}

\newcommand{\ueps}[2]{u^{#1}_{#2}}
\newcommand{\rsw}[2]{r^{#1}_{#2}}
\newcommand{\Qsq}{Q^{2}}
\newcommand{\sT}{\sigma_T}\newcommand{\sL}{\sigma_L}
\newcommand{\RT}[2]{\mathrm{Re}\,T_{#1}T_{#2}^{*}}
\newcommand{\RU}[2]{\mathrm{Re}\,U_{#1}U_{#2}^{*}}
\newcommand{\IT}[2]{\mathrm{Im}\,T_{#1}T_{#2}^{*}}
\newcommand{\IU}[2]{\mathrm{Im}\,U_{#1}U_{#2}^{*}}
\newcommand{\aT}[1]{|T_{#1}|^{2}}
\newcommand{\aU}[1]{|U_{#1}|^{2}}
\newcommand{\dmo}{1\text{-}1}
\newcommand{\dash}{\scalebox{0.75}[1]{\textemdash}}

\begin{document}

\title{Direct Bayesian Inference of Helicity Amplitudes from Detector-Level Scattering Data}

\author{B.~Singh}
\email{singh@jlab.org}
\thanks{All authors other than the first contributed equally and are listed in alphabetical order.}
\affiliation{Thomas Jefferson National Accelerator Facility, Newport News, Virginia 23606, USA}
\author{H.~Avakian}
\affiliation{Thomas Jefferson National Accelerator Facility, Newport News, Virginia 23606, USA}
\author{S.~Bhattacharya}
\affiliation{Department of Physics, University of Connecticut, Storrs, Connecticut, USA}
\author{V.~Burkert}
\affiliation{Thomas Jefferson National Accelerator Facility, Newport News, Virginia 23606, USA}
\author{L.~Elouadrhiri}
\affiliation{Thomas Jefferson National Accelerator Facility, Newport News, Virginia 23606, USA}
\author{D.I.~Glazier}
\affiliation{School of Physics and Astronomy, University of Glasgow, Glasgow, United Kingdom}
\author{V.~Kubarovsky}
\affiliation{Thomas Jefferson National Accelerator Facility, Newport News, Virginia 23606, USA}
\author{D.~Leahy}
\affiliation{School of Physics and Astronomy, University of Glasgow, Glasgow, United Kingdom}
\author{Y.~Li}
\affiliation{Department of Computer Science, Old Dominion University, Norfolk, Virginia, USA}
\author{R.G.~Milner}
\affiliation{Massachusetts Institute of Technology, Cambridge, Massachusetts, USA}
\author{Y.~Wang}
\affiliation{Massachusetts Institute of Technology, Cambridge, Massachusetts, USA}

\begin{abstract}
Helicity amplitudes give the most complete description of a variety of scattering reactions used in studies of strong interactions, but they cannot be
measured directly: experiments record bilinear combinations of them folded through a detector
response, and conventional analyses recover them through multiple stages that introduce discrete
ambiguities and require a separate extraction of the absolute cross sections. Here we replace that
chain with a single Bayesian inference that determines the experimentally identifiable amplitude
parameters directly from detector-level measurements. A score-based diffusion model, trained on a
forward simulator, provides the full posterior in each kinematic bin, with the detector response
carried by the forward model and positivity of the spin-density matrix guaranteed by the
parameterization. The observable amplitude content in electroproduction of final particles grows with polarization of beams and targets, providing additional sensitivity to underlying phases. In simulation, the posterior achieves empirical coverage at or above the nominal level and delivers the angular
observables and the separated contributions from longitudinal and transverse photons with their correlations
retained. It transfers without retraining to a realistic detector response absent from training and
remains reliable in the weakly constrained nucleon-helicity-flip sector, where per-bin likelihood
maximization degrades. The result is a general framework for a broad class of inverse problems, those in which
parameters enter the observables nonlinearly, are subject to exact ambiguities and physical
constraints, and are measured only through an instrument response; phase retrieval, quantum-state
tomography, and partial-wave analysis are further instances. Its core requires only a forward
simulation of the complete measurement process; ambiguities, physical constraints, and instrumental
effects are handled within one statistically consistent posterior, applicable across exclusive
vector-meson programs at Jefferson Lab, COMPASS, HERMES, and the future Electron-Ion Collider.
\end{abstract}

\maketitle

\section{Introduction}

Experiments often seek quantities that cannot be observed directly. Instead, detectors record indirect signatures, such as event rates, angular distributions or particle trajectories, from which the underlying physical process must be inferred. This task becomes especially difficult when the measured signal depends nonlinearly on the quantities of interest, is altered by the detector, and must obey exact physical constraints. Such inverse problems arise across science, from imaging and quantum-state reconstruction to particle physics.

Exclusive vector-meson electroproduction provides a particularly demanding example. An electron scatters from a proton by exchanging a virtual photon, which converts into a vector meson, a short-lived quark--antiquark state carrying the quantum numbers of the photon, while the proton recoils intact. The reaction is exclusive in that every final-state particle is detected, so each event is fully reconstructed. The meson itself is never seen directly: the $\phi$ and $\rho^0$ considered here decay almost immediately into $K^+K^-$ and $\pi^+\pi^-$ pairs, and it is the angular distribution of those decay products, together with the event rate, that records how the meson was produced.

The quantities that describe such a reaction most completely are the helicity amplitudes: complex numbers that specify how it proceeds for each allowed combination of particle spin states. They connect the measured distributions of final-state particles to the internal structure of the nucleon, constraining how quarks and gluons are distributed, correlated and polarized inside the proton: within QCD factorization they give access to generalized parton distributions (GPDs), which describe the spatial distribution of partons in the transverse plane, and more generally to generalized transverse-momentum distributions (GTMDs), which additionally encode correlations involving parton momentum, position and spin~\citep{Collins:1996fb,Diehl:2007jy,Goloskokov:2006hr,Bhattacharya:2022vvo}. Polarization measurements are particularly valuable because they reveal interference effects and helicity-changing processes that are inaccessible in unpolarized data alone, among them the GPD $E$, which is closely connected to the decomposition of the nucleon's angular momentum.

Reconstructing the vector meson from its decay is also what makes the measurement demanding: the extra angular variables inform on the production process but enlarge the space in which the data must be described. Realistic Monte Carlo generators that describe both production and decay are therefore essential for developing and validating extraction methods~\citep{Avakian:2015vha}. Such modeling matters beyond exclusive measurements: vector mesons constitute a significant contribution to observed final states in semi-inclusive deep-inelastic scattering (SIDIS), and their decay products modify measured kinematic distributions as well as spin and azimuthal asymmetries~\citep{HERMES:2006pof,COMPASS:2019lcm,Avakian:2024xrr}, so interpreting SIDIS data requires a quantitative understanding of exclusive vector-meson production~\citep{Avakian:2026ozw}.

Yet helicity amplitudes are not themselves observables. Experiments measure rates and decay-angle distributions that depend on products of amplitudes, so distinct amplitude configurations can generate identical data. The measured distributions are also shaped by the acceptance, efficiency and finite resolution of the detector. In addition, any physically allowed solution must correspond to a positive semidefinite spin-density matrix. These features make amplitude extraction intrinsically ambiguous and poorly suited to a simple point-by-point inversion.

Conventional analyses address this problem through a sequence of intermediate measurements. Angular distributions are first fitted to obtain spin-density matrix elements (SDMEs)~\citep{Schilling:1973ag,HERMES:2009oim,COMPASS:2022xig,H1:2009cml,ZEUS:2005bhf,CLAS:2005nkx}, while the absolute longitudinal and transverse cross sections are obtained separately, often through measurements at multiple beam energies, and a polarized nucleon adds further structure functions~\citep{Diehl:2007jy}. The helicity amplitudes are then reconstructed from these quantities. This strategy has been highly successful, but distributing the inference across several stages can obscure correlations, complicate uncertainty propagation and require separate treatments of detector effects, physical constraints and discrete ambiguities. In parallel, Ref.~\citep{leahy2026} has shown that amplitudes can be extracted directly from experimentally measured decay angle distributions in the case of produced mesons of any spin.

\begin{figure*}[!t]\centering
\includegraphics[width=\linewidth]{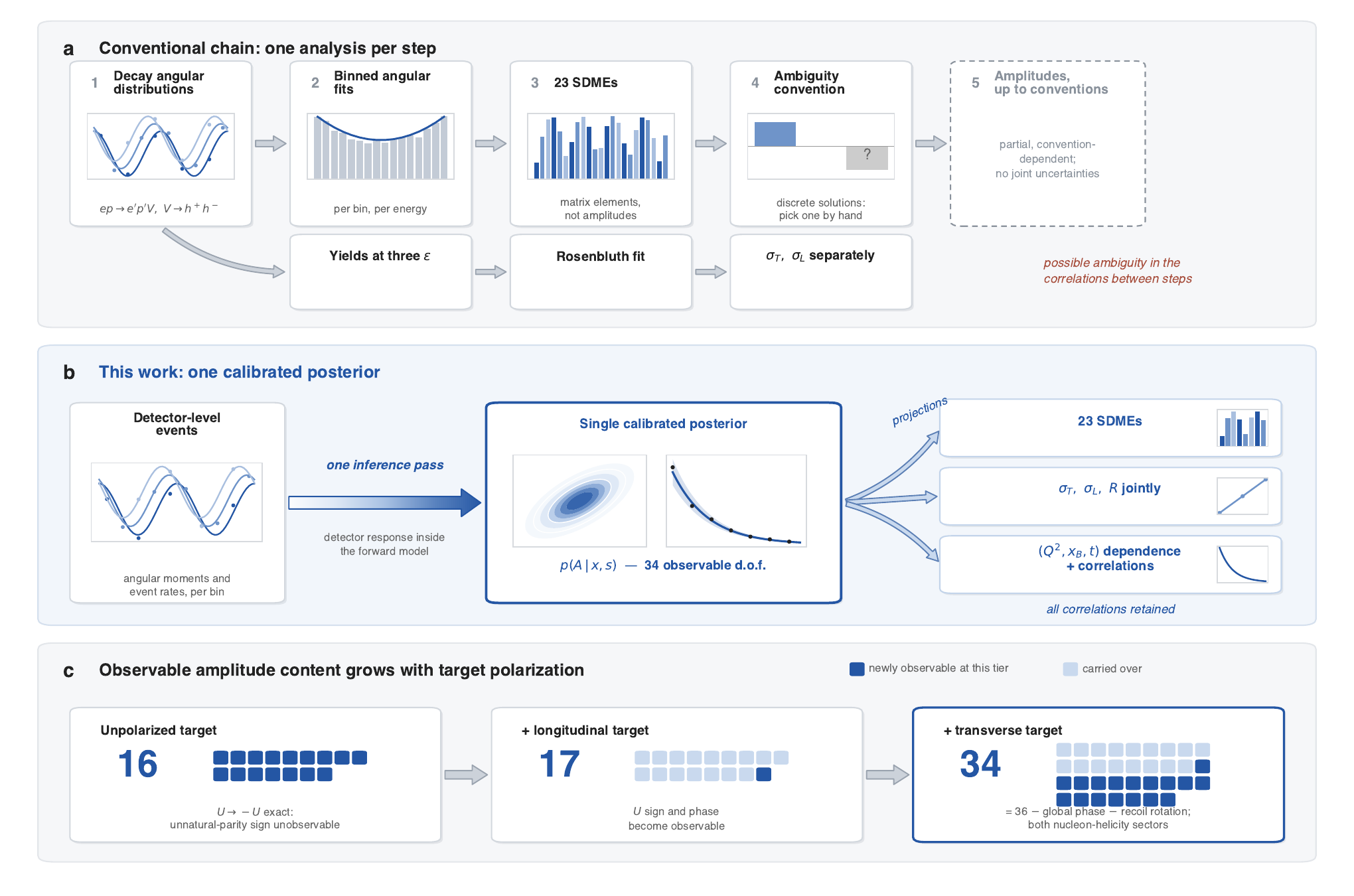}
\caption{\textbf{From detector events to the complete observable amplitude posterior.}
(a) The conventional chain: binned angular fits yield SDMEs, a convention resolves the discrete
ambiguities, and a separate Rosenbluth fit across beam energies supplies $\sT$ and $\sL$; each
step carries its own uncertainties, and propagating the correlations between steps requires a
dedicated procedure, such as a bootstrap over the full chain.
(b) This work: the same detector-level information, compressed into angular moments and event
rates with the detector response carried by the forward model, determines a single calibrated
posterior over the amplitudes; the SDMEs, the separated cross sections, and their dependence on
$Q^2$, $x_B$, and $t$ follow as projections of that one object, with all correlations retained.
(c) The observable amplitude content by polarization tier; each cell is one amplitude parameter,
filled when it first becomes observable and pale once carried over: 16 parameters from an unpolarized
target, where the natural--unnatural relative phase is unobservable (leaving $U\to-U$ as the
residual of the $U_{11}$-real gauge); 17 with
a longitudinal target, which promotes that sign and phase to observables; 34 with the combined
longitudinal and transverse program, the complete content after the global phase and the exact
recoil rotation $\mathcal R(\alpha)$ are removed. The 36 of panel~(c) is the number of real components carried by the 18 complex
amplitudes, before the two unobservable directions are removed. Symbols: $A$ denotes the amplitude
parameters, $x$ the detector-level features of one kinematic bin (its angular moments and event
rates) and $s$ the measured detector response, so $p(A\,|\,x,s)$ in panel~(b) is the posterior of
the amplitudes given the data and the instrument; $\varepsilon$ is the virtual-photon polarization
parameter, $Q^2$ the photon virtuality, $x_B$ the Bjorken variable and $t$ the squared momentum
transfer; $\sT$ and $\sL$ are the transverse and longitudinal cross sections and $R=\sL/\sT$ their
ratio; SDME abbreviates spin-density matrix element.}
\label{fig:concept}
\end{figure*}

Here we introduce a detector-level Bayesian framework that infers helicity amplitudes directly from measured angular distributions and event rates. A forward simulator incorporates detector acceptance and resolution, while a conditional score-based diffusion model~\citep{Ho:2020epu,Song:2020hus} learns the posterior over the experimentally identifiable amplitudes (Sec.~\ref{sec:xsec}) in each kinematic bin by simulation-based inference~\citep{Cranmer:2019eaq,Papamakarios:2018zoy,Brehmer:2018kdj}. By parameterizing the amplitudes directly, the framework guarantees a positive-semidefinite spin-density matrix (Sec.~\ref{sec:sdm}). Spin-density-matrix elements and separated longitudinal and transverse cross sections then emerge as correlated projections of a single posterior, rather than as products of separate fits. Figure~\ref{fig:concept} contrasts the two approaches.

We demonstrate the framework in simulated exclusive $\phi$ and $\rho^0$ electroproduction for a CLAS12-like detector~\citep{Burkert:2020akg}. Across the combined longitudinal--transverse polarized-target program, it recovers all 34 experimentally identifiable amplitude parameters and yields calibrated posterior uncertainties (Sec.~\ref{sec:results}). It remains reliable in the weakly constrained nucleon-helicity-flip sector, where per-bin likelihood maximization degrades. A single model trained across a family of detector responses transfers without retraining to a realistic response absent from training, enabling one training to serve every kinematic bin and detector within that family. That economy is central to the programs now being prepared: the polarized-target and multi-energy running at Jefferson Lab, and the far wider range of channels, kinematics and detector configurations expected at the Electron-Ion Collider (EIC)~\citep{AbdulKhalek:2021gbh,Bhattacharya:2026qnd}, where a separate fit in every bin, or a network retrained for every detector, would dominate the cost of the analysis. More broadly, the approach applies to constrained inverse problems in which nonlinear, ambiguity-prone parameters are observed through an instrument response, including partial-wave analysis, phase retrieval and quantum-state reconstruction (Sec.~\ref{sec:discussion}).

The paper is organized as follows. Section~\ref{sec:physics} fixes the physics inputs, Sec.~\ref{sec:framework} builds the detector-conditioned diffusion posterior, Sec.~\ref{sec:validation} sets out the validation, Sec.~\ref{sec:results} collects the results, and Sec.~\ref{sec:discussion} treats scope, systematics and generality. Three appendices give the complete formalism, including the polarized-target and nucleon-helicity-flip case (App.~\ref{app:polarized}); the method; and the supporting results, among them the lower polarization tiers (App.~\ref{app:modeAB}), the maximum-likelihood cross-check (App.~\ref{app:uml}) and the uncertainty budget (App.~\ref{app:unc}).

\section{Physics formulation of helicity-amplitude extraction}
\label{sec:physics}

The inference framework of Sec.~\ref{sec:framework} is fixed by three physics inputs: the helicity
amplitudes that describe the reaction, the observables that constrain them, and the positivity
constraint that any physical solution must satisfy. This section sets out the three in turn and
then specializes to the demonstration channels.

\subsection{Helicity amplitudes and experimental observables}\label{sec:xsec}
We consider exclusive electroproduction of a vector meson, $ep\to e'p'V$. In the one-photon
approximation the reaction proceeds by the exchange of a virtual photon of virtuality $Q^2=-q^2$,
with Bjorken variable $x_B=Q^2/(2\,p\!\cdot\!q)$,
which produces the meson off the proton (mass $M_p$) with squared momentum transfer $t=(p-p')^2$.
The virtual photon polarization is
\begin{equation}
\varepsilon=\frac{1-y-\tfrac14\gamma^2y^2}{1-y+\tfrac12 y^2+\tfrac14\gamma^2y^2},\qquad
y=\frac{p\!\cdot\! q}{p\!\cdot\! k},\quad \gamma=\frac{2M_p x_B}{Q},
\label{eq:eps}
\end{equation}
with $k$ the incoming lepton momentum,
and the unpolarized cross section splits as $d\sigma/dt = d\sT/dt+\varepsilon\,d\sL/dt$ with the
longitudinal-to-transverse ratio $R=\sL/\sT$.

\begin{figure}[!ht]\centering
\includegraphics[width=\linewidth]{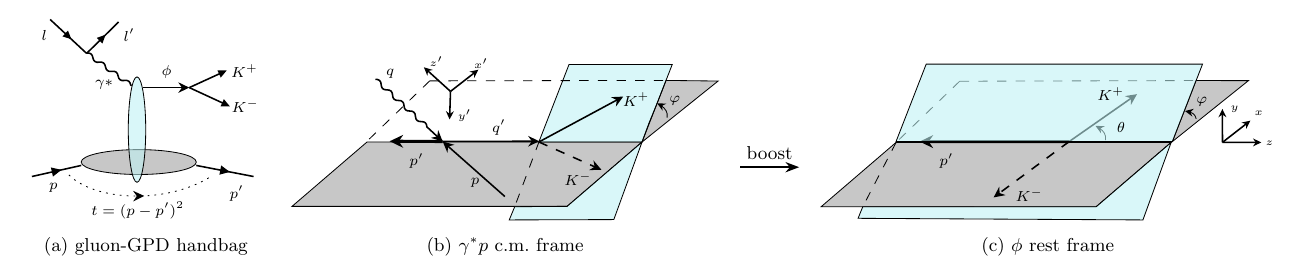}
\caption{Exclusive vector-meson electroproduction and its decay kinematics, shown for $\phi\to K^+K^-$ (identical definitions apply to $\rho^0\to\pi^+\pi^-$). (a) The factorized production amplitude
for $\gamma^*p\to\phi\,p$: the virtual photon couples to the proton through a $t$-channel exchange,
drawn as a shaded blob to represent the full exchange, dominantly two-gluon (parameterized by the gluon
GPD) but including multi-gluon and other contributions still under discussion; the $\phi$ ($V$) is
formed from a $q\bar q$ pair, and $t=(p-p')^2$ probes the transverse gluon distribution. (b) The $\gamma^*p$ center-of-mass frame, with the helicity axes
$(x',y',z')$: the production plane $(q,p',p)$ (grey) and the $K^+K^-$ decay plane (blue) meet along the
$\phi$ direction, and the azimuth $\varphi$ between them fixes the decay-plane orientation. (c)
The $\phi$ rest frame, reached by a boost along $z$: the $K^+K^-$ decay fixes the polar angle $\theta$
(of the $K^+$ from $z$) and the azimuth $\varphi$ (between the production and decay planes). Together
with $\Phi$, which orients the production plane relative to the lepton plane (defined in the text), these
are the helicity-frame variables $(\theta,\varphi,\Phi)$ used throughout, in the conventions of
Sec.~\ref{sec:xsec}. Momenta: $l$ and $l'$ are the incoming and scattered lepton, with the
incoming lepton momentum written $k$ in Eq.~\eqref{eq:eps}; $q=l-l'$ is the virtual photon, $p$ and
$p'$ the incoming and recoiling proton, and $q'$ the produced meson. The unprimed axes $(x,y,z)$ in
panel~(c) are those of the $\gamma^*p$ frame from which the boost is taken.}
\label{fig:process}
\end{figure}

The decay angular distribution is described in the vector-meson helicity rest frame by the
angles~\citep{Diehl:2007jy} $\Omega=(\cos\theta,\varphi,\Phi)$: the polar angle $\theta$ of the positive decay daughter is measured
from the helicity axis (the meson direction in the $\gamma^* p$ center of mass), the azimuth
$\varphi$ is the angle of the decay plane relative to the production plane, and $\Phi$ orients the
production plane relative to the lepton plane (Fig.~\ref{fig:process}). Both azimuths are
right-handed about their respective axes ($\hat z$ along the meson direction for $\varphi$, along
$\vec q$ for $\Phi$), with $\Phi$ in the Trento convention, which for this angle agrees with the
Schilling--Wolf convention used in the experimental SDME literature. Relative to Ref.~\citep{Diehl:2007jy}, our $\Phi$ coincides with
Diehl's production azimuth while the decay azimuth and the beam-helicity sign are opposite
($\varphi=-\varphi_{\rm D}$, $P_b=-P_b^{\rm D}$); all expressions in this paper are given in our
conventions. Together with $\varepsilon$ and the beam helicity, these variables fully specify the
measured angular distribution.

The reaction is described by the complex production helicity amplitudes
$T_{\mu\nu}(x_B,\Qsq,t)$ (meson helicity $\mu$, photon helicity $\nu$). For the nucleon-helicity-non-flip sector the production amplitude decomposes into
natural- and unnatural-parity exchange, $F=T+U$, with $T_{-\mu,-\nu}=+(-1)^{\mu-\nu}T_{\mu\nu}$ and
$U_{-\mu,-\nu}=-(-1)^{\mu-\nu}U_{\mu\nu}$. This leaves five natural amplitudes
$\{T_{11},T_{00},T_{01},T_{10},T_{1\text{-}1}\}$ and four unnatural amplitudes
$\{U_{11},U_{01},U_{10},U_{1\text{-}1}\}$ ($U_{00}=0$ by parity): $T_{11}\!:\gT\!\to\VT$,
$T_{00}\!:\gL\!\to\VL$ (the natural-parity $s$-channel helicity-conserving, SCHC, amplitudes), $T_{01}\!:\gT\!\to\VL$, $T_{10}\!:\gL\!\to\VT$, and
the double-flip $T_{1\text{-}1}\!:\gamma^*_{-T}\!\to\VT$.

These nine complex amplitudes contain 18 real degrees of freedom. One overall phase is
unobservable and is fixed by choosing $T_{11}\!\in\!\mathbb R_{\ge0}$. Unpolarized observables
leave a second phase undetermined, which we fix by taking $U_{11}$ real. The unpolarized
non-flip sector therefore carries 16 independent real amplitude parameters, the minimal complete
set for its decay distribution (App.~\ref{app:amp}).

Separately from these continuous phase choices, the unpolarized intensity is strictly quadratic in
the unnatural sector and is therefore exactly invariant under the global sign flip
$U_{\mu\nu}\to-U_{\mu\nu}$. This sign is not determined by unpolarized data, regardless of
statistics, and we quote the unnatural amplitudes from unpolarized data in the convention $U_{11}\ge0$. Target
polarization introduces $T$--$U$ interference terms that make the sign
observable~\citep{Diehl:2007jy}, and the polarized extraction of Sec.~\ref{sec:results} measures it
(App.~\ref{app:polarized}).

Both $T_{\mu\nu}$ and $U_{\mu\nu}$ are functions of all three kinematic variables,
$(x_B,\Qsq,t)$. In the closure studies below we use a fixed $x_B$ window, $0.08<x_B<0.5$, and
extract them as functions of $\Qsq$ and $t$ within it. The cross-section weighting concentrates the
accepted sample at $\langle x_B\rangle\simeq0.23$ (16--84\% range $0.15$--$0.31$), and the injected
truth is taken to be independent of $x_B$ across the window, so what is quoted at each $(\Qsq,t)$
point is the cross-section-weighted average of the amplitudes over that $x_B$ coverage.
Because $x_B$ is recoverable exactly from $(\Qsq,\varepsilon,E)$ event by event, the training and
calibration sample can be composed in whatever $x_B$ window the extraction uses; here that is the
full window above, and matching the two removes any train-versus-extraction composition mismatch.
The same analysis runs in narrower $x_B$ bins on real data, with the training composition moved with
them; the detector-conditioned acceptance already carries the corresponding $x_B$ dependence.

The sixfold differential cross section factorizes into the virtual-photon flux and the decay angular
distribution,
\begin{widetext}
\begin{equation}
\frac{d\sigma}{dx_B\,dQ^2\,dt\,d\Omega}=\Gamma(x_B,Q^2,y)\,W(\Omega;\varepsilon,P_b),\qquad
\Gamma=\frac{\alpha_{\rm em}}{2\pi}\,\frac{y^2}{1-\varepsilon}\,\frac{1-x_B}{x_B}\,\frac{1}{Q^2},
\qquad d\Omega=d\cos\theta\,d\varphi\,d\Phi/2\pi,
\label{eq:dsigma}
\end{equation}
\end{widetext}
where $\Gamma$ is the flux of transverse virtual photons and $\alpha_{\rm em}$ the fine-structure
constant. The intensity $W$ is bilinear in the amplitudes and organizes into azimuthal modulations,
\begin{widetext}
\begin{equation}
W(\Omega;\varepsilon,P_b)=\frac{3}{4\pi}\Big[\,\mathcal U
+\varepsilon\,\mathcal L_2+\sqrt{2\varepsilon(1{+}\varepsilon)}\,\mathcal L_1
+P_b\big(\sqrt{1{-}\varepsilon^2}\,\mathcal B+\sqrt{2\varepsilon(1{-}\varepsilon)}\,\mathcal B_1\big)\Big],
\label{eq:W}
\end{equation}
\end{widetext}
where $\mathcal U$ is the unpolarized term independent of the production-plane azimuth $\Phi$,
$\mathcal L_2$ carries the $\cos2\Phi,\sin2\Phi$ linear-polarization modulations, $\mathcal L_1$ the
$\cos\Phi,\sin\Phi$ longitudinal--transverse interference, and $\mathcal B,\mathcal B_1$ the
beam-polarized terms ($P_b$ the beam polarization).

With a polarized target the angular distribution gains one angle and three structure-function
families. For target spin $(S_L,S_T,\Phi_S)$, with $S_L$ and $S_T$ the longitudinal and transverse
polarization components and $\Phi_S$ the azimuth of the transverse spin direction
(App.~\ref{app:polarized}), the intensity generalizes to
$W = W_{UU}+P_bW_{LU}+S_L(W_{UL}+P_bW_{LL})+S_T(W_{UT}+P_bW_{LT})$. The longitudinal blocks carry
the natural--unnatural interference structure functions $l$, which are linear in the unnatural
sector where the unpolarized ones are quadratic, and the transverse blocks the
nucleon-helicity-flip--sensitive families $s$ and $n$. The longitudinal-target terms therefore make
the unnatural-sector sign and phase observable, raising the identifiable content from 16 to 17
parameters, and sharpen its magnitude wherever that magnitude is small; the transverse-target terms
open the nucleon-helicity-flip sector and bring the total to 34. Both enlarged observable spaces,
and the mechanism behind the $l$ blocks, are constructed in App.~\ref{app:polarized}. The polarized
observables themselves are the $(\Phi-\Phi_S)$-weighted moments of $W$ over $\Omega$ together with
the spin-state rate asymmetries. The target polarization magnitude is written $P$ throughout, and
the beam polarization $P_b$; the closure studies of Sec.~\ref{sec:results} use $P=0.8$ and
$P_b=0.85$.

Each block is a sum of azimuthal $\times$ decay-angle functions whose coefficients
are bilinear in $T$ and $U$; the full set is given in App.~\ref{app:modulations}. The leading
term is the meson decay distribution, the transverse-meson ($\sin^2\!\theta$) and longitudinal-meson
($\cos^2\!\theta$) diagonals plus the $\cos\varphi,\cos2\varphi$ modulations,
\begin{widetext}
\begin{equation}
\mathcal U=\Big[\tfrac12\big(|T_{11}|^2{+}|T_{1\text{-}1}|^2{+}|U_{11}|^2{+}|U_{1\text{-}1}|^2\big)
+\varepsilon\big(|T_{10}|^2{+}|U_{10}|^2\big)\Big]\sin^2\!\theta
+\big(|T_{01}|^2{+}\varepsilon|T_{00}|^2{+}|U_{01}|^2\big)\cos^2\!\theta+\dots,
\label{eq:Udecay}
\end{equation}
\end{widetext}
where the displayed terms are illustrative; the complete angular decomposition is given in
App.~\ref{app:modulations}. The transverse and longitudinal cross sections, written throughout
without the explicit $t$ derivative where no ambiguity arises, are the two amplitude sums
\begin{widetext}
\begin{equation}
\sT=|T_{11}|^2+|T_{01}|^2+|T_{1\text{-}1}|^2+|U_{11}|^2+|U_{01}|^2+|U_{1\text{-}1}|^2,\qquad
\sL=|T_{00}|^2+2|T_{10}|^2+2|U_{10}|^2,
\label{eq:trace}
\end{equation}
\end{widetext}
reproducing the decomposition quoted after Eq.~\eqref{eq:eps}.

\subsection{Spin-density matrix representation and physical constraints}\label{sec:sdm}
The amplitudes assemble into the vector-meson spin-density matrix
$\rho$~\citep{Diehl:2007jy}. With $T$ and $U$ the $3\times3$ matrices of
natural- and unnatural-parity amplitudes in the meson and photon helicities,
\begin{equation}
\rho=TT^{\dagger}+UU^{\dagger},
\label{eq:rho}
\end{equation}
where the sum is incoherent for an unpolarized target
(App.~\ref{app:modulations}); target polarization turns on precisely
that interference, which is why it enlarges the identifiable content.
Each term is of the form $MM^{\dagger}$ and is therefore positive
semidefinite, as is their sum. Thus every $(T,U)$ yields an admissible
$\rho$: the inference is parameterized by the amplitudes, and $\rho$ is
constructed from them, with no positivity constraint to impose.
With both nucleon-helicity sectors the same construction holds
with an additional sum over the flip index (App.~\ref{app:polarized}).

The conventional angular description reports the normalized spin-density
matrix information through the 23 spin-density matrix elements
(SDMEs)~\citep{Schilling:1973ag} $\rsw{\alpha}{ij}$. These are derived
quantities, functions of $T$ and $U$, not independent inference
parameters; the absolute rate supplies the overall normalization. The
modulation blocks and the complete dictionary mapping SDMEs to amplitudes
are collected in Apps.~\ref{app:modulations},~\ref{app:sdme} and
Table~\ref{tab:dict}.

Each amplitude carries a distinct physics role. Collinear factorization is
established at leading twist for longitudinal photons: $T_{00}$
($\gL\!\to\!\VL$) is the leading-twist amplitude described by the GPD
factorization framework, for the $s\bar s$-dominated $\phi$ dominantly
through gluon exchange~\citep{Collins:1996fb}. The transverse $T_{11}$
($\gT\!\to\!\VT$) is formally power-suppressed in $1/Q$ yet sizable at
moderate $Q^2$; in phenomenological models both SCHC amplitudes proceed
through two-gluon exchange, with $T_{00}$ growing relative to $T_{11}$ with
$Q^2$~\citep{Goloskokov:2006hr}. The helicity-changing $T_{01}$, $T_{10}$,
and $T_{1\text{-}1}$ violate SCHC and are suppressed, entering through
transverse-momentum and higher-twist effects, with $T_{10}$ vanishing in
the leading gluon-GPD mechanism considered here (a falsifiable prediction
tested in Sec.~\ref{sec:results}); the unnatural-parity amplitudes
$U_{\mu\nu}$ are associated with quark-exchange mechanisms and are expected
to be small for the $s\bar s$-dominated $\phi$.

We extract the full nine-amplitude set without assuming natural parity.
Because $\phi$ production is expected to be dominated by natural-parity
(Pomeron) exchange, the unnatural amplitudes should be small; fitting them
keeps this expectation a test, not an assumption. In the closure test
below, the posterior correctly bounds them at their injected near-zero
values.

\subsection{Demonstration channels: exclusive \texorpdfstring{$\phi$ and $\rho^0$}{phi and rho0} electroproduction}\label{sec:channel}
We consider two vector-meson channels as complementary tests of the inference framework. In
$ep\to e'p'\phi$ with $\phi\to K^+K^-$ the $\phi$ is
nearly a pure $s\bar s$ state (branching ratio $\sim49.1\%$~\citep{ParticleDataGroup:2022pth}) whose production is
expected to be dominated by natural-parity exchange, so the unnatural-parity amplitudes should come
out small; this is the null test. In $ep\to e'p'\rho^0$ with $\rho^0\to\pi^+\pi^-$ the
unnatural-parity sector of the simulated truth is non-negligible and has to be recovered. A null
test on its own cannot separate an extraction that finds small amplitudes from one biased toward
zero, so both are used.

The positive decay daughter is the $K^+$ for the $\phi$ and the $\pi^+$ for the $\rho^0$, with the
angle conventions of Sec.~\ref{sec:xsec}. Nothing in the inference is tied to either channel, since
it uses only the vector-meson spin-density matrix and the two-pseudoscalar decay distribution;
another vector meson needs only its own forward model, with the meson and daughter masses, the
decay-angle definition and the acceptance re-specified. The main text presents the $\phi$; the
$\rho^0$ results are in App.~\ref{app:modeAB}.

\section{Bayesian inference framework}
\label{sec:framework}

With the physics fixed, the extraction reduces to a single question, which this section formulates
as a Bayesian inverse problem and then solves: given one kinematic bin's measured features and the
instrument that produced them, which amplitude sets remain admissible?

\subsection{Formulation of the inverse problem}
\label{sec:framework:inverse}

The physics is described by a finite vector of parameters $\theta$ (the generic symbol of the
inference literature, distinct from the decay angle of Sec.~\ref{sec:xsec}); in the present application
$\theta$ is the vector $A$ of 16 real amplitude parameters of Sec.~\ref{sec:xsec}, the real and
imaginary parts of the nine complex non-flip amplitudes with the global phase fixed by
$T_{11}\!\in\!\mathbb R_{\ge0}$ and $U_{11}$ real (App.~\ref{app:amp}). This is the precise
inference target, and the basis defining it rests on stated assumptions: the one-photon-exchange
approximation, the nucleon-helicity-non-flip sector probed by an unpolarized target, and a fixed
$x_B$ window. Within this basis the 16 parameters are exactly the number of directions the
observables determine: the Jacobian $\partial u/\partial A$ has numerical rank 16 at randomly
sampled points across the prior (App.~\ref{app:amp}). In the polarized-target modes of
Sec.~\ref{sec:results} the same construction enlarges $\theta$ to the 17- and 34-parameter sets of
App.~\ref{app:polarized} and the features gain the target-spin-dependent moments and per-spin-state
rates; nothing else in the analysis changes. The experiment does not
observe $\theta$ directly. The underlying distribution of events is a known function of $\theta$,
bilinear in the amplitudes, and every recorded event has passed through the response of the
apparatus: a kinematics-dependent acceptance and efficiency, and a finite resolution that migrates
events between measured values. We denote this instrument response collectively by $s$. The
recorded events in a given measurement (a kinematic bin, a beam energy) are compressed into a
vector of summary features $x$, here the event-averaged angular moments together with normalized
event rates defined below.

Stated in these terms, the three difficulties of the Introduction become precise. First, the map
$\theta \mapsto x$ is many-to-one: bilinear observables determine the parameters only up to
discrete and continuous ambiguities, so a point inversion is ill-posed and the honest answer is a
probability distribution over $\theta$. Second, the instrument response is entangled with the
physics: the observables are functions of both $\theta$ and $s$, and explicit unfolding of $s$ from
binned data is numerically unstable. Third, the parameters are subject to exact physical
constraints, here the positivity of the spin-density matrix of Sec.~\ref{sec:sdm}, that a generic
fitting procedure does not respect automatically.

The approach addresses all three. It requires only:
(i)~a validated forward simulation that maps a parameter point
$\theta$ and an instrument response $s$ to the distribution of summary
features $x$, including all instrumental effects;
(ii)~a finite parameter vector $\theta$ whose dimension matches the
information content of the observables; and
(iii)~measurable summary features $x$ computable identically on
simulation and on data. No likelihood needs to be tractable, no unfolding is
performed, and no invertibility of the forward map is assumed.

We call the extraction \emph{direct} in the sense that the amplitudes themselves are the
inferred parameters, not quantities reconstructed downstream from fitted observables. The
inference nonetheless proceeds through the explicit chain just enumerated: an amplitude
parameterization, a forward simulator, detector-level summary features, a prior, and a posterior
model. Each is stated and tested below, and each is a place where an assumption enters.

\subsection{Detector simulation}
\label{sec:toydet}

The formulation places no restriction on the detector geometry or design; to exercise it on a
concrete case we model the CLAS12 detector, whose several sub-systems combine into a
kinematics-dependent acceptance and resolution. Official detector simulations are not publicly
releasable; we use a structured parameterized acceptance $\eta$ that captures the salient
CLAS12-like features: an energy-dependent overall efficiency, a forward/backward $\cos\theta$ loss,
six-fold $\varphi$ sector gaps, and a mild production-plane modulation,
together with a realistic angular resolution that smears the helicity-frame angles
($\sigma_{\cos\theta}=0.05$, $\sigma_\varphi=\sigma_\Phi\approx5^\circ$). Both effects are folded
into the forward model, so the entire analysis below runs on a detector with acceptance and
resolution. A fast deeply-virtual-meson-production (DVMP) event generator produces full lab
four-vectors and the exclusive-production variables with
cross-section-weighted Poisson yields; the three beam energies populate
distinct $\varepsilon$ ranges, providing the Rosenbluth lever.

We summarize each bin by the 23 angular moments $\langle f_k\rangle\equiv\frac1N\sum_i f_k(\Omega_i)$,
the event-averages of known angular basis functions $f_k(\Omega)$ over the decay angles
$\Omega=(\cos\theta,\varphi,\Phi)$. The decay distribution is the intensity $W(\Omega)$ of Eq.~\eqref{eq:W}, a finite sum
$W(\Omega)\propto\sum_k c_k\,f_k(\Omega)$ over this basis whose coefficients $c_k$ are bilinear in the
amplitudes $T,U$ (linear combinations of the structure functions $u$); each moment is therefore a directly measurable
projection fixing one linear combination of these amplitude products (23 moments for the 23
independent combinations the angular shape determines, equivalently the normalized SDMEs reported by
experiments). The angular information is used unbinned: each $\langle f_k\rangle$ is a sample mean over individual
events, with no angular histogram. The extraction is performed within the fixed $x_B$ window ($0.08<x_B<0.5$), in $(\Qsq,t)$ bins at the three beam energies, exactly as in the unbinned-likelihood SDME analyses of
COMPASS~\citep{COMPASS:2022xig} and HERMES~\citep{HERMES:2009oim}. One caveat is stated explicitly: once the
acceptance is folded in, the basis functions lose their orthogonality and the 23 reconstructed
moments are no longer guaranteed to be sufficient statistics for the amplitudes. The empirical
measure of this information loss is the comparison with the likelihood fit of the same events,
unbinned in the decay angles within the same kinematic bins: the posterior matches its central
values, at a measured cost of a factor $2$--$3$ in single-bin interval width
(App.~\ref{app:uml}).

The choice of summary is not a structural feature of the formulation. Nothing in the formulation of
Sec.~\ref{sec:framework:inverse} requires a compression step: the conditioning input $x$ may be any
statistic computable identically on simulation and data, and the moments are used here because they
are the quantities the SDME literature already reports, they make the forward model analytic and
therefore fast, and their information content can be audited against the unbinned likelihood. A
richer summary, or event-level conditioning through a permutation-invariant encoder, substitutes
for $x$ without altering the parameterization, the calibration, or the detector conditioning. The
demonstration is thus at the moment level by construction and by choice; the inference is at the
amplitude level, and closing the residual gap between the two is a matter of the encoder, not of
the method.

Both detector effects, acceptance and resolution, are folded into the simulator;
neither is corrected out of the data. Taking the acceptance first: it is a property of the apparatus
alone, and following the likelihood convention we fold the
same $\eta$ into the simulator,
\begin{equation}
\langle f_k\rangle_{\rm acc}(u)=
\frac{\int f_k\,W(\Omega;u,\varepsilon)\,\eta(\Omega)\,d\Omega}
     {\int W(\Omega;u,\varepsilon)\,\eta(\Omega)\,d\Omega},
\label{eq:facc}
\end{equation}
evaluated from a flat (phase-space) sample passed through $\eta$, which enters only as an accept/reject
weight and need not be known analytically (on real data $\eta$ is the full detector Monte Carlo);
the measured moments are raw. Since $W$ is linear in the 28 structure functions $u$ (the bilinear
products of the amplitudes $T,U$), $\langle f_k\rangle_{\rm acc}=(Mu)_k/(b\!\cdot\!u)$ with
precomputed $M,b$, giving a fast forward model whose
acceptance-folded moments match the raw data moments to RMS $\approx10^{-3}$ (the method uses only
forward samples, not gradients; App.~\ref{app:impl}).

Finite resolution smears the measured decay angles
$(\cos\theta,\varphi,\Phi)$ and the kinematics $(\Qsq,x_B,t)$, and hence $\varepsilon$. It is handled exactly as the
acceptance, as part of the forward model, never deconvolved from the data. In the analytic moment
map (Eq.~\ref{eq:facc}) a resolution kernel simply smears the angular basis functions $f_k$, which
leaves the linear, differentiable form $(Mu)/(b\!\cdot\!u)$ intact; equivalently, and as on real
data, one draws truth amplitudes, applies the full detector response (acceptance, smearing,
and efficiency), and builds the conditioning features from the reconstructed quantities. Either way
the network is trained on (truth amplitude, reconstructed-level feature) pairs and inverts acceptance
and resolution together by construction, with none of the bin-migration instabilities of
explicit unfolding, the same property that makes the acceptance handling stable.
Accordingly, the nominal CLAS12-like detector model already includes a realistic angular resolution;
the main closure (Fig.~\ref{fig:polC}; App.~\ref{app:modeAB}) is
therefore obtained in the presence of both acceptance and resolution. For real data the same
treatment is implemented through the full detector Monte Carlo, which folds acceptance, efficiency, and
resolution together. Kinematic ($\Qsq,x_B,t$, and hence $\varepsilon$) resolution enters the same simulation as bin migration and is handled by binning
in the reconstructed variables.

The detector, finally, enters the analysis as a variable, no longer a fixed assumption. Instead of
training a network for one fixed detector, we draw the training detectors from a broad,
non-parametric family: random smooth fields (the log-acceptance is a random low-order harmonic
expansion in $(\cos\theta,\varphi,\Phi)$ passed through a logistic, times a random angular
resolution) and hard-edged multi-sector detectors with random sector counts ($4,5,7,8$), dead-gap
widths, and forward cuts, completed by a third class of extreme stress geometries whose sector
counts and cut depths are pushed beyond the nominal range. The family therefore spans smooth,
sharply-structured, and deliberately hostile acceptances, not a fixed-form parametric slice. The network is informed of each detector not
through a hand-built parameterization but through its measured response $s$: how it folds a pair of
fixed probe amplitudes into moments and rates, a quantity obtainable from Monte Carlo for any
detector. A single network thus learns the posterior $q_\psi(A\,|\,x,s)$ over the whole family, and
its transfer to detectors absent from training is a validation target in its own right
(Secs.~\ref{sec:val-det} and~\ref{sec:detamort}).

Two features make this conditioning workable at all; neither amounts to unfolding with an unknown
instrument. First, the response is not inferred blindly: $s$ identifies the detector to the
network, so the task is interpolation within a measured family. Second, the observables
over-constrain the amplitudes, 69 moments and three rates for 16 parameters (95 azimuthal
harmonics for 34 in the polarized program, App.~\ref{app:polarized}), so a response assumption
that drifts from the truth tends to surface as inconsistency among the redundant observables
instead of being absorbed silently into the extracted amplitudes; for a detector far outside the
family the ensemble is expected to register that inconsistency as an inflated model width
(Sec.~\ref{sec:val-det}). This redundancy is specific to working at the amplitude
level; a direct fit of individual SDMEs, with as many parameters as observables, would not have
it. On measured data the training family would in addition be centered on the experiment's own
simulation, with distortions of the GEMC response (acceptance cut-offs, harmonic modulations,
resolution scalings) in place of fully synthetic fields, so that the true instrument lies near
the middle of the trained family.

The family varies resolution as well as acceptance: every training detector carries its own
angular smearing, drawn together with its acceptance field, so the network is trained on the joint
fluctuation of the two, not on acceptance alone at a fixed resolution. This is what makes
resolution dilution tractable here. Finite resolution damps the measured angular coefficients, and
recovering the undamped amplitudes would ordinarily require unfolding against a known response.
The response is not unknown in this construction: it is supplied to the network as the measured
signature $s$, and the network has been trained across many acceptance$\,\times\,$resolution
realizations to invert the corresponding damping conditioned on $s$. The claim is therefore not
that an accurate instrument model is unnecessary, but that it enters as a conditioning input to a
single trained inference, not as a pre-fit correction to the measured events. What lies
outside this is error in $s$ itself, which is the simulation-fidelity systematic of
Sec.~\ref{sec:discussion} and is not carried by the quoted bands.

\subsection{Simulation-based Bayesian inference}
\label{sec:framework:sbi}

The inference target is the Bayesian posterior $p(A \mid x, s)$: the distribution of amplitude
values consistent with the measured features of one bin, given the instrument that produced them.
We learn this posterior by simulation-based
inference~\citep{Cranmer:2019eaq,Papamakarios:2018zoy,Brehmer:2018kdj}: the forward simulation generates
training pairs $(A, x)$ by drawing $A$ from a prior, simulating the experiment, and computing the
features, and the posterior of $A$ given $(x, s)$ is fit to these pairs.

The moments alone cannot fix the absolute scale: rescaling the overall magnitude of the cross
section ($u\to cu$, or equivalently $A\to\sqrt{c}\,A$, for any constant $c>0$) leaves every
$\langle f_k\rangle$ unchanged. The absolute scale is carried by the event rate. We append rate
features, the per-$\varepsilon$ reduced yield
$Y(\varepsilon)=N_{\rm obs}/(K_{\rm lumi}\bar\eta)=\sT+\varepsilon\sL$, with $K_{\rm lumi}$ the
luminosity normalization and $\bar\eta$ the mean acceptance; the identity holds up to an
acceptance-shape term carried identically by the forward model on simulation and data. The three
beam energies are $6.535$, $7.546$, and $10.6$~GeV, those of the existing CLAS12 RG-K and RG-A run
periods, so the multi-energy lever assumed here corresponds to recorded datasets rather than to a
hypothetical program. Together they trace out the Rosenbluth line: a straight line in $\varepsilon$
whose intercept is $\sT$ and whose slope is $\sL$. The conditioning vector is therefore
$x=[\,23\times3=69\text{ moments}\,|\,3\text{ rates}\,|\,3\ \varepsilon\text{ values}\,|\,P_b\,]$,
76 components in the unpolarized case, so the $L/T$
separation is obtained from the rates jointly with the angular fit, with no separate multi-energy
fitting step. The claim is stated precisely: the rate features carry the
information of the conventional Rosenbluth line, folded through the acceptance identically on
simulation and data, and no more; the network does not create
information beyond it, but combines the separation and the angular analysis into one coherent
posterior, so that the angular and rate uncertainties propagate jointly into every derived
quantity.

The last four entries are what free the trained network from the conditions of any particular run.
The virtual-photon polarization $\varepsilon$ and the beam polarization $P_b$ enter as continuous
conditioning inputs rather than as fixed properties of a bin: the acceptance-folded moment map is
stored exactly as a quadratic in $P_b$, a $P_b$-independent part plus a $P_b^{2}$ part, and is
assembled analytically at any $\varepsilon$, so every training example is drawn at its own
$(\varepsilon,P_b)$ instead of at a tabulated value. One trained ensemble therefore applies to a
dataset at any beam polarization and at any position along the $\varepsilon$ lever, in place of
silently assuming a single setting. In the polarized-target modes the target polarization $P$ joins
them, amortized over $0.5\le P\le0.9$, and the vector gains the spin-difference moments and the
per-spin-state rate asymmetries: 27 signed moments and one asymmetry per beam energy with a
longitudinal target (161 components), and a further 56 signed moments and one asymmetry per energy
for the combined longitudinal and transverse program (332 components; App.~\ref{app:polarized}).
What the network is never given is the bin's own $(\Qsq,t)$ coordinates: the amplitudes, and hence
the kinematic trends they trace, come from the per-bin observables alone.

Simulation-based inference of this kind has been applied to inverse problems in other fields, from
gravitational-wave parameter estimation~\citep{Dax:2021tsq} to dynamical-system
identification~\citep{zhu2026}; in hadronic physics, machine learning has been used for Compton
form factor extraction in deeply virtual Compton scattering~\citep{Almaeen:2024guo,Grigsby:2020auv} and for
partial-wave fitting in photoproduction~\citep{Glazier:2025emr}. To our knowledge the present work
is the first demonstration of simulation-based Bayesian inference of vector-meson helicity
amplitudes in which the detector response is folded into the forward model, the posterior
uncertainties are calibrated against simulation, and positivity is guaranteed by construction.

\subsection{Score-based diffusion implementation}
\label{sec:framework:diffusion}

\begin{figure*}[t]\centering
\includegraphics[width=\linewidth]{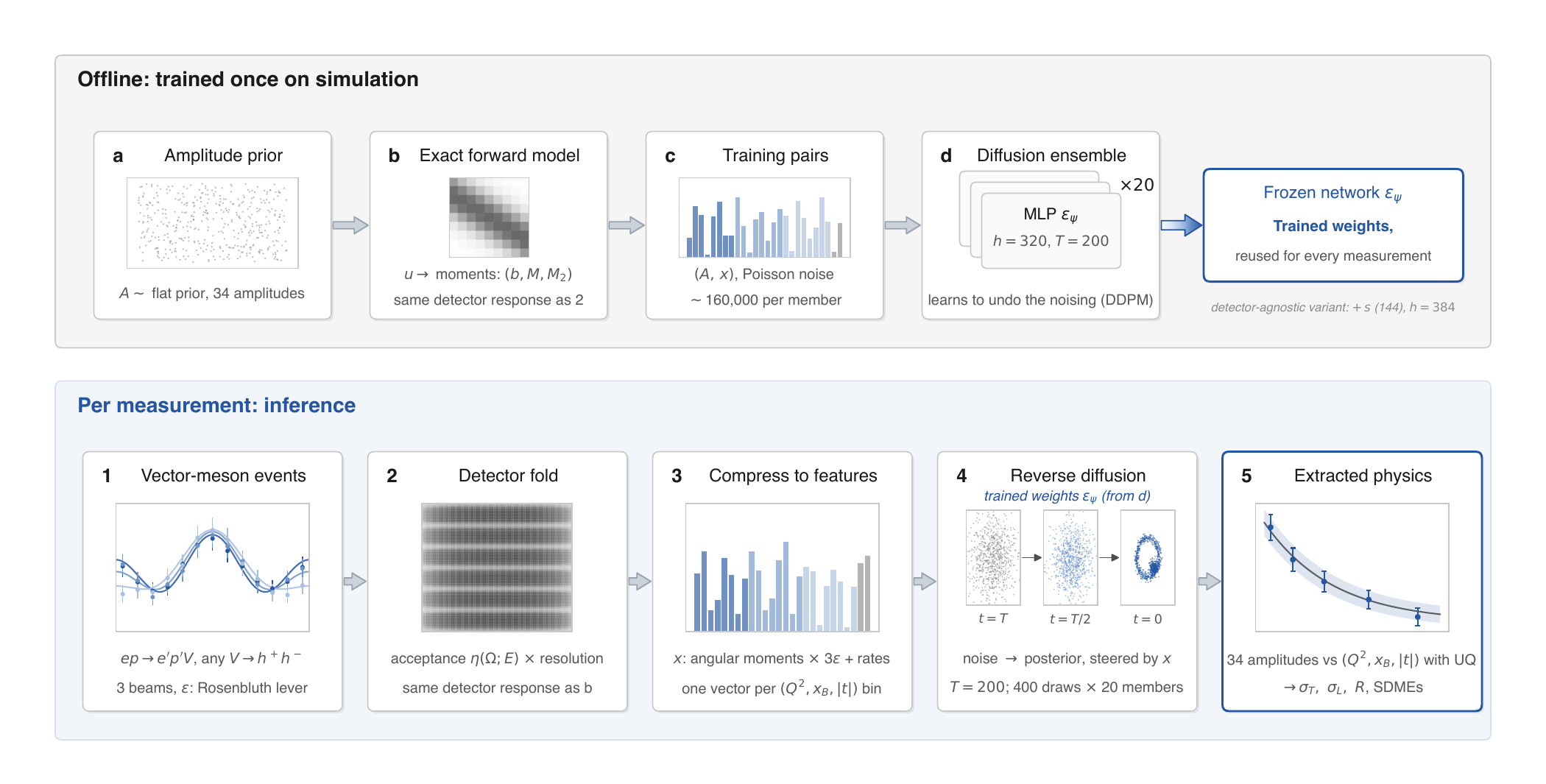}
\caption{Overview of the inference chain, illustrated for the vector-meson application of
Sec.~\ref{sec:physics}. Top row, offline stage, performed once: amplitude vectors (a), drawn from
a flat prior over the 34 real amplitude parameters of the full polarized program, both
nucleon-helicity sectors (App.~\ref{app:polarized}; 16 in the unpolarized case,
Sec.~\ref{sec:xsec}), are passed through the acceptance- and resolution-folded
forward model (b) into training pairs of amplitudes and detector-level features (c), on which the
conditional diffusion ensemble is trained (d); the product is a frozen network whose weights are
reused for every measurement. Bottom row, per-measurement stage: measured events (1) are folded
through the same detector response (2) and compressed into the conditioning features, the angular
moments at three $\varepsilon$ together with the event rates, one vector per $(\Qsq,x_B,t)$
bin (3); reverse diffusion steered by
these features maps noise to the amplitude posterior (4), which is propagated to
$\sT,\sL,R$ and the 23 SDMEs with calibrated uncertainties (5). The moments are scale-free, so the
absolute scale enters through the rate features.}
\label{fig:pipeline}
\end{figure*}

For the conditional posterior we use a score-based diffusion
model~\citep{Ho:2020epu,Song:2020hus}; Fig.~\ref{fig:pipeline} summarizes the resulting workflow, from the
one-time training stage to the per-measurement posterior. Three properties recommend it here. The training is a
stable regression, with no invertible architecture or tractable Jacobian required as in normalizing
flows; the representable posteriors are unrestricted in form and may be multimodal, unlike
mixture-density networks with a fixed mode count; and, because the network is trained once over the
whole prior, each new measurement costs a single conditional sampling pass, in contrast to per-bin
likelihood maximization or Markov-chain Monte Carlo, which must be rerun for every bin, with
Markov chains additionally mixing poorly when phases are nearly
degenerate~\citep{Betancourt:2017ebh}. (This train-once, apply-everywhere
property is known in the machine-learning literature as amortization.) Relative to an extended unbinned likelihood fit, the decisive addition is therefore not
statistical power in a single bin. The central values agree (Sec.~\ref{sec:results}), and the
compression of the events into moments and rates carries the interval-width cost quantified in
Sec.~\ref{sec:toydet}. The addition is the
combination: one training serves every bin and every detector in the family, the posterior is free
to be multimodal where the inversion is ambiguous, and the angular, rate, and detector information
propagate jointly in a single object. The method is not proposed as a replacement for maximum
likelihood on grounds of statistical efficiency; its advantage is representational, a calibrated
posterior over the physical amplitude space that retains the information the conventional chain
distributes across separate analysis stages.

Exact constraints are imposed by parameterization, never by penalty: the network operates on
the unconstrained parameter vector $A$ from which the physical objects are built in a manifestly
allowed form, the spin-density matrix assembled as $\rho = TT^{\dagger} + UU^{\dagger}$, positive
semidefinite for every $A$. Every posterior sample is therefore a physical state by construction,
and constraint boundaries deform the posterior instead of truncating it.

The model is trained on $(A_0,x)$ pairs from the forward model above (settings in Table~\ref{tab:diffusion}).
For the fixed nominal detector the instrument response $s$ is held constant and is suppressed in
the notation below; it re-enters as an explicit conditioning input, $z_\psi(A_t,t,x,s)$, in
the cross-detector application of Sec.~\ref{sec:detamort}. The forward process (fixed, used only
for training) corrupts a true amplitude vector $A_0$ over $T$ steps by repeatedly adding Gaussian
noise ($\mathcal N$ denotes the normal distribution),
\begin{widetext}
\begin{equation}
q(A_t\,|\,A_{t-1})=\mathcal N\!\left(\sqrt{1-\beta_t}\,A_{t-1},\,\beta_t I\right),\qquad
A_t=\sqrt{\bar\alpha_t}\,A_0+\sqrt{1-\bar\alpha_t}\,z,\;\;z\sim\mathcal N(0,I),
\label{eq:forward}
\end{equation}
\end{widetext}
with $\alpha_t\equiv1-\beta_t$ and $\bar\alpha_t=\prod_{s'\le t}\alpha_{s'}$, so $A_T$ is strongly
noised (for the schedule used, $\bar\alpha_T\approx0.13$; the residual signal fraction is absorbed
by the learned reverse process, and the calibration of Sec.~\ref{sec:framework:uq} covers the full
chain end to end) and the injected noise
$z$ is the regression target (we write the diffusion noise $z$ throughout, keeping $\varepsilon$
for the virtual-photon polarization of Sec.~\ref{sec:xsec}). The reverse process (learned) trains a network
$z_\psi(A_t,t,x)$ to predict that noise, conditioned on the measured features $x$, by minimizing
\begin{equation}
\mathcal L(\psi)=\mathbb E_{A_0\sim p(A),\;t,\;z}
\big\lVert\,z-z_\psi(A_t,t,x)\,\big\rVert^2
\label{eq:loss}
\end{equation}
over $(A_0,x)$ pairs produced by the acceptance- and resolution-folded simulator above; Eq.~\eqref{eq:loss}
is the (simplified) denoising score-matching objective. This objective is agnostic to the
underlying physics: it requires only joint samples $(A, x)$ from prior and forward map, with no need for a
tractable likelihood, invertibility, or analytic gradients of the simulator, and applies
unchanged whether the simulator describes meson production, phase-retrieval optics, or
quantum-state reconstruction; the physics enters through the construction of $x$ and the prior,
not through the objective. Knowing the noise is equivalent to knowing the
score, $\nabla_{\!A_t}\!\log q(A_t\,|\,x)=-z_\psi(A_t,t,x)/\sqrt{1-\bar\alpha_t}$, so the trained
network defines the learned reverse step $p_\psi(A_{t-1}\,|\,A_t,x)$. At inference we take the measured $x$ of a real bin, start from Gaussian
noise, and apply the reverse step $T$ times; the endpoint is one posterior sample, and many noise
seeds give the full posterior, propagated through $u=u(A)$ to the SDMEs and $\sT,\sL,R$. Trained once
over the whole prior, the network is then evaluated per bin in a single forward pass;
a flat prior augmented by a random overall magnitude ensures the training data cover the
small-amplitude (large-$|t|$) regime (Fig.~\ref{fig:priors}), with architecture and schedule in
App.~\ref{app:diffusion}.

\subsection{Posterior uncertainty quantification}
\label{sec:framework:uq}

A learned posterior carries no a priori guarantee that its credible intervals have the stated
coverage, so calibration is a required component of the analysis, not an afterthought. We validate
the posteriors with simulation-based calibration~\citep{Talts:2018zdk}; its construction, the
complementary fixed-truth coverage test, and the resulting coverage values are given in
Sec.~\ref{sec:val-sbc} and App.~\ref{app:unc}.

The reported uncertainty is decomposed with a deep
ensemble~\citep{Lakshminarayanan:2017tnw} of independently trained networks through the law of
total variance: the mean within-member variance is the statistical
component, driven by the finite event count entering the
features, while the variance of the member means bounds the model
component from training stochasticity and finite simulation
budget. The scope of this model term is stated precisely: the ensemble spread quantifies
optimization variability, finite training statistics, and network approximation error; it does not
represent unmodeled detector or physics effects, which enter separately as the simulation-fidelity
systematic (Sec.~\ref{sec:discussion}). Both the calibration and the decomposition are performed entirely
in simulation, before any data are analyzed, and the same machinery
quantifies where the posterior can be trusted as the instrument or the
parameter point moves away from the trained family.

\section{Validation strategy}
\label{sec:validation}

The analysis is validated along three independent axes before any result is quoted: the physics
input (a realistic Monte Carlo truth with known amplitudes), the instrument (transfer to detector
simulations absent from training), and the statistics (calibration of the posterior widths). Each
axis is described here; the quantitative outcomes are collected in Sec.~\ref{sec:results}.

\subsection{Physics-based Monte Carlo}\label{sec:physmc}
The physics truth model used for validation is built in the nucleon-helicity-non-flip amplitude
basis; the extraction itself assumes no production model, the network being trained on a flat
prior over the amplitudes. A realistic Monte Carlo
generator is nonetheless needed to validate the procedure against known input amplitudes with realistic
kinematic dependences, and is in this sense a component of the extraction framework rather than an
accessory to it~\citep{Avakian:2015vha}. For this purpose the amplitudes are modeled in a modified perturbative approach
that retains the transverse momentum of the quark--antiquark pair in the meson light-cone wave
function, replacing the collinear distribution amplitude, within the otherwise leading-twist
treatment~\citep{Bhattacharya:2026qnd}.
This description provides realistic kinematic dependences for the helicity amplitudes. Which
proton GPD and GTMD structures enter which amplitudes, and hence how sensitive the extraction is
to parton orbital angular momentum and spin--orbit correlations, remains the subject of ongoing
theoretical work and is not assumed here. The dominant
transitions, $T_{00}$ ($\gL\!\to\!\VL$) and $T_{11}$ ($\gT\!\to\!\VT$), together with the
helicity-changing $T_{01}$ ($\gT\!\to\!\VL$) and $T_{10}$ ($\gL\!\to\!\VT$), are assigned the
kinematic dependences of
Refs.~\citep{Lautenschlager:2013uya,Cuic:2023mki,Guo:2025muf,Guo:2024wxy,Anikin:2002wg,Goloskokov:2005sd,Goloskokov:2007nt,Goloskokov:2013mba,Bhattacharya:2026qnd}.
The extracted amplitudes then constrain the widths of the parton momentum distributions in the initial
state and in the final-state hadronization.

\subsection{Held-out detector simulations}\label{sec:val-det}
The decisive instrumental test is whether the single trained ensemble of Sec.~\ref{sec:toydet}
transfers, with no retraining, to detectors absent from its training family. Four levels of
difficulty are used. First, held-out random smooth acceptance fields drawn from the same family as
the training detectors. Second, CLAS12-like six-sector detectors with discrete azimuthal dead gaps
and sharp $\cos\theta$ and forward cuts: a structurally different geometry class, since the
training set contains $4,5,7,8$-sector detectors but no six-sector ones. Third, extreme $3,10,12$-sector stress
geometries far outside the trained sector range. Fourth, and most realistic, events folded through
a fast Monte Carlo (FastMC) surrogate of the official CLAS12 GEMC simulation. It follows the
factorized machine-learned detector-simulation strategy of Ref.~\citep{Darulis:2022brn}, in which
the acceptance is learned by probability classification and the reconstruction is modeled
separately: here a classifier network trained by binary cross-entropy returns the acceptance
probability over the full seven production and decay variables ($Q^2$, $x_B$, $t$, $\varepsilon$,
and the three decay angles $\cos\theta,\varphi,\Phi$), conditioned in addition on the run
configuration and the laboratory kinematics (App.~\ref{app:impl}), and a conditional mixture-density network
models the reconstructed-momentum residuals of the final-state particles, both trained on GEMC
events and reproducing their response. Folding through the surrogate in place of GEMC event
by event is a computational choice, not an idealization.

The detector response is complemented by a resolution stress test: the angular smearing is raised to $\approx14^\circ$, about three times the nominal value, and the network retrained (a changed
resolution defines a different forward model, so this is the one place a new network is required;
the transfer tests above reuse a single network across detectors).

The trained network is not expected to be universal: for a detector whose measured response falls
well outside the trained family, the ensemble (model) posterior width inflates, signaling reduced reliability instead of returning an over-confident estimate, and the case is addressed by adding
its structural class to the one-time training set.

\subsection{Simulation-based calibration}\label{sec:val-sbc}
The posterior widths are validated by simulation-based calibration~\citep{Talts:2018zdk}: truths drawn
from the prior are simulated, inferred, and ranked among their own posterior samples, and the rank
distribution is uniform only if the posterior is correctly calibrated, with the empirical coverage
of every credible level read off directly. This is complemented by a fixed-truth coverage test, in
which many independent statistical realizations of a single physics configuration are analyzed and
the pull distribution of the estimates against the known truth is examined. The statistical and
model components of the reported uncertainty are separated with the deep ensemble of
Sec.~\ref{sec:framework:uq}, and the resulting calibration factors and coverage values are reported
in App.~\ref{app:unc}.

\section{Results}
\label{sec:results}

\subsection{Main result: the complete observable polarized amplitude set}
\label{sec:modeC}
\begin{figure*}[!t]\centering
\includegraphics[width=\linewidth,height=0.92\textheight,keepaspectratio]{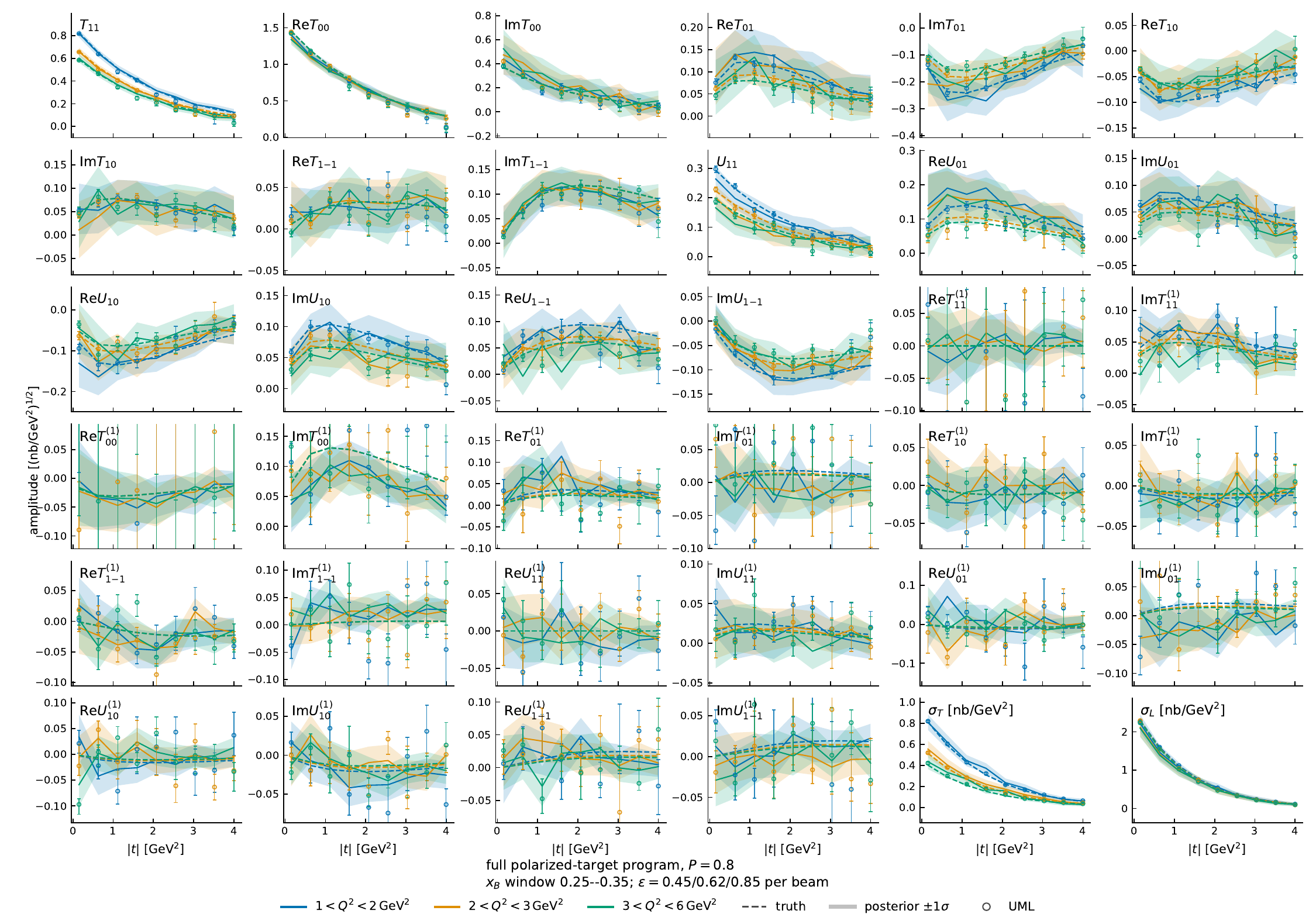}
\caption{\textbf{Full polarized-target extraction, $\phi\to K^+K^-$.} The 34-parameter helicity-amplitude set (the
16 nucleon-helicity-non-flip parameters and the 18 flip parameters $T^{(1)},U^{(1)}$;
App.~\ref{app:polarized}) and the derived $\sT,\sL$, extracted vs.\ $|t|$ for three $Q^2$ windows from the full polarized-target program (longitudinal and transverse settings at $P=0.8$, half the luminosity each, equal-luminosity spin states within each setting) in a full simulation closure. Signed posterior mean $\pm1\sigma$ (band) against the injected truth (dashed);
open circles are an independent unbinned maximum-likelihood fit of the same events, with
bootstrap percentile intervals (App.~\ref{app:uml}); the two agree point-by-point in the
non-flip sector, while in the flip sector the per-bin likelihood fit is the weaker estimator
for the reasons set out there. The sign and phase of the unnatural sector are measured.}
\label{fig:polC}
\end{figure*}

The central result is the extraction of the complete experimentally identifiable amplitude
content of exclusive vector-meson electroproduction\dash 34 parameters\dash from the decay angular
distributions and rates of the full polarized-target program, with half the luminosity on a
longitudinally polarized target and half transverse, in a full simulation closure. The count is the
36 real parameters of both nucleon-helicity sectors less the two unobservable directions, the
overall phase and the exact recoil rotation: the Jacobian of the $(u,l,s,n)$ structure functions
has rank 34 on the 36 components (Sec.~\ref{sec:xsec}; App.~\ref{app:polarized}). The longitudinal setting supplies the natural--unnatural interference (linear in the non-flip $U$ sector), the transverse setting the flip-sensitive structure functions, so the full 34-parameter set is constrained simultaneously, with all but the softest flip combination well determined (below). Figure~\ref{fig:polC} shows all 34 parameters and
the derived $\sT,\sL$ as functions of $|t|$ for three $Q^2$ windows: the signed posterior (band)
tracks the injected truth (dashed). The unnatural-parity
sector, its phase, and the nucleon-helicity-flip block are here measured quantities, and an
independent unbinned maximum-likelihood fit of the same events, the method used by COMPASS,
agrees with the posterior point-by-point (App.~\ref{app:uml}). The nucleon-flip amplitudes are determined within the exact recoil-rotation
ambiguity of App.~\ref{app:polarized}; the rotation-invariant combinations, in particular the
antisymmetric flip contractions ($\epsilon$-contractions, App.~\ref{app:polarized}), are determined
outright. Those amplitude combinations are what is measured; their identification with GPD-$E$
sensitivity is an interpretation that rests on the
factorization framework of App.~\ref{app:gpd} and is not part of the extraction. All 34 degrees of freedom are identifiable in principle; the posterior precision across them is
highly nonuniform, and the softest flip combination, $\mathrm{Re}\,T^{(1)}_{00}$, remains weakly
constrained (App.~\ref{app:uml}). The flip sector is
also where working at the level of the posterior matters most. Most of its 18 parameters are
suppressed by powers of $\sqrt{t'}/M_p$ (App.~\ref{app:polarized}) and all are weakly constrained
within any single bin, so the likelihood
geometry there is far from the quadratic regime in which a Gaussian error from the Hessian at the
optimum is a faithful summary; a posterior, which integrates over the poorly constrained directions
instead of maximizing along them, remains meaningful exactly where that approximation degrades
(App.~\ref{app:uml}). This is the
end point of a mode ladder whose earlier rungs are validated separately: the unpolarized 16-parameter
extraction and the longitudinally polarized 17-parameter extraction (where the
$U$-sector sign and phase first become observable) are presented in App.~\ref{app:modeAB}, and every
capability claimed there is a strict subset of the full extraction shown here.

\subsection{Consistency with the standard analysis}
The spin-density matrix elements of the standard analysis follow from the amplitude posterior as
derived quantities (errors propagated through $u=u(A)$), with SDME closure RMS $0.014$ over the bin
grid, and agree with the unbinned maximum-likelihood fit across the full grid: the per-SDME pull
between the two methods over all 23 SDMEs
and 15 bins has mean $-0.12$ and width $0.55$ (Fig.~\ref{fig:valextra}a, App.~\ref{app:uml}). The SCHC
consistency relations [Eq.~(27) of Ref.~\citep{COMPASS:2022xig}] are satisfied by the extracted set. The
angular likelihood is scale-free: it fixes the amplitude direction but not the absolute
$\sT,\sL$, which is exactly the information the rate features supply (App.~\ref{app:uml}).

\subsection{Transfer to a realistic detector}
\label{sec:detamort}
\begin{figure*}[t]\centering
\includegraphics[width=\linewidth]{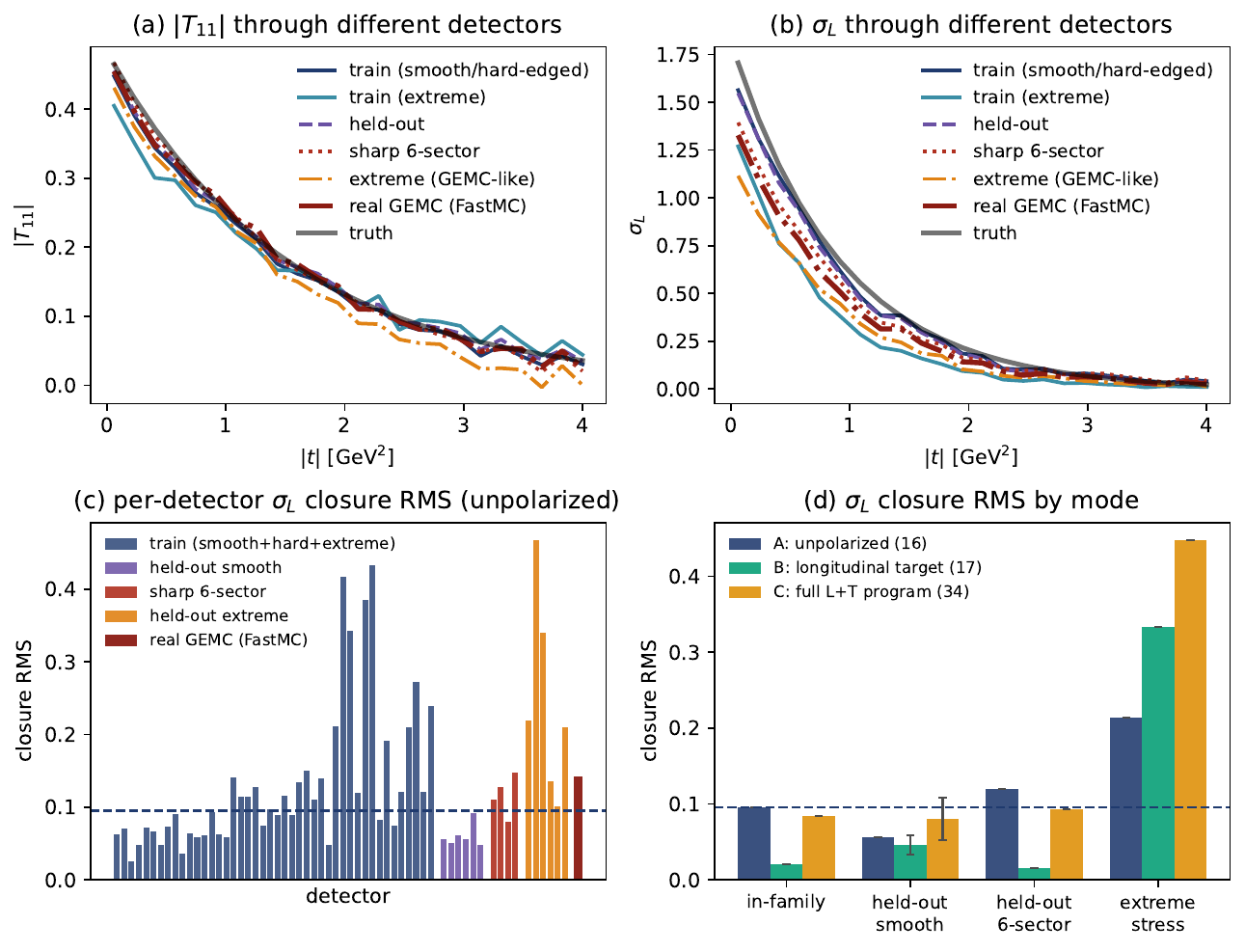}
\caption{Transfer of a single trained ensemble to a realistic detector simulation. One
$20$-member deep ensemble of diffusion posteriors $q_\psi(A\,|\,x,s)$ is trained over $16$ random
smooth acceptance$\,\times\,$resolution fields, $12$ hard-edged multi-sector detectors
(random $4,5,7,8$-sector geometries), and $16$ extreme stress geometries, conditioned on each
detector's measured response $s$ (its effect on fixed probe amplitudes), and then applied
unchanged to held-out smooth fields, held-out CLAS12-like six-sector detectors, extreme
$3,10,12$-sector stress geometries, and events folded through a fast Monte Carlo (FastMC) trained
on the official CLAS12 GEMC simulation (``real GEMC (FastMC)''), a detector absent from training.
(a),(b) Extracted $|T_{11}|$ and $\sL$ vs.\ $|t|$: all detectors, including the real-GEMC case (dark
red), track the injected truth (grey) (per-detector posterior bands are suppressed for legibility with
several detectors overlaid; the calibrated width is shown in Fig.~\ref{fig:closure} and
App.~\ref{app:unc}).
(c) Per-detector $\sL$ closure RMS: the real-GEMC case ($0.14$) lies at the level of the held-out
hard-edged six-sector detectors ($0.08$--$0.15$), above the held-out smooth fields
($0.048$--$0.091$) and well below the held-out extreme stress geometries (up to $0.47$); the dashed
line is the training-family median ($0.095$). The corresponding test for the full polarized
extraction is summarized in App.~\ref{app:modeAB}.}
\label{fig:detagnostic}
\end{figure*}

The detector enters the extraction only as a conditioning variable (Sec.~\ref{sec:toydet}); the
held-out tests of Sec.~\ref{sec:val-det} probe how far that conditioning carries.
Trained on the family of $16$ smooth fields, $12$ hard-edged multi-sector detectors, and $16$
extreme stress geometries, the same network extracts amplitudes without retraining through $6$
held-out smooth fields with per-detector $\sL$ closure RMS $0.048$--$0.091$ (mean $0.060$),
against a training-family median of $0.095$ (Fig.~\ref{fig:detagnostic}). The more stringent test
is structural: applied, with no retraining, to hard-edged CLAS12-like six-sector detectors held
out of training, it recovers $|T_{11}|$ and $\sL$ across the full $|t|$ range with closure RMS
$0.08$--$0.15$ (mean $0.12$), within a factor of two of the held-out smooth level and well below
the held-out extreme stress geometries (mean $0.25$). A detector geometry qualitatively different
from the smooth fields, and absent from training, therefore costs about a factor of two, well
short of an order of magnitude (Fig.~\ref{fig:detagnostic}c).

The most realistic test replaces the synthetic detectors with the GEMC-derived FastMC of
Sec.~\ref{sec:val-det}. Although this detector never appears in training, the extracted $|T_{11}|$ and
$\sL$ track the injected truth across the full $|t|$ range with closure RMS $0.14$, at the level
of the held-out hard-edged detectors and well below the extreme stress geometries
(Fig.~\ref{fig:detagnostic}): the network, trained only on parameterized detectors,
generalizes to a realistic detector simulation; the corresponding complete 16-parameter closure
through the GEMC-trained response, at the family-median level, is Fig.~\ref{fig:fmcfull}. A single network thus serves
the whole family with no per-detector retraining, applying to detectors whose measured response lies
within or near the trained family. The polarized ensembles are amortized over a detector family in
the same way, and the transfer carries over (App.~\ref{app:modeAB}).

We quantify this generalization on a large held-out set of $70$ detectors spanning the family,
including extreme $3,10,12$-sector stress geometries: the $\sL$ closure RMS has median $10.4\%$ of the
mean $\sL$ and stays bounded even for the most structurally distant detectors, demonstrating
generalization over the tested family in simulation (Fig.~\ref{fig:genconf}, App.~\ref{app:unc}). This
does not address the simulation-versus-data systematic, which sets the real-data accuracy
(Sec.~\ref{sec:discussion}).

\subsection{A falsifiable physics test}
Beyond reproducing the likelihood fit at the level of the standard observables, the extraction
returns the production amplitudes directly and with them an amplitude-level test, within
the nucleon-helicity-non-flip basis, of the helicity-selection rule associated with the leading
gluon-GPD mechanism (App.~\ref{app:gpd}), $\mathcal M^{LT}=0\Leftrightarrow T_{10}\equiv0$. The
extraction is discriminating, not merely consistent: injecting a truth that obeys the rule returns
$T_{10}$ consistent with zero (mean pull $0.4$, maximum $1.4\sigma$ across the grid), while injecting
one that violates it recovers the nonzero $T_{10}$ with correlation $0.97$. The same analysis
therefore both respects the prediction and would falsify it, which is the property an amplitude-level
test must have to be worth performing. The test is amplitude-level in that the extracted $T_{10}$ is
obtained without reference to a production model, though the rule it tests is itself a prediction of
one.
The assumptions behind the prediction are stated explicitly: $T_{10}=0$ is not a consequence of QCD
in general but of leading-twist collinear factorization with natural-parity two-gluon exchange in
the nucleon-helicity-non-flip sector, neglecting finite-mass and intrinsic-transverse-momentum
corrections; a measured nonzero $T_{10}$ falsifies this set of assumptions, not QCD. More broadly,
the extraction identifies the physical amplitude, not its microscopic origin: a nonzero
helicity-changing amplitude may receive contributions from intrinsic transverse momentum, the meson
wave function, higher-twist terms, finite-$t$ corrections, or residual nucleon-helicity-flip
contamination, and distinguishing among these requires the extracted kinematic dependence across $x_B$, $\Qsq$, and $t$ together with theory input.

\section{Discussion}
\label{sec:discussion}

The method does not replace the established analysis chain; it carries out its steps in a single inference. Where the angular
likelihood is well posed the posteriors coincide with the standard maximum-likelihood fit at the
level of the SDMEs (Sec.~\ref{sec:results}), while additionally delivering the amplitudes
themselves, the absolute $\sT,\sL$ scale from the same fit, and uncertainties checked by
simulation-based calibration. The established observables emerge as projections of a single
posterior, no longer as separate extractions. The method does not add information to the
experiment; it reorganizes the information already recorded into one object in which every
correlation is explicit.

Systematics that act on the data enter the analysis exactly as in a conventional measurement. The
non-exclusive and hyperon-production backgrounds are suppressed by exclusivity cuts, and radiative
corrections are applied by the standard procedure; both act on the measured yields and kinematics
and leave the extraction procedure unchanged. One channel-specific effect is different in kind: the
$K^+K^-$ S-wave continuum interfering with the $\phi$ under the mass peak (and, for $\rho^0$, the
analogous skewing of the resonance shape) is not an incoherent background to be subtracted but a
modification of the amplitude model itself. It is incorporated by adding the corresponding S-wave
amplitude to the same forward model; the inference is unchanged, and the contamination is fit
jointly instead of being assumed away.

On real data the reduced yields that carry the absolute scale depend on the luminosity, the
average efficiency, the decay branching fraction, radiative corrections, and background
subtraction, and their uncertainties are correlated across beam energies; a correlated
normalization offset between energies enters the Rosenbluth lever directly and is the classic
systematic of any $L/T$ separation. In the polarized-target settings the target polarization
magnitude and the dilution factor enter the same way, as normalization-like nuisance parameters on
the spin-dependent yields. These inputs will be introduced into the forward simulator as
nuisance parameters, perturbing the rate features within their uncertainties so that the posterior
inherits them; establishing that budget is part of the real-data analysis.

It is worth stating plainly which uncertainties the reported bands do and do not contain. Three
distinct terms enter an extraction of this kind. The statistical term, from the finite event count
of each bin, is in the bands and is calibrated by simulation-based calibration. The model term,
from network initialization, finite training statistics, and the approximation error of the learned
posterior, is in the bands through the deep ensemble and is measured directly (App.~\ref{app:unc}).
The third term, the fidelity of the forward simulation itself, is \emph{not} in the bands, and no
closure test can put it there: any mismatch between the simulated and the true instrument response
propagates into the extraction, exactly as it does in a conventional acceptance-corrected analysis,
and a closure test cannot see it because truth and instrument are supplied by the same simulator.
This systematic is external to the inference; it is not a defect of it. The method does not remove
simulation dependence, and we do not claim it does: it moves the detector model into the generative
likelihood, where its uncertainty can be stated and propagated through the inference, in place of a
fixed acceptance correction applied to the data before any fit. What the
detector-transfer studies of Sec.~\ref{sec:detamort} bound is the generalization of the network
across a family of detector responses, a different quantity. The most stringent of them, however,
deserves to be stated in measurement terms. GEMC is the official Geant4 simulation of the CLAS12
spectrometer, developed independently of this work and validated against beam data by the
collaboration's standard program; it is the response model on which published CLAS12 acceptance
corrections rest. The FastMC test of Sec.~\ref{sec:detamort} is therefore a synthetic-data
experiment: the events are produced through a detector response derived from GEMC, with its
acceptance and resolutions in all seven kinematic and decay variables, while the inference was trained only
on the parameterized family and never saw this response. The instrument that produces the data and
the instrument model implicit in the analysis are distinct objects, which is the separation that
defines a real measurement, realized here at the fidelity level the collaboration certifies for
physics analyses, up to the surrogate's percent-level reproduction of the GEMC response
(App.~\ref{app:impl}). What remains outside the test is the residual difference between GEMC and the
apparatus itself: the same residual carried by every published acceptance-corrected CLAS12
measurement, bounded there by the experiment's simulation-validation program, which applies to
this method unchanged. On real data that program supplies the third term explicitly, and its full
quantification remains the decisive step.

All results presented here are closure tests in simulation. Within that scope the validation is
comprehensive: realistic physics truths, independent statistical realizations, detectors absent
from training, and calibrated uncertainties. One of these tests survives the move to real data:
inferring the amplitudes, regenerating events from them, and comparing those events with the
measured ones requires no truth, and it closes here for both channels
(App.~\ref{app:closureloop}). Two stress tests remain natural next steps: extraction
under a deliberately misspecified detector response containing an effect absent from the training
family (a localized inefficiency, a non-Gaussian resolution tail, a correlated angular bias), and
a scan of the training prior, to which the smallest amplitudes and the phases are the most
sensitive. The enlarged parameter space of the polarized program also invites an adaptive training
strategy that concentrates simulations where they most sharpen the
posterior~\citep{jitao2026}, which controls the cost of the additional tiers.
Finally, the amplitude extraction is production-model independent within the chosen
basis, but its interpretation in terms of GPD or GTMD structures depends additionally on
factorization assumptions, the hard coefficients, the meson wave function, and evolution
(Sec.~\ref{sec:physmc}); the two steps are logically separate, and only the first is performed here.

The inverse problem solved here is not specific to vector-meson production, and this is the
method's principal claim to generality. Generality of the objective does not, however, exempt a
deployment from problem-specific requirements. Three must hold and are verified here: the summary
observables must retain the parameter information (the harmonic moments, near-sufficient because
the intensity is bilinear in the amplitudes); the prior's support and sampling density must cover
the physically realized range; and the training noise model must match the experimental
statistics. The calibration suite tests all three. Whenever the experimentally accessible quantities are
bilinear, or more generally nonlinear, functions of complex amplitudes, are observed only after an
instrument response, and describe a state confined to a constrained manifold, the same three
difficulties recur together: exact ambiguities that no amount of data removes, a response that
cannot be inverted before fitting, and a positivity constraint that a generic fit does not respect.
Inferring the constrained parameters as a posterior from a forward simulator addresses all three at
once, and requires of a new application only that its forward model be simulable. Partial-wave
analysis in photoproduction and spectroscopy shares the structure almost
exactly~\citep{GlueX:2023fcq,GlueX:2023pev}; phase retrieval and quantum-state reconstruction share
the ambiguity and positivity structure~\citep{fienup1982,shechtman2015}; the extraction of
parton distributions shares the folding of parameters through both a hard kernel and an apparatus.
Semi-inclusive deep-inelastic scattering is the nearest experimental neighbor: its azimuthal
asymmetries are instrument-folded projections of transverse-momentum-dependent distributions, and
the same forward-simulator-plus-posterior construction carries over with only the forward model
exchanged, as it does across exclusive channels, beam energies, and detector configurations at
Jefferson Lab and the EIC.
The vector meson is the stringent test case here, not the boundary of the method.

The step this establishes is methodological, a single
statistically consistent inference in place of the multi-step chain; the application to measured
data, with the systematic budget above, is where the method becomes a measurement.

\section{Conclusions}

Interpreting observables in hard exclusive vector-meson production in terms of 3D partonic distributions requires separating the underlying helicity amplitudes, in particular the amplitude for longitudinal vector-meson production by longitudinal virtual photons.
We have determined, in a full simulation closure, the complete experimentally identifiable
amplitude content of exclusive vector-meson electroproduction directly from the measured decay
angular distributions and event rates, using the combined longitudinal- and transverse-target
program. The counting is as follows. The reaction is described by 18 complex amplitudes, nine in
each of the two nucleon-helicity sectors, and therefore by 36 real components. Two of those
components are not observable. Because the recoil-nucleon helicity is summed over and never
measured, every observable bilinear is invariant under an overall phase and under one real
rotation $\mathcal R(\alpha)$ that mixes the flip and non-flip sectors. The remaining $36-2=34$ parameters
are what the polarized program determines, and what we extract here
(App.~\ref{app:polarized}). Positivity of the vector-meson spin-density matrix is imposed by construction, and the credible intervals are calibrated by simulation-based calibration, not asserted. The amplitudes are the primary quantity: the
separated cross sections $\sT$ and $\sL$, their ratio $R$, and the 23 spin-density matrix elements
follow as projections of one posterior, agreeing bin by bin with the injected truth and, wherever
the per-bin angular likelihood is well posed, with an independent unbinned maximum-likelihood
analysis of the same events. A realistic Monte Carlo
generator for the hadronic decays of vector mesons was a critical element of that validation: it
supplies the known input amplitudes against which the extracted ones are compared.
The longitudinal-to-transverse separation comes from the variation of the event rate with the
virtual-photon polarization, obtained jointly with the angular analysis and with no separate
Rosenbluth fit. The detector enters as an input, not as a correction: a single trained
ensemble, conditioned on the measured detector response, transfers without retraining to a fast
Monte Carlo derived from the official CLAS12 GEMC simulation, a response it never saw in training.
The unpolarized 16-parameter and longitudinally polarized 17-parameter extractions are exact
sub-sectors of this result, reproduced in closure (App.~\ref{app:modeAB}).

The same procedure carries to any vector meson with a two-pseudoscalar decay, with only the
forward model re-specified, as shown here for $\rho^0\!\to\pi^+\pi^-$ and applicable to
$K^{*}\!\to K\pi$ in the same way; the three tiers map directly onto unpolarized,
longitudinal-target, and combined longitudinal--transverse running. Beyond this channel, the
inference core is unchanged for any problem in which nonlinear, ambiguity-prone parameters are
measured through an instrument response, provided the deployment conditions of
Sec.~\ref{sec:discussion} are met. The immediate step is the test against direct GEMC simulation,
and application to real multi-energy CLAS12 data, which turns the closure demonstrated here into a
measurement, and through it into a map of the helicity amplitudes across $x_B$, $\Qsq$, and $t$.


\begin{acknowledgments}
We acknowledge the support of the ExtreMe Matter Institute (EMMI), whose Rapid Reaction Task Force
\emph{Impact of Vector Mesons on the Studies of the 3D Structure of the Nucleon} (CERN, June 2026)
is where the amplitude-level formulation and the inference machinery combined in this work were
brought together; we thank its organizers and participants.
We also thank Bakur Parsamyan for valuable discussions on the spin-density matrix elements. The schematic
figures and several figure refinements were prepared with the assistance of large language models.

The work of B.~Singh, H.~Avakian, V.~Burkert, L.~Elouadrhiri, and
V.~Kubarovsky was supported by the U.S.\ Department of Energy, Office of Science, Office of Nuclear
Physics under Contract No.\ 89243126CSC000213.
S.~B. acknowledges support from the U.S. National Science Foundation under Grant No.~PHY-2609761.
D.I.~Glazier is supported by the UK
Science and Technology Facilities Council under grant ST/V00106X/1.
The work of Y.~Li is partially supported by the U.S.\ Department of Energy, Office of Science, Office
of Nuclear Physics, Office of Advanced Scientific Computing Research through the Scientific Discovery
through Advanced Computing (SciDAC) program, under contracts DE-AC02-06CH11357, DE-AC05-06OR23177, and
DE-SC0023472, for the award \emph{Femtoscale Imaging of Nuclei using Exascale Platforms}. R.G. Milner and Y. Wang acknowledge support from the U.S. Department of Energy, Office of Nuclear Physics under Grant No.
DE-FG02-94ER40818. 

The code used for training, extraction, and the figures in this
work, together with machine-readable (CSV/NPZ) amplitude tables accompanying every closure figure,
will be made available at \url{https://github.com/sbhawani/diffusion-amplitude-extraction} upon
publication.
\end{acknowledgments}

\bibliography{bibliography}

\clearpage
\appendix
\section{Formalism}
This appendix fixes the conventions and gives the amplitude decomposition of the observables: the spin-density matrix elements, the full non-flip helicity-amplitude set, the azimuthal modulation blocks, the polarized-target and nucleon-helicity-flip formalism, and the gluon-GPD framework, which underlies only the physics interpretation and is used nowhere in the extraction.

Table~\ref{tab:symbols} collects the symbols used throughout, including the few glyphs whose
meaning is scoped to a single section.

\begin{table*}[t]\centering\footnotesize
\caption{Symbols used in this paper. Where a symbol is scoped to one section that scope is stated;
elsewhere the meaning is global. Note in particular that $\theta$ denotes the meson decay polar
angle everywhere except in Sec.~\ref{sec:framework:inverse}, and that $\mathcal M^{ab}$ of
App.~\ref{app:gpd} orders its helicity labels (photon, meson), the reverse of $T_{\mu\nu}$.}
\label{tab:symbols}
\begin{ruledtabular}
\begin{tabular}{@{}lp{0.78\textwidth}@{}}
\multicolumn{2}{@{}l}{\emph{Kinematics and angles}}\\
$\Qsq,\;x_B,\;t$ & photon virtuality, Bjorken variable, squared momentum transfer; $t'=t_{\rm min}-t$\\
$\varepsilon$ & virtual-photon polarization parameter, Eq.~\eqref{eq:eps}\\
$\theta,\;\varphi,\;\Phi$ & meson decay polar angle, decay azimuth, production-plane azimuth; $\Omega=(\cos\theta,\varphi,\Phi)$\\
$\Phi_S$ & azimuth of the transverse target spin about the virtual photon, measured from the lepton plane like $\Phi$\\
$\Gamma$ & flux of transverse virtual photons, Eq.~\eqref{eq:dsigma}\\
\\
\multicolumn{2}{@{}l}{\emph{Polarization}}\\
$P_b$ & beam polarization ($0.85$ in the closure studies)\\
$P$; $S_L,S_T$ & target polarization magnitude ($0.8$); its longitudinal and transverse components\\
\\
\multicolumn{2}{@{}l}{\emph{Amplitudes and observables}}\\
$T_{\mu\nu},\;U_{\mu\nu}$ & natural- and unnatural-parity helicity amplitudes; $\mu$ meson helicity, $\nu$ photon helicity\\
$\sigma,\;\lambda$; $f$ & final and initial nucleon helicity; flip index $f=|\sigma-\lambda|$, giving $T^{(0)},T^{(1)}$ (App.~\ref{app:polarized})\\
$\mathcal R(\alpha)$ & the exact recoil rotation mixing the $f{=}0$ and $f{=}1$ sectors (App.~\ref{app:polarized})\\
$u,\;l,\;s,\;n$ & the four bilinear structure-function families, Eq.~\eqref{eq:ulsn}\\
$\rho$; $\rsw{\alpha}{\lambda\lambda'}$ & vector-meson spin-density matrix; the 23 Schilling--Wolf SDMEs\\
$\sT,\;\sL$; $R$ & transverse and longitudinal cross sections; their ratio $R=\sL/\sT$\\
$\mathcal M^{ab}$ & leading-order amplitudes of App.~\ref{app:gpd} only; superscripts ordered (photon, meson)\\
\\
\multicolumn{2}{@{}l}{\emph{Inference}}\\
$A$ & the amplitude parameter vector (16, 17 or 34 real components)\\
$\theta$ & generic parameter vector of the inference literature, Sec.~\ref{sec:framework:inverse} only\\
$x$; $s$ & detector-level conditioning features (69 moments and 3 rates); measured instrument response\\
$f_k$; $\langle f_k\rangle$ & angular basis functions; their event averages, the angular moments\\
$\eta$; $M,\,b$ & detector acceptance; the acceptance-folded moment tensors of Eq.~\eqref{eq:facc}\\
$\psi$; $z,\,z_\psi$ & score-network weights; the diffusion noise and the network predicting it, Sec.~\ref{sec:framework:diffusion}\\
$T$; $\beta_t,\alpha_t,\bar\alpha_t$ & number of diffusion steps (Sec.~\ref{sec:framework:diffusion} and Table~\ref{tab:diffusion}), never the natural-parity amplitudes; the noise schedule\\
$K$; $N_{\rm cal}$ & deep-ensemble members ($K=20$); the calibration Monte Carlo sample size\\
\end{tabular}
\end{ruledtabular}
\end{table*}

\subsection{The 23 SDMEs: convention, classes, and amplitude decomposition}\label{app:sdme}
We follow the Schilling--Wolf formalism~\citep{Schilling:1973ag} in the notation of Diehl and
Sapeta~\citep{Diehl:2005pc} and HERMES/COMPASS~\citep{HERMES:2009oim,COMPASS:2022xig}. The vector-meson spin
density matrix $\rho_{\lambda_V\lambda_V'}$ is built from the virtual-photon density matrix
$\varrho^{U+L}=\varrho^U+P_b\,\varrho^L$ (unpolarized $U$ plus beam-polarized $L$) and decomposed into
nine matrices $\rho^{\alpha}$ with $\alpha=0$ (unpolarized $T$), $1,2$ (linear polarization), $3$
(circular), $4$ (longitudinal), and $5$--$8$ ($T$/$L$ interference). Since $\sT$ and $\sL$ cannot be
separated within one beam energy, one reports the normalized combinations
\begin{widetext}
\begin{equation}
\rsw{04}{\lambda\lambda'}=\frac{\rho^{0}_{\lambda\lambda'}+\varepsilon R\,\rho^{4}_{\lambda\lambda'}}
{1+\varepsilon R},\qquad
\rsw{\alpha}{\lambda\lambda'}=\begin{cases}
\rho^{\alpha}_{\lambda\lambda'}/(1+\varepsilon R), & \alpha=1,2,3,\\[2pt]
\sqrt R\,\rho^{\alpha}_{\lambda\lambda'}/(1+\varepsilon R), & \alpha=5,6,7,8,
\end{cases}
\label{eq:rdef}
\end{equation}
\end{widetext}
with $R=\sL/\sT$. There are 23 such SDMEs for an unpolarized target: 15 from the unpolarized beam and 8
coupled to the beam polarization (a polarized target enlarges the set, App.~\ref{app:polarized}). In our convention they follow from the 28 unnormalized structure functions
$\ueps{\mu\mu'}{\nu\nu'}=T_{\mu\nu}T_{\mu'\nu'}^{*}+U_{\mu\nu}U_{\mu'\nu'}^{*}$ (meson helicities $\mu\mu'$ as superscripts, photon helicities $\nu\nu'$ as subscripts) as
$\rsw{\alpha}{\lambda\lambda'}=(u\text{-combination})_\alpha/(\sT+\varepsilon\sL)$ (the
meson-diagonal transverse pairs enter only through the sums $\ueps{++}{\nu\nu}{+}\ueps{--}{\nu\nu}$
appearing below); representative
entries are $\rsw{04}{00}=(\ueps{00}{++}+\varepsilon\ueps{00}{00})/(\sT+\varepsilon\sL)$,
$\rsw{1}{00}=\ueps{00}{-+}/(\cdot)$, $\rsw{5}{00}=-\sqrt2\,\mathrm{Re}\,\ueps{00}{0+}/(\cdot)$, and the
complete map is implemented (and unit-tested) in the released code.

Each helicity amplitude decomposes into natural- and unnatural-parity exchange,
$F=T+U$~\citep{Schilling:1973ag,Diehl:2005pc}; in the sector used here (nucleon-helicity non-flip) the independent amplitudes are the five
natural-parity $T_{11},T_{00},T_{01},T_{10},T_{1\text{-}1}$ and the four unnatural-parity
$U_{11},U_{01},U_{10},U_{1\text{-}1}$ (App.~\ref{app:amp}). Table~\ref{tab:classes} lists
all 23 SDMEs in the five COMPASS classes together with the amplitude product that dominates each.

\begin{table*}[t]\centering\small
\caption{The 23 SDMEs grouped into the COMPASS/HERMES classes A--E by helicity transition, with the
dominant helicity-amplitude product. Starred (${}^{\star}$) elements are the eight beam-polarized
SDMEs ($\alpha=3,7,8$). $T_{11}\!:\gamma^*_T\!\to V_T$, $T_{00}\!:\gamma^*_L\!\to V_L$ (the SCHC
amplitudes); $T_{01}\!:\gamma^*_T\!\to V_L$, $T_{10}\!:\gamma^*_L\!\to V_T$,
$T_{1\text{-}1}\!:\gamma^*_{-T}\!\to V_T$.}
\label{tab:classes}
\resizebox{\textwidth}{!}{%
\begin{tabular}{@{}llll@{}}
\toprule
Class & Transition & SDMEs & Dominant amplitude product \\
\midrule
A & $\gamma^*_L\!\to V_L,\ \gamma^*_T\!\to V_T$ &
  $\rsw{04}{00},\ \rsw{1}{1\text{-}1},\ \mathrm{Im}\,\rsw{2}{1\text{-}1}$ &
  $|T_{00}|^2,\ |T_{11}|^2$ \\
B & interference of the two above &
  $\mathrm{Re}\,\rsw{5}{10},\ \mathrm{Im}\,\rsw{6}{10},\ \mathrm{Im}\,\rsw{7}{10}{}^{\star},\ \mathrm{Re}\,\rsw{8}{10}{}^{\star}$ &
  $\mathrm{Re/Im}\,(T_{11}T_{00}^{*})$ \\
C & $\gamma^*_T\!\to V_L$ &
  $\mathrm{Re}\,\rsw{04}{10},\ \mathrm{Re}\,\rsw{1}{10},\ \mathrm{Im}\,\rsw{2}{10},\ \rsw{5}{00},\ \rsw{1}{00},\ \mathrm{Im}\,\rsw{3}{10}{}^{\star},\ \rsw{8}{00}{}^{\star}$ &
  $T_{01}T_{11}^{*},\ T_{01}T_{00}^{*}$ \\
D & $\gamma^*_L\!\to V_T$ &
  $\rsw{5}{11},\ \rsw{5}{1\text{-}1},\ \mathrm{Im}\,\rsw{6}{1\text{-}1},\ \mathrm{Im}\,\rsw{7}{1\text{-}1}{}^{\star},\ \rsw{8}{11}{}^{\star},\ \rsw{8}{1\text{-}1}{}^{\star}$ &
  $T_{10}T_{11}^{*}$ \\
E & $\gamma^*_{-T}\!\to V_T$ &
  $\rsw{04}{1\text{-}1},\ \rsw{1}{11},\ \mathrm{Im}\,\rsw{3}{1\text{-}1}{}^{\star}$ &
  $T_{1\text{-}1}T_{11}^{*}$ \\
\bottomrule
\end{tabular}}
\end{table*}

Under SCHC only the diagonal transitions survive ($T_{00},T_{11}$ nonzero), so classes C, D, E vanish
and the class-A/B elements obey, e.g.,
$\rsw{1}{1\text{-}1}=-\,\mathrm{Im}\,\rsw{2}{1\text{-}1}$,
$\mathrm{Re}\,\rsw{5}{10}=-\,\mathrm{Im}\,\rsw{6}{10}$,
$\mathrm{Im}\,\rsw{7}{10}=\mathrm{Re}\,\rsw{8}{10}$~\citep{Schilling:1973ag,COMPASS:2022xig}. Nonzero class-C
elements (in particular for $\gamma^*_T\!\to V_L$) are the standard signature of SCHC violation and, in
the Goloskokov--Kroll picture, probe chiral-odd (transversity) GPDs~\citep{Goloskokov:2006hr}. The longitudinal-to-transverse
ratio is obtained from the same amplitudes,
{\small
\begin{equation}
R=\frac{\sL}{\sT}=\frac{|T_{00}|^2+2|T_{10}|^2+2|U_{10}|^2}
{|T_{11}|^2{+}|T_{01}|^2{+}|T_{1\text{-}1}|^2{+}|U_{11}|^2{+}|U_{01}|^2{+}|U_{1\text{-}1}|^2},
\end{equation}}
and the $T_{11}$--$T_{00}$ phase from the class-B interference SDMEs.

\subsection{Helicity amplitudes: the full non-flip set}\label{app:amp}
For nucleon-helicity non-flip there are $3\times3=9$ complex amplitudes $F_{\mu\nu}(x_B,\Qsq,t)$ (meson helicity
$\mu$, photon helicity $\nu\in\{-1,0,1\}$). Parity separates them into natural- and unnatural-parity
exchange, $F=T+U$, with $T_{-\mu,-\nu}=+(-1)^{\mu-\nu}T_{\mu\nu}$ (natural) and
$U_{-\mu,-\nu}=-(-1)^{\mu-\nu}U_{\mu\nu}$ (unnatural). The independent set is therefore the five natural
amplitudes $\{T_{11},T_{00},T_{01},T_{10},T_{1\text{-}1}\}$ and the four unnatural amplitudes
$\{U_{11},U_{01},U_{10},U_{1\text{-}1}\}$, with $U_{00}=0$ forced by the unnatural relation. After fixing
the global phase ($T_{11}\!\in\!\mathbb R_{\ge0}$), one phase direction remains undetermined by the
unpolarized observables: $u=TT^{\dagger}+UU^{\dagger}$ is exactly invariant under a global phase
rotation of the entire $U$ block, $U_{\mu\nu}\to e^{i\beta}U_{\mu\nu}$, so precisely one relative
phase is unobservable, which we fix
by taking $U_{11}$ real; the resulting 16 real parameters are exactly the number of directions
identifiable from the unnormalized structure functions, i.e.\ the normalized SDMEs together with
the overall rate (the Jacobian $\partial u/\partial A$ has numerical rank 16 at randomly sampled
points across the prior). For an
unpolarized target the structure functions are the incoherent sum over the two parities,
$u^{\mu\mu'}_{\nu\nu'}=T_{\mu\nu}T_{\mu'\nu'}^{*}+U_{\mu\nu}U_{\mu'\nu'}^{*}$ (no $T$--$U$
interference), and the cross sections are those of Eq.~\eqref{eq:trace}. The map is verified against
the SCHC limit $\rsw{04}{00}=\varepsilon R/(1+\varepsilon R)$, positivity, global-phase invariance, and
the rank test above.

\subsection{Cross section: azimuthal modulation blocks}\label{app:modulations}
The intensity of Eq.~\eqref{eq:W} splits into a leading decay term and azimuthal modulations, each a
product of a $\Phi$-azimuthal function, a decay-angle function in $(\theta,\varphi)$, and a coefficient
bilinear in the amplitudes. There is no natural--unnatural interference for an unpolarized target, so
every coefficient is a sum of a $T$-bilinear and a $U$-bilinear part; because $U$ obeys the opposite
parity relation, the $U$-part is not obtained from the $T$-part by $T\!\to\!U$ (several signs
flip, and terms with $T_{00}$ have no $U$ partner since $U_{00}=0$). An immediate consequence is an
exact discrete ambiguity: every coefficient is quadratic in the unnatural sector with no linear
term, so the intensity is invariant under a global sign flip of the whole unnatural block,
$U_{\mu\nu}\to-U_{\mu\nu}$. Relative signs and phases within $U$ are measurable, but the
overall $U$ sign is not an observable of unpolarized data (fixing $U_{11}$ real removes the continuous
phase freedom, not this discrete one); we therefore quote all unnatural amplitudes in the convention
$U_{11}\ge 0$, folding the exactly bimodal posterior onto this branch before any $U$ statistic is
taken. The ambiguity is physical, not a limitation of the method: for a polarized target the
single-spin structure functions are the natural--unnatural interferences
$l^{\mu\mu'}_{\nu\nu'}=T_{\mu\nu}U^{*}_{\mu'\nu'}+U_{\mu\nu}T^{*}_{\mu'\nu'}$, linear in $U$
and odd under $U\to-U$, so target polarization measures the sign itself~\citep{Diehl:2007jy} (App.~\ref{app:polarized}). The blocks are ($T_{11},U_{11}$
real throughout):
\begin{widetext}
\begin{align}
\mathcal U=&\Big[\tfrac12\big(\aT{11}{+}\aT{\dmo}{+}\aU{11}{+}\aU{\dmo}\big)+\varepsilon\big(\aT{10}{+}\aU{10}\big)\Big]\sin^2\!\theta
+\big(\aT{01}{+}\varepsilon\aT{00}{+}\aU{01}\big)\cos^2\!\theta\notag\\
&-\sqrt2\,A_{10}\sin2\theta\cos\varphi-A_{\dmo}\sin^2\!\theta\cos2\varphi,\label{eq:blockU}\\[2pt]
\mathcal L_2=&\cos2\Phi\big[-B^c_{11}\sin^2\!\theta-B^c_{00}\cos^2\!\theta+\sqrt2 B^c_{10}\sin2\theta\cos\varphi+B^c_{\dmo}\sin^2\!\theta\cos2\varphi\big]\notag\\
&-\sin2\Phi\big[\sqrt2 B^s_{10}\sin2\theta\sin\varphi+B^s_{\dmo}\sin^2\!\theta\sin2\varphi\big],\\[2pt]
\mathcal L_1=&\cos\Phi\big[C^c_{11}\sin^2\!\theta+C^c_{00}\cos^2\!\theta-\sqrt2 C^c_{10}\sin2\theta\cos\varphi-C^c_{\dmo}\sin^2\!\theta\cos2\varphi\big]\notag\\
&+\sin\Phi\big[\sqrt2 C^s_{10}\sin2\theta\sin\varphi+C^s_{\dmo}\sin^2\!\theta\sin2\varphi\big],\\[2pt]
\mathcal B=&\sqrt2\,D_{10}\sin2\theta\sin\varphi+D_{\dmo}\sin^2\!\theta\sin2\varphi,\\[2pt]
\mathcal B_1=&\cos\Phi\big[\sqrt2 E^c_{10}\sin2\theta\sin\varphi+E^c_{\dmo}\sin^2\!\theta\sin2\varphi\big]\notag\\
&+\sin\Phi\big[E^s_{11}\sin^2\!\theta+E^s_{00}\cos^2\!\theta-\sqrt2 E^s_{10}\sin2\theta\cos\varphi-E^s_{\dmo}\sin^2\!\theta\cos2\varphi\big].
\end{align}
\end{widetext}
The coefficients are the unnormalized Diehl structure-function combinations, so the Schilling--Wolf SDMEs
are $\rsw{\alpha}{ij}=(\text{coefficient})/(\sT+\varepsilon\sL)$; the leading $\sin^2\!\theta$ and
$\cos^2\!\theta$ coefficients are the transverse- and longitudinal-meson diagonals
$\tfrac12[(\ueps{++}{++}{+}\ueps{--}{++})+\varepsilon(\ueps{++}{00}{+}\ueps{--}{00})]$ and
$\ueps{00}{++}{+}\varepsilon\ueps{00}{00}$. Table~\ref{tab:dict} gives every coefficient as its
Schilling--Wolf SDME, its unnormalized $u$-combination, and its decomposition in amplitude products of $T$ and $U$.

\begin{table*}[t]\centering\footnotesize
\caption{Map relating each Schilling--Wolf SDME to its unnormalized structure-function combination $u$
and to its amplitude products $(T,U)$, for all 23 modulation coefficients (block/coefficient of
App.~\ref{app:modulations}; $c/s$ superscripts are the $\cos/\sin$ terms). The normalized SDME is
$\rsw{\alpha}{ij}=(\text{$u$-combination})/(\sT+\varepsilon\sL)$. Starred rows ($\alpha=3,7,8$) are
beam-polarized. $T_{11},U_{11}$ real.}
\label{tab:dict}
\setlength{\tabcolsep}{10pt}\renewcommand{\arraystretch}{1.25}
\begin{tabular}{@{}lll p{8.5cm}@{}}
\toprule
coeff. & SW SDME & $u$-combination & amplitude products $(T,U)$\\
\midrule
\multicolumn{4}{@{}l}{$\mathcal U$: unpolarized, $\Phi$-independent}\\
$A_{00}$ & $\rsw{04}{00}$ & $\ueps{00}{++}{+}\varepsilon\ueps{00}{00}$ & $\aT{01}+\varepsilon\aT{00}+\aU{01}$\\
$A_{10}$ & $\mathrm{Re}\,\rsw{04}{10}$ & $\tfrac12\mathrm{Re}(\ueps{0+}{++}{-}\ueps{-0}{++})+\varepsilon\mathrm{Re}\,\ueps{0+}{00}$ & $\varepsilon\RT{00}{10}+\tfrac12\RT{01}{11}-\tfrac12\RT{01}{\dmo}+\tfrac12\RU{01}{11}+\tfrac12\RU{01}{\dmo}$\\
$A_{\dmo}$ & $\rsw{04}{\dmo}$ & $\mathrm{Re}\,\ueps{-+}{++}+\varepsilon\mathrm{Re}\,\ueps{-+}{00}$ & $\RT{11}{\dmo}-\varepsilon\aT{10}-\RU{11}{\dmo}+\varepsilon\aU{10}$\\
\midrule
\multicolumn{4}{@{}l}{$\mathcal L_2$: linear polarization ($\cos2\Phi,\sin2\Phi$)}\\
$B^c_{11}$ & $\rsw{1}{11}$ & $\mathrm{Re}\,\ueps{++}{-+}$ & $\RT{11}{\dmo}+\RU{11}{\dmo}$\\
$B^c_{00}$ & $\rsw{1}{00}$ & $\ueps{00}{-+}$ & $-\aT{01}+\aU{01}$\\
$B^c_{10}$ & $\mathrm{Re}\,\rsw{1}{10}$ & $\tfrac12\mathrm{Re}(\ueps{0+}{-+}{+}\ueps{+0}{-+})$ & $-\tfrac12\RT{01}{11}+\tfrac12\RT{01}{\dmo}+\tfrac12\RU{01}{11}+\tfrac12\RU{01}{\dmo}$\\
$B^c_{\dmo}$ & $\rsw{1}{\dmo}$ & $\tfrac12(\ueps{-+}{-+}{+}\ueps{+-}{-+})$ & $\tfrac12(\aT{11}{+}\aT{\dmo})-\tfrac12(\aU{11}{+}\aU{\dmo})$\\
$B^s_{10}$ & $\mathrm{Im}\,\rsw{2}{10}$ & $\tfrac12\mathrm{Re}(\ueps{+0}{-+}{-}\ueps{0+}{-+})$ & $\tfrac12\RT{01}{11}+\tfrac12\RT{01}{\dmo}-\tfrac12\RU{01}{11}+\tfrac12\RU{01}{\dmo}$\\
$B^s_{\dmo}$ & $\mathrm{Im}\,\rsw{2}{\dmo}$ & $\tfrac12(\ueps{+-}{-+}{-}\ueps{-+}{-+})$ & $\tfrac12(\aT{\dmo}{-}\aT{11})+\tfrac12(\aU{11}{-}\aU{\dmo})$\\
\midrule
\multicolumn{4}{@{}l}{$\mathcal L_1$: L--T interference ($\cos\Phi,\sin\Phi$)}\\
$C^c_{11}$ & $\rsw{5}{11}$ & $-\tfrac{1}{\sqrt2}\,\mathrm{Re}(\ueps{++}{0+}{+}\ueps{--}{0+})$ & $\tfrac1{\sqrt2}(\RT{10}{\dmo}{-}\RT{10}{11})+\tfrac1{\sqrt2}(\RU{10}{\dmo}{-}\RU{10}{11})$\\
$C^c_{00}$ & $\rsw{5}{00}$ & $-\sqrt2\,\mathrm{Re}\,\ueps{00}{0+}$ & $-\sqrt2\,\RT{00}{01}$\\
$C^c_{10}$ & $\mathrm{Re}\,\rsw{5}{10}$ & $\tfrac{1}{2\sqrt2}\mathrm{Re}\big[(\ueps{0-}{0+}{-}\ueps{+0}{0+})-(\ueps{0+}{0+}{-}\ueps{-0}{0+})\big]$ & $\tfrac{1}{2\sqrt2}(\RT{00}{\dmo}{-}\RT{00}{11}{-}2\RT{01}{10})$\\
$C^c_{\dmo}$ & $\rsw{5}{\dmo}$ & $-\tfrac{1}{\sqrt2}\,\mathrm{Re}(\ueps{-+}{0+}{+}\ueps{+-}{0+})$ & $\tfrac1{\sqrt2}(\RT{10}{11}{-}\RT{10}{\dmo})+\tfrac1{\sqrt2}(\RU{10}{\dmo}{-}\RU{10}{11})$\\
$C^s_{10}$ & $\mathrm{Im}\,\rsw{6}{10}$ & $\tfrac{1}{2\sqrt2}\mathrm{Re}\big[(\ueps{0+}{0+}{-}\ueps{-0}{0+})+(\ueps{0-}{0+}{-}\ueps{+0}{0+})\big]$ & $\tfrac{1}{2\sqrt2}(\RT{00}{11}{+}\RT{00}{\dmo})-\tfrac1{\sqrt2}\RU{01}{10}$\\
$C^s_{\dmo}$ & $\mathrm{Im}\,\rsw{6}{\dmo}$ & $\tfrac{1}{\sqrt2}\,\mathrm{Re}(\ueps{-+}{0+}{-}\ueps{+-}{0+})$ & $-\tfrac1{\sqrt2}(\RT{10}{11}{+}\RT{10}{\dmo})+\tfrac1{\sqrt2}(\RU{10}{11}{+}\RU{10}{\dmo})$\\
\midrule
\multicolumn{4}{@{}l}{$\mathcal B$: beam spin, $\Phi$-independent ${}^{\star}$}\\
$D_{10}$ & $\mathrm{Im}\,\rsw{3}{10}{}^{\star}$ & $-\tfrac12\mathrm{Im}(\ueps{0+}{++}{-}\ueps{-0}{++})$ & $-\tfrac12(\IT{01}{11}{+}\IT{01}{\dmo})-\tfrac12\IU{01}{11}+\tfrac12\IU{01}{\dmo}$\\
$D_{\dmo}$ & $\mathrm{Im}\,\rsw{3}{\dmo}{}^{\star}$ & $-\mathrm{Im}\,\ueps{-+}{++}$ & $\IT{11}{\dmo}-\IU{11}{\dmo}$\\
\midrule
\multicolumn{4}{@{}l}{$\mathcal B_1$: beam spin $\times$ L--T interference ${}^{\star}$}\\
$E^c_{10}$ & $\mathrm{Im}\,\rsw{7}{10}{}^{\star}$ & $\tfrac{1}{2\sqrt2}\mathrm{Im}\big[(\ueps{0+}{0+}{-}\ueps{-0}{0+})+(\ueps{0-}{0+}{-}\ueps{+0}{0+})\big]$ & $\tfrac{1}{2\sqrt2}(\IT{00}{11}{+}\IT{00}{\dmo})+\tfrac1{\sqrt2}\IU{01}{10}$\\
$E^c_{\dmo}$ & $\mathrm{Im}\,\rsw{7}{\dmo}{}^{\star}$ & $\tfrac{1}{\sqrt2}\,\mathrm{Im}(\ueps{-+}{0+}{-}\ueps{+-}{0+})$ & $-\tfrac1{\sqrt2}(\IT{10}{11}{+}\IT{10}{\dmo})+\tfrac1{\sqrt2}(\IU{10}{11}{+}\IU{10}{\dmo})$\\
$E^s_{11}$ & $\rsw{8}{11}{}^{\star}$ & $\tfrac{1}{\sqrt2}\mathrm{Im}(\ueps{++}{0+}{+}\ueps{--}{0+})$ & $\tfrac1{\sqrt2}(\IT{10}{11}{-}\IT{10}{\dmo})+\tfrac1{\sqrt2}(\IU{10}{11}{-}\IU{10}{\dmo})$\\
$E^s_{00}$ & $\rsw{8}{00}{}^{\star}$ & $\sqrt2\,\mathrm{Im}\,\ueps{00}{0+}$ & $\sqrt2\,\IT{00}{01}$\\
$E^s_{10}$ & $\mathrm{Re}\,\rsw{8}{10}{}^{\star}$ & $\tfrac{1}{2\sqrt2}\mathrm{Im}\big[(\ueps{0+}{0+}{-}\ueps{-0}{0+})-(\ueps{0-}{0+}{-}\ueps{+0}{0+})\big]$ & $\tfrac{1}{2\sqrt2}(\IT{00}{11}{-}\IT{00}{\dmo}{-}2\IT{01}{10})$\\
$E^s_{\dmo}$ & $\rsw{8}{\dmo}{}^{\star}$ & $\tfrac{1}{\sqrt2}\,\mathrm{Im}(\ueps{-+}{0+}{+}\ueps{+-}{0+})$ & $-\tfrac1{\sqrt2}(\IT{10}{11}{-}\IT{10}{\dmo})+\tfrac1{\sqrt2}(\IU{10}{11}{-}\IU{10}{\dmo})$\\
\bottomrule
\end{tabular}
\end{table*}

\subsection{Polarized target and nucleon-helicity flip: the complete formalism}\label{app:polarized}
The main result of Sec.~\ref{sec:modeC} is the full polarized-target extraction; this appendix
collects the complete formalism behind it (both nucleon-spin sectors and every
target-polarization state) and makes the sign structure of the unnatural sector fully explicit.
The unpolarized and longitudinal-target tiers of App.~\ref{app:modeAB} are its exact reductions,
and the published analysis code implements the complete case.

\emph{Amplitudes.} Restoring both nucleon helicities explicitly, write $T^{\sigma\lambda}_{\mu\nu}$ and
$U^{\sigma\lambda}_{\mu\nu}$ for the natural- and unnatural-parity amplitudes with initial ($\lambda$) and
final ($\sigma$) nucleon helicities, $\sigma,\lambda=\pm\tfrac12$ (written $\pm$), each nucleon configuration carrying the same
$(\mu,\nu)$ parity completion as in App.~\ref{app:amp} (in particular $U^{\sigma\lambda}_{00}=0$
throughout). Parity relates the four nucleon configurations pairwise,
\begin{align}
T^{--}_{\mu\nu}&=+T^{++}_{\mu\nu}, & U^{--}_{\mu\nu}&=-U^{++}_{\mu\nu}, &&\text{(non-flip)}\notag\\
T^{+-}_{\mu\nu}&=-T^{-+}_{\mu\nu}, & U^{+-}_{\mu\nu}&=+U^{-+}_{\mu\nu}, &&\text{(flip)}
\label{eq:pairs}
\end{align}
so exactly one representative survives per pair, labeled by the flip index $f=|\sigma-\lambda|$:
$T^{(0)}\!\equiv\!T^{++}$, $U^{(0)}\!\equiv\!U^{++}$ (non-flip, the amplitudes of the main text) and
$T^{(1)}\!\equiv\!T^{-+}$, $U^{(1)}\!\equiv\!U^{-+}$ (flip). The $\lambda=-$ configurations are not
discarded: they enter every helicity sum with full weight through Eq.~\eqref{eq:pairs}. The $f{=}0$ set
carries the 16 real parameters of the main text; the $f{=}1$ set adds five natural and four unnatural
amplitudes ($U^{(1)}_{00}=0$ again), 18 further real parameters for 34 in total. Near threshold each
unit of overall helicity change costs one power of $\sqrt{t'}/M_p$ with $t'=t_{\rm min}-t$
[$F\propto(\sqrt{t'}/M_p)^{|(\mu-\nu)-(\sigma-\lambda)|}$]~\citep{COMPASS:2022xig}: at the same
$(\mu,\nu)$ a nucleon-helicity flip adds $\sqrt{t'}$ suppression except where it compensates the
photon--meson helicity change ($\mu-\nu=\sigma-\lambda$), as for the $(0,1)$ amplitudes, so most,
though not all, of the flip sector is additionally $t'$-suppressed.

\emph{Residual symmetry and gauge fixing.} Because the recoil-nucleon helicity is summed and never
observed, the observable bilinears are invariant under any $U(2)$ rotation of the recoil index that
commutes with the parity relations of Eq.~\eqref{eq:pairs}. That commutant is two-dimensional: the
global phase, and the real rotation
$\mathcal R(\alpha):(X^{(0)},X^{(1)})\to(\cos\alpha\,X^{(0)}-\sin\alpha\,X^{(1)},\,
\sin\alpha\,X^{(0)}+\cos\alpha\,X^{(1)})$ applied identically to $X=T$ and $X=U$. Every
family of Eq.~\eqref{eq:ulsn} is exactly invariant under $\mathcal R(\alpha)$. The invariance is
structural in the flip index: $u$ and $l$ are \emph{$f$-traces}, contractions of the amplitude
bilinears with $\delta_{ff'}$ over the flip index $f$, while $s$ and $n$ are
\emph{$\epsilon$-contractions}, contractions with the antisymmetric symbol $\epsilon_{ff'}$
($\epsilon_{01}=-\epsilon_{10}=1$); both invariants of the real rotation acting on the doublet
$(X^{(0)},X^{(1)})$. We verify the invariance numerically to $\mathcal{O}(10^{-16})$.
The 36 real amplitude components therefore carry $36-2=34$ observable parameters. Our two conventions fix
both flat directions: $T^{(0)}_{11}\in\mathbb{R}_{\ge0}$ fixes the phase, and, once the flip sector
is present, $\mathrm{Im}\,U^{(0)}_{11}=0$ acts as the rotation gauge. Numerically, the Jacobian of
the full $(u,l,s,n)$ set has rank 34 on the 36 free components and rank 34 on the 34-parameter
slice, confirming that no physical direction is frozen in the full polarized extraction. The same algebraic condition thus plays three different roles across the tiers defined below: an
unobservable reporting convention for unpolarized data, released once a longitudinal
target makes the $U$ phase physical, and re-imposed in the full polarized extraction as
the rotation gauge (a different flat
direction), so the exact reduction between the three data configurations is preserved. The rotation gauge anchored on
$\mathrm{Im}\,U^{(0)}_{11}$ does degenerate as the flip block vanishes (the slice rank drops to 33 at
exactly zero flip, with conditioning shrinking with the flip size); in the physically expected
small-flip regime the robust choice is to anchor on a flip amplitude that does not itself vanish
there, e.g.\ $\mathrm{Im}\,T^{(1)}_{00}=0$; this is a statement about the gauge direction, not about
how well any component is determined (App.~\ref{app:uml}).

\emph{Structure functions.} Four bilinear families exhaust the spin-dependent cross
section~\citep{Diehl:2007jy} (element indices $(\mu\nu),(\mu'\nu')$ as in $\ueps{\mu\mu'}{\nu\nu'}$,
suppressed here):
\begin{align}
u &= \textstyle\sum_f\big(T^{(f)}T^{(f)*}+U^{(f)}U^{(f)*}\big),\notag\\
l &= \textstyle\sum_f\big(T^{(f)}U^{(f)*}+U^{(f)}T^{(f)*}\big),\notag\\
s &= T^{(0)}U^{(1)*}-U^{(0)}T^{(1)*}-T^{(1)}U^{(0)*}+U^{(1)}T^{(0)*},\notag\\
n &= -T^{(0)}T^{(1)*}+U^{(0)}U^{(1)*}+T^{(1)}T^{(0)*}-U^{(1)}U^{(0)*}.
\label{eq:ulsn}
\end{align}
$u$ alone builds the unpolarized intensity, and Eq.~\eqref{eq:ulsn} is its exact generalization of the
main analysis: restricted to the non-flip sector ($f{=}0$) it reduces to
$u^{\mu\mu'}_{\nu\nu'}=T_{\mu\nu}T^{*}_{\mu'\nu'}+U_{\mu\nu}U^{*}_{\mu'\nu'}$ of
App.~\ref{app:amp} (precisely the $u$-combinations out of which every coefficient of
Eq.~\eqref{eq:W} is built in Table~\ref{tab:dict}), while the flip sector enters $u$ only
additively, enlarging each $|\cdot|^{2}$ sum without interference. In the same way the elements of
$l$ build the longitudinal-target blocks below, and those of $s$ and $n$ the transverse ones:
$l$ accompanies longitudinal target polarization; $s$ and $n$ accompany transverse polarization.

\emph{Angular distribution.} With beam polarization $P_b$ and target spin $(S_L,S_T,\Phi_S)$, where $\Phi_S$ is the azimuth of
the transverse spin direction about the virtual photon, measured from the lepton plane like
$\Phi$~\citep{Diehl:2007jy},
\begin{align}
W \;=\; W_{UU}+P_b W_{LU}
&\;+\;S_L\big(W_{UL}+P_b W_{LL}\big)\notag\\
&\;+\;S_T\big(W_{UT}+P_b W_{LT}\big),
\label{eq:Wfull}
\end{align}
where $W_{UU}+P_bW_{LU}$ is exactly Eq.~\eqref{eq:W}, and every block decomposes over the meson
polarization as $\tfrac{3}{4\pi}\big[\cos^2\!\theta\,W^{LL}+\sqrt2\cos\theta\sin\theta\,W^{LT}
+\sin^2\!\theta\,W^{TT}\big]$. Written in that decomposition, so that the substitution rules of
Eq.~\eqref{eq:replace} act on them term by term, the unpolarized-target blocks are
\begin{widetext}
\begin{align}
W^{LL}_{UU} &= \ueps{00}{++} + \varepsilon\,\ueps{00}{00}
 - 2\sqrt{\varepsilon(1{+}\varepsilon)}\,\cos\Phi\,\mathrm{Re}\,\ueps{00}{0+}
 - \varepsilon\,\cos2\Phi\,\ueps{00}{-+},\label{eq:WUU}\\[2pt]
W^{LT}_{UU} &= \sqrt{\varepsilon(1{+}\varepsilon)}\,\cos(\Phi-\varphi)\,\mathrm{Re}\,\big(\ueps{0+}{0+}{-}\ueps{-0}{0+}\big)
 - \cos\varphi\,\mathrm{Re}\,\big[\big(\ueps{0+}{++}{-}\ueps{-0}{++}\big){+}2\varepsilon\,\ueps{0+}{00}\big]
 + \varepsilon\,\cos(2\Phi-\varphi)\,\mathrm{Re}\,\ueps{0+}{-+}\notag\\
&\quad - \sqrt{\varepsilon(1{+}\varepsilon)}\,\cos(\Phi+\varphi)\,\mathrm{Re}\,\big(\ueps{0-}{0+}{-}\ueps{+0}{0+}\big)
 + \varepsilon\,\cos(2\Phi+\varphi)\,\mathrm{Re}\,\ueps{+0}{-+},\\[2pt]
W^{TT}_{UU} &= \tfrac12\big(\ueps{++}{++}{+}\ueps{--}{++}\big) + \tfrac{\varepsilon}{2}\big(\ueps{++}{00}{+}\ueps{--}{00}\big)
 + \tfrac{\varepsilon}{2}\,\cos(2\Phi-2\varphi)\,\ueps{-+}{-+}
 - \sqrt{\varepsilon(1{+}\varepsilon)}\,\cos\Phi\,\mathrm{Re}\,\big(\ueps{++}{0+}{+}\ueps{--}{0+}\big)\notag\\
&\quad + \sqrt{\varepsilon(1{+}\varepsilon)}\,\cos(\Phi-2\varphi)\,\mathrm{Re}\,\ueps{-+}{0+}
 - \cos 2\varphi\,\mathrm{Re}\,\big[\ueps{-+}{++}{+}\varepsilon\,\ueps{-+}{00}\big]
 - \varepsilon\,\cos 2\Phi\,\mathrm{Re}\,\ueps{++}{-+}\notag\\
&\quad + \sqrt{\varepsilon(1{+}\varepsilon)}\,\cos(\Phi+2\varphi)\,\mathrm{Re}\,\ueps{+-}{0+}
 + \tfrac{\varepsilon}{2}\,\cos(2\Phi+2\varphi)\,\ueps{+-}{-+},\\[6pt]
W^{LL}_{LU} &= 2\sqrt{\varepsilon(1{-}\varepsilon)}\,\sin\Phi\,\mathrm{Im}\,\ueps{00}{0+},\\[2pt]
W^{LT}_{LU} &= -\sqrt{\varepsilon(1{-}\varepsilon)}\,\sin(\Phi-\varphi)\,\mathrm{Im}\,\big(\ueps{0+}{0+}{-}\ueps{-0}{0+}\big)
 - \sqrt{1{-}\varepsilon^{2}}\,\sin\varphi\,\mathrm{Im}\,\big(\ueps{0+}{++}{-}\ueps{-0}{++}\big)\notag\\
&\quad + \sqrt{\varepsilon(1{-}\varepsilon)}\,\sin(\Phi+\varphi)\,\mathrm{Im}\,\big(\ueps{0-}{0+}{-}\ueps{+0}{0+}\big),\\[2pt]
W^{TT}_{LU} &= \sqrt{\varepsilon(1{-}\varepsilon)}\,\sin\Phi\,\mathrm{Im}\,\big(\ueps{++}{0+}{+}\ueps{--}{0+}\big)
 - \sqrt{\varepsilon(1{-}\varepsilon)}\,\sin(\Phi-2\varphi)\,\mathrm{Im}\,\ueps{-+}{0+}\notag\\
&\quad - \sqrt{1{-}\varepsilon^{2}}\,\sin 2\varphi\,\mathrm{Im}\,\ueps{-+}{++}
 - \sqrt{\varepsilon(1{-}\varepsilon)}\,\sin(\Phi+2\varphi)\,\mathrm{Im}\,\ueps{+-}{0+}.\label{eq:WLU}
\end{align}
\end{widetext}
These are Eq.~\eqref{eq:W} regrouped, nothing new: they reproduce it, and the implementation of the
forward model, to $\mathcal{O}(10^{-15})$ over random amplitudes, $\varepsilon$, angles and beam
helicities. Comparing them with the blocks below makes each substitution rule explicit\dash
$\mathrm{Re}\,u\!\to\!\mathrm{Im}\,l$ with $\cos\!\to\!\sin$ for a longitudinal target, and
$\mathrm{Re}\,u\!\to\!\mathrm{Im}\,n$ at fixed harmonic for a transverse one\dash and shows directly
which terms have no polarized counterpart: the two constant ($k{=}m{=}0$) terms of $W^{LL}_{UU}$ and
$W^{TT}_{UU}$, which carry the normalization.
In the conventions of Eq.~\eqref{eq:W} (same azimuths, same beam sign $P_b$; the
$\varepsilon$ weights stay inside the blocks; relative to
Ref.~\citep{Diehl:2007jy} this means $\varphi=-\varphi_{\rm D}$ and $P_b=-P_b^{\rm D}$ with $\Phi$
unchanged, so the $\sin$-harmonic signs below differ from Diehl's Eqs.~(4.13)--(4.18)
accordingly), the
longitudinal-target blocks read as follows.
Every coefficient below is a linear combination of the interference structure functions
$l^{\mu\mu'}_{\nu\nu'}$ of Eq.~\eqref{eq:ulsn}, linear in the unnatural sector in contrast to
every entry of Table~\ref{tab:dict}. No further angular
machinery is needed: the longitudinal blocks follow from Eqs.~\eqref{eq:WUU}--\eqref{eq:WLU} by the
first substitution rule of Eq.~\eqref{eq:replace}, and the transverse-target blocks,
Eqs.~\eqref{eq:WLT}, follow from the same unpolarized blocks and the longitudinal ones by the
remaining rules; the transverse blocks use the same meson-polarization decomposition.
The first rule says something simple about the physics. The unpolarized elements $u$ are built
from same-sign target-helicity bilinears and contain $TT^{*}+UU^{*}$, so each parity sector enters
only through squared moduli. The longitudinal elements $l$ come from the opposite-sign bilinears
and contain $TU^{*}+UT^{*}$, the interference between the two sectors. Trading $\mathrm{Re}\,u$ for
$\mathrm{Im}\,l$, and a cosine for a sine, therefore means that this interference is measured as a
relative phase and not as a magnitude. The double-helicity-flip pair shows it most directly.
$\ueps{-+}{-+}$ is a sum of squared moduli and is therefore real, while the parity relations
$T_{-\mu,-\nu}=+(-1)^{\mu-\nu}T_{\mu\nu}$ and $U_{-\mu,-\nu}=-(-1)^{\mu-\nu}U_{\mu\nu}$ make the two
terms of $l^{-+}_{-+}$ negatives of each other's conjugate, so that element is purely imaginary.
The unpolarized member of the pair therefore carries a magnitude and its longitudinal partner
carries a phase, and only $\mathrm{Im}\,l^{-+}_{-+}$ reaches $W^{TT}_{UL}$. This elementwise
statement should not be read as saying that a longitudinal target adds nothing to the unnatural
magnitudes: because $u$ is quadratic in $U$ while $l$ is linear, the sensitivity of $u$ to $|U|$
falls away as $|U|\to0$ while that of $l$ does not, so the longitudinal blocks dominate the
magnitude information precisely in the small-$|U|$ regime expected for the $\phi$. A longitudinal target therefore fixes the sign and phase of the unnatural sector, and sharpens its magnitude wherever that magnitude is small.
\begin{widetext}
\begin{align}
W^{LL}_{UL} &= -2\sqrt{\varepsilon(1{+}\varepsilon)}\,\sin\Phi\,\mathrm{Im}\,l^{00}_{0+} - \varepsilon\,\sin 2\Phi\,\mathrm{Im}\,l^{00}_{-+},\label{eq:WUL}\\[2pt]
W^{LT}_{UL} &= \sqrt{\varepsilon(1{+}\varepsilon)}\,\sin(\Phi-\varphi)\,\mathrm{Im}\,\big(l^{0+}_{0+}{-}l^{-0}_{0+}\big) + \sin\varphi\,\mathrm{Im}\,\big[\big(l^{0+}_{++}{-}l^{-0}_{++}\big){+}2\varepsilon\,l^{0+}_{00}\big] + \varepsilon\,\sin(2\Phi-\varphi)\,\mathrm{Im}\,l^{0+}_{-+}\notag\\
&\quad -\sqrt{\varepsilon(1{+}\varepsilon)}\,\sin(\Phi+\varphi)\,\mathrm{Im}\,\big(l^{0-}_{0+}{-}l^{+0}_{0+}\big) + \varepsilon\,\sin(2\Phi+\varphi)\,\mathrm{Im}\,l^{+0}_{-+},\\[2pt]
W^{TT}_{UL} &= \tfrac{\varepsilon}{2}\,\sin(2\Phi-2\varphi)\,\mathrm{Im}\,l^{-+}_{-+} - \sqrt{\varepsilon(1{+}\varepsilon)}\,\sin\Phi\,\mathrm{Im}\,\big(l^{++}_{0+}{+}l^{--}_{0+}\big) + \sqrt{\varepsilon(1{+}\varepsilon)}\,\sin(\Phi-2\varphi)\,\mathrm{Im}\,l^{-+}_{0+}\notag\\
&\quad + \sin 2\varphi\,\mathrm{Im}\,\big[l^{-+}_{++}{+}\varepsilon\,l^{-+}_{00}\big] - \varepsilon\,\sin 2\Phi\,\mathrm{Im}\,l^{++}_{-+} + \sqrt{\varepsilon(1{+}\varepsilon)}\,\sin(\Phi+2\varphi)\,\mathrm{Im}\,l^{+-}_{0+}\notag\\
&\quad + \tfrac{\varepsilon}{2}\,\sin(2\Phi+2\varphi)\,\mathrm{Im}\,l^{+-}_{-+},\\[6pt]
W^{LL}_{LL} &= 2\sqrt{\varepsilon(1{-}\varepsilon)}\,\cos\Phi\,\mathrm{Re}\,l^{00}_{0+} - \sqrt{1{-}\varepsilon^{2}}\,\mathrm{Re}\,l^{00}_{++},\\[2pt]
W^{LT}_{LL} &= -\sqrt{\varepsilon(1{-}\varepsilon)}\,\cos(\Phi-\varphi)\,\mathrm{Re}\,\big(l^{0+}_{0+}{-}l^{-0}_{0+}\big) + \sqrt{1{-}\varepsilon^{2}}\,\cos\varphi\,\mathrm{Re}\,\big(l^{0+}_{++}{-}l^{-0}_{++}\big) + \sqrt{\varepsilon(1{-}\varepsilon)}\,\cos(\Phi+\varphi)\,\mathrm{Re}\,\big(l^{0-}_{0+}{-}l^{+0}_{0+}\big),\\[2pt]
W^{TT}_{LL} &= -\tfrac{\sqrt{1-\varepsilon^{2}}}{2}\,\mathrm{Re}\,\big[l^{++}_{++}{+}l^{--}_{++}\big] + \sqrt{\varepsilon(1{-}\varepsilon)}\,\cos\Phi\,\mathrm{Re}\,\big(l^{++}_{0+}{+}l^{--}_{0+}\big) - \sqrt{\varepsilon(1{-}\varepsilon)}\,\cos(\Phi-2\varphi)\,\mathrm{Re}\,l^{-+}_{0+}\notag\\
&\quad + \sqrt{1{-}\varepsilon^{2}}\,\cos 2\varphi\,\mathrm{Re}\,l^{-+}_{++} - \sqrt{\varepsilon(1{-}\varepsilon)}\,\cos(\Phi+2\varphi)\,\mathrm{Re}\,l^{+-}_{0+}.\label{eq:WLL}
\end{align}
%
\begin{align}
&\cos(k\Phi{+}m\varphi)\,\mathrm{Re}\,u \to \sin(k\Phi{+}m\varphi)\,\mathrm{Im}\,l,\qquad
\sin(k\Phi{+}m\varphi)\,\mathrm{Im}\,u \to \cos(k\Phi{+}m\varphi)\,\mathrm{Re}\,l
\quad(\text{longitudinal}),\notag\\
&\mathrm{Re}\,u \to S_T\sin(\Phi{-}\Phi_S)\,\mathrm{Im}\,n,\quad
\mathrm{Im}\,u \to -S_T\sin(\Phi{-}\Phi_S)\,\mathrm{Re}\,n,\notag\\
&S_L\,\mathrm{Im}\,l \to S_T\cos(\Phi{-}\Phi_S)\,\mathrm{Im}\,s,\quad
S_L\,\mathrm{Re}\,l \to S_T\cos(\Phi{-}\Phi_S)\,\mathrm{Re}\,s
\quad(\text{transverse, harmonics unchanged}),
\label{eq:replace}
\end{align}
%
\begin{align}
W^{LL}_{UT} &= \sin(\Phi{-}\Phi_S)\big[\,\mathrm{Im}\,n^{00}_{++} + \varepsilon\,\mathrm{Im}\,n^{00}_{00}\notag\\
&\qquad\qquad\quad - 2\sqrt{\varepsilon(1{+}\varepsilon)}\,\cos\Phi\,\mathrm{Im}\,n^{00}_{0+} - \varepsilon\,\cos 2\Phi\,\mathrm{Im}\,n^{00}_{-+}\,\big]\notag\\
&\quad + \cos(\Phi{-}\Phi_S)\big[\,- 2\sqrt{\varepsilon(1{+}\varepsilon)}\,\sin\Phi\,\mathrm{Im}\,s^{00}_{0+} - \varepsilon\,\sin 2\Phi\,\mathrm{Im}\,s^{00}_{-+}\,\big],\\[4pt]
W^{LT}_{UT} &= \sin(\Phi{-}\Phi_S)\big[\,\sqrt{\varepsilon(1{+}\varepsilon)}\,\cos(\Phi{-}\varphi)\,\mathrm{Im}\,\big(n^{0+}_{0+}{-}n^{-0}_{0+}\big) - \cos\varphi\,\mathrm{Im}\,\big(n^{0+}_{++}{-}n^{-0}_{++}\big)\notag\\
&\qquad\qquad\quad - 2\varepsilon\,\cos\varphi\,\mathrm{Im}\,n^{0+}_{00} + \varepsilon\,\cos(2\Phi{-}\varphi)\,\mathrm{Im}\,n^{0+}_{-+}\notag\\
&\qquad\qquad\quad - \sqrt{\varepsilon(1{+}\varepsilon)}\,\cos(\Phi{+}\varphi)\,\mathrm{Im}\,\big(n^{0-}_{0+}{-}n^{+0}_{0+}\big) + \varepsilon\,\cos(2\Phi{+}\varphi)\,\mathrm{Im}\,n^{+0}_{-+}\,\big]\notag\\
&\quad + \cos(\Phi{-}\Phi_S)\big[\,\sqrt{\varepsilon(1{+}\varepsilon)}\,\sin(\Phi{-}\varphi)\,\mathrm{Im}\,\big(s^{0+}_{0+}{-}s^{-0}_{0+}\big) + \sin\varphi\,\mathrm{Im}\,\big(s^{0+}_{++}{-}s^{-0}_{++}\big)\notag\\
&\qquad\qquad\quad + 2\varepsilon\,\sin\varphi\,\mathrm{Im}\,s^{0+}_{00} + \varepsilon\,\sin(2\Phi{-}\varphi)\,\mathrm{Im}\,s^{0+}_{-+}\notag\\
&\qquad\qquad\quad - \sqrt{\varepsilon(1{+}\varepsilon)}\,\sin(\Phi{+}\varphi)\,\mathrm{Im}\,\big(s^{0-}_{0+}{-}s^{+0}_{0+}\big) + \varepsilon\,\sin(2\Phi{+}\varphi)\,\mathrm{Im}\,s^{+0}_{-+}\,\big],\\[4pt]
W^{TT}_{UT} &= \sin(\Phi{-}\Phi_S)\big[\,\tfrac12\,\mathrm{Im}\,n^{++}_{++} + \tfrac12\,\mathrm{Im}\,n^{--}_{++}\notag\\
&\qquad\qquad\quad + \varepsilon\,\mathrm{Im}\,n^{++}_{00} + \tfrac{\varepsilon}{2}\,\cos(2\Phi{-}2\varphi)\,\mathrm{Im}\,n^{-+}_{-+}\notag\\
&\qquad\qquad\quad - \sqrt{\varepsilon(1{+}\varepsilon)}\,\cos\Phi\,\mathrm{Im}\,\big(n^{++}_{0+}{+}n^{--}_{0+}\big) + \sqrt{\varepsilon(1{+}\varepsilon)}\,\cos(\Phi{-}2\varphi)\,\mathrm{Im}\,n^{-+}_{0+}\notag\\
&\qquad\qquad\quad - \cos 2\varphi\,\mathrm{Im}\,n^{-+}_{++} - \varepsilon\,\cos 2\varphi\,\mathrm{Im}\,n^{-+}_{00}\notag\\
&\qquad\qquad\quad - \varepsilon\,\cos 2\Phi\,\mathrm{Im}\,n^{++}_{-+} + \sqrt{\varepsilon(1{+}\varepsilon)}\,\cos(\Phi{+}2\varphi)\,\mathrm{Im}\,n^{+-}_{0+}\notag\\
&\qquad\qquad\quad + \tfrac{\varepsilon}{2}\,\cos(2\Phi{+}2\varphi)\,\mathrm{Im}\,n^{+-}_{-+}\,\big]\notag\\
&\quad + \cos(\Phi{-}\Phi_S)\big[\,\tfrac{\varepsilon}{2}\,\sin(2\Phi{-}2\varphi)\,\mathrm{Im}\,s^{-+}_{-+} - \sqrt{\varepsilon(1{+}\varepsilon)}\,\sin\Phi\,\mathrm{Im}\,\big(s^{++}_{0+}{+}s^{--}_{0+}\big)\notag\\
&\qquad\qquad\quad + \sqrt{\varepsilon(1{+}\varepsilon)}\,\sin(\Phi{-}2\varphi)\,\mathrm{Im}\,s^{-+}_{0+} + \sin 2\varphi\,\mathrm{Im}\,s^{-+}_{++}\notag\\
&\qquad\qquad\quad + \varepsilon\,\sin 2\varphi\,\mathrm{Im}\,s^{-+}_{00} - \varepsilon\,\sin 2\Phi\,\mathrm{Im}\,s^{++}_{-+}\notag\\
&\qquad\qquad\quad + \sqrt{\varepsilon(1{+}\varepsilon)}\,\sin(\Phi{+}2\varphi)\,\mathrm{Im}\,s^{+-}_{0+} + \tfrac{\varepsilon}{2}\,\sin(2\Phi{+}2\varphi)\,\mathrm{Im}\,s^{+-}_{-+}\,\big],\\[4pt]
W^{LL}_{LT} &= \sin(\Phi{-}\Phi_S)\big[\,- 2\sqrt{\varepsilon(1{-}\varepsilon)}\,\sin\Phi\,\mathrm{Re}\,n^{00}_{0+}\,\big]\notag\\
&\quad + \cos(\Phi{-}\Phi_S)\big[\,2\sqrt{\varepsilon(1{-}\varepsilon)}\,\cos\Phi\,\mathrm{Re}\,s^{00}_{0+} - \sqrt{1{-}\varepsilon^{2}}\,\mathrm{Re}\,s^{00}_{++}\,\big],\\[4pt]
W^{LT}_{LT} &= \sin(\Phi{-}\Phi_S)\big[\,\sqrt{\varepsilon(1{-}\varepsilon)}\,\sin(\Phi{-}\varphi)\,\mathrm{Re}\,\big(n^{0+}_{0+}{-}n^{-0}_{0+}\big) + \sqrt{1{-}\varepsilon^{2}}\,\sin\varphi\,\mathrm{Re}\,\big(n^{0+}_{++}{-}n^{-0}_{++}\big)\notag\\
&\qquad\qquad\quad - \sqrt{\varepsilon(1{-}\varepsilon)}\,\sin(\Phi{+}\varphi)\,\mathrm{Re}\,\big(n^{0-}_{0+}{-}n^{+0}_{0+}\big)\,\big]\notag\\
&\quad + \cos(\Phi{-}\Phi_S)\big[\,- \sqrt{\varepsilon(1{-}\varepsilon)}\,\cos(\Phi{-}\varphi)\,\mathrm{Re}\,\big(s^{0+}_{0+}{-}s^{-0}_{0+}\big) + \sqrt{1{-}\varepsilon^{2}}\,\cos\varphi\,\mathrm{Re}\,\big(s^{0+}_{++}{-}s^{-0}_{++}\big)\notag\\
&\qquad\qquad\quad + \sqrt{\varepsilon(1{-}\varepsilon)}\,\cos(\Phi{+}\varphi)\,\mathrm{Re}\,\big(s^{0-}_{0+}{-}s^{+0}_{0+}\big)\,\big],\\[4pt]
W^{TT}_{LT} &= \sin(\Phi{-}\Phi_S)\big[\,- \sqrt{\varepsilon(1{-}\varepsilon)}\,\sin\Phi\,\mathrm{Re}\,\big(n^{++}_{0+}{+}n^{--}_{0+}\big) + \sqrt{\varepsilon(1{-}\varepsilon)}\,\sin(\Phi{-}2\varphi)\,\mathrm{Re}\,n^{-+}_{0+}\notag\\
&\qquad\qquad\quad + \sqrt{1{-}\varepsilon^{2}}\,\sin 2\varphi\,\mathrm{Re}\,n^{-+}_{++} + \sqrt{\varepsilon(1{-}\varepsilon)}\,\sin(\Phi{+}2\varphi)\,\mathrm{Re}\,n^{+-}_{0+}\,\big]\notag\\
&\quad + \cos(\Phi{-}\Phi_S)\big[\,- \tfrac{\sqrt{1-\varepsilon^{2}}}{2}\,\mathrm{Re}\,s^{++}_{++} - \tfrac{\sqrt{1-\varepsilon^{2}}}{2}\,\mathrm{Re}\,s^{--}_{++}\notag\\
&\qquad\qquad\quad + \sqrt{\varepsilon(1{-}\varepsilon)}\,\cos\Phi\,\mathrm{Re}\,\big(s^{++}_{0+}{+}s^{--}_{0+}\big) - \sqrt{\varepsilon(1{-}\varepsilon)}\,\cos(\Phi{-}2\varphi)\,\mathrm{Re}\,s^{-+}_{0+}\notag\\
&\qquad\qquad\quad + \sqrt{1{-}\varepsilon^{2}}\,\cos 2\varphi\,\mathrm{Re}\,s^{-+}_{++} - \sqrt{\varepsilon(1{-}\varepsilon)}\,\cos(\Phi{+}2\varphi)\,\mathrm{Re}\,s^{+-}_{0+}\,\big].
\label{eq:WLT}
\end{align}
\end{widetext}
Here the $\sin(\Phi{-}\Phi_S)$ harmonics multiply the $n$ family and the $\cos(\Phi{-}\Phi_S)$
harmonics the $s$ family of Eq.~\eqref{eq:ulsn}, both linear in the nucleon-helicity-flip
amplitudes; these expressions are likewise generated from, and verified numerically to
$\mathcal{O}(10^{-16})$ against, the implementation of the forward model. Together with
Eq.~\eqref{eq:W} and Eqs.~\eqref{eq:WUL}--\eqref{eq:WLL} they specify every block of
Eq.~\eqref{eq:Wfull}.

\emph{Observability ladder.} Equations~\eqref{eq:ulsn}--\eqref{eq:replace} organize what each dataset
can determine. Unpolarized data see only $u$, quadratic in both the unnatural sector and the nucleon-flip
sector: the global sign change $U\to-U$ is exact (App.~\ref{app:modulations}), and the flip block
enters only incoherently. Longitudinal polarization adds $l$, linear in $U^{(0)}$ and odd under
$U\to-U$: the overall unnatural sign and phase become observable. In the strict non-flip
truncation the only residual symmetry is the global phase, so the longitudinal-target extraction carries 17 parameters
(the Jacobian of $(u,l)$ has rank 17 on the 18 non-flip components), and the direction frozen by the
unpolarized-only convention ($\mathrm{Im}\,U^{(0)}_{11}=0$) is precisely the $U$-sector phase that
$S_L$ promotes to an observable, so that constraint must be released once longitudinal-target data are fitted. Transverse
polarization adds $s$ and $n$, linear in the flip amplitudes: the flip sector becomes
observable up to the exact $\mathcal R(\alpha)$ rotation, which mixes flip and non-flip, so per-bin flip
labels are fixed by the gauge choice (or by external input: the $\sqrt{t'}/M_p$ threshold behavior
across bins, or recoil polarimetry), and are never measured outright. Rotation-invariant combinations
are observable unconditionally, in particular the $\epsilon$-contractions such as
$\mathrm{Im}\,n^{00}_{00}$ that carry the GPD-$E$ sensitivity, and those are the quantities we
recommend reporting. Three nested data configurations follow: the unpolarized non-flip extraction (16 parameters),
the longitudinal-target extraction (17 parameters), and the full polarized program (34
parameters, $S_L$ and $S_T$ data together), each reducing exactly to the previous one as the
polarization is switched off. All three are strongly over-determined: the azimuthal
structures of $W_{UU},W_{LU},W_{UL},W_{LL},W_{UT},W_{LT}$ provide $15+8+14+10+30+18=95$ independent
measurements~\citep{Diehl:2007jy} for at most 34 parameters ($W_{UT}$ counts $14$ $s$- and $16$
$n$-harmonics, the $n$ sector gaining one because the counterpart of the normalization-fixed
constant of $W_{UU}$ is an independent quantity). The measured Jacobian ranks over the free
amplitude components ($16/18$ on $u$; $17/18$ on $u,l$; $34/36$ on $u,l,s,n$) make the ladder
quantitative. The implementation takes the polarized blocks term by term from Ref.~\citep{Diehl:2007jy}, mapped
to the conventions of Eq.~\eqref{eq:W}, and is verified numerically: the $u$-sector of
Eq.~\eqref{eq:Wfull} reproduces Eq.~\eqref{eq:W} to double precision, the blocks obey their parity/time-reversal symmetries, the $S$-dependent
sectors vanish identically in the appropriate limits, and $W>0$ for pure polarization states.

\subsection{Gluon-GPD framework}\label{app:gpd}
This framework is model-dependent and underlies only the physics interpretation in the main text; the
amplitude/SDME extraction makes no use of it. Throughout this subsection the superscripts on
$\mathcal M^{ab}$ label (photon, meson) helicities in that order, the reverse of the
(meson, photon) order of $T_{\mu\nu}$: $\mathcal M^{LT}$ is thus $\gL\!\to\!\VT$, i.e.\ $T_{10}$,
and $\mathcal M^{TL}$ is $\gT\!\to\!\VL$, i.e.\ $T_{01}$. The leading-order helicity amplitudes are
$\mathcal M^{LL}=C\,\delta_{\lambda\lambda'}
\{\mathcal H^{LL}_{\rm eff}\}$, $\mathcal M^{TT}=C\,\delta_{\lambda\lambda'}
(\boldsymbol\epsilon^{\gamma^*}_\perp\!\cdot\boldsymbol\epsilon^V_\perp)\{\mathcal H^{TT}_{\rm eff}\}$,
$\mathcal M^{LT}=0$, and a twist-3 $\mathcal M^{TL}$ carrying the gluon orbital angular momentum
($\mathcal F^{TL}_{1,1}$) and spin--orbit ($\mathcal G^{TL}_{1,1}$) terms, with
$C\propto e\,e_q\,\alpha_s f_V m_V/(N_c(m_V^2+Q^2))$ and $H^g_{\rm eff}=H^g-\tfrac{\xi^2}{1-\xi^2}
E^g$~\citep{Bhattacharya:2026qnd,Bhattacharya:2022vvo}. For $\phi$ and $\rho$ the structure is identical; only $C$ differs.

\section{Method}
This appendix describes the forward model and the inference machinery: the acceptance- and resolution-folded simulator and its validation, the diffusion-model architecture and training, and the calibration and acceptance-correction studies.

\subsection{Implementation and forward-model validation}\label{app:impl}
The acceptance-folded moment map $(M,b)$ is calibrated from a flat sample through $\eta$ per energy;
$\langle f\rangle_{\rm acc}(u_{\rm truth})$ matches the raw data moments to RMS $\approx10^{-3}$.
Yields are Poisson with mean $\propto(\sT+\varepsilon\sL)\times$acceptance, so the rate feature is
data-driven and the moment noise scales with the expected per-bin count.

\begin{figure*}[!t]\centering
\includegraphics[width=0.9\linewidth]{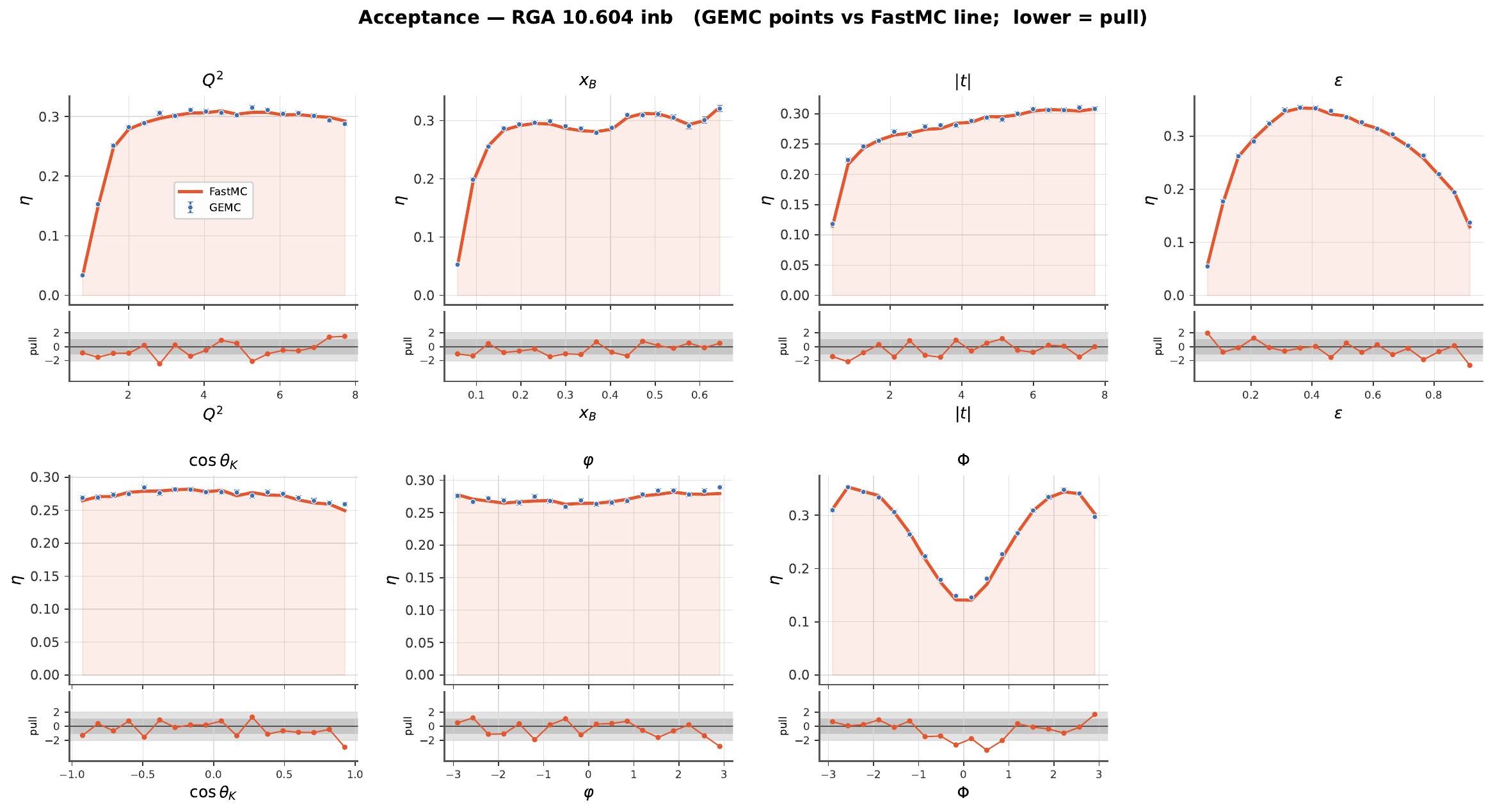}
\caption{\textbf{The GEMC-trained response surrogate against held-out GEMC events}
($10.604$~GeV, inbending torus; the surrogate's own validation suite). Acceptance versus each
kinematic and decay variable: GEMC points with binomial errors against the surrogate's prediction,
with the pull beneath each panel (shaded $\pm1$, $\pm2$ bands). On the full
$8\times10^{6}$-event sample of this configuration the binned projections agree to $1.3\%$ on
average ($0.2748$ vs.\ $0.2704$ overall); the remaining six (beam energy, torus) configurations
validate at the same level.}
\label{fig:fmcval}
\end{figure*}

\begin{figure}[!ht]\centering
\includegraphics[width=0.94\linewidth]{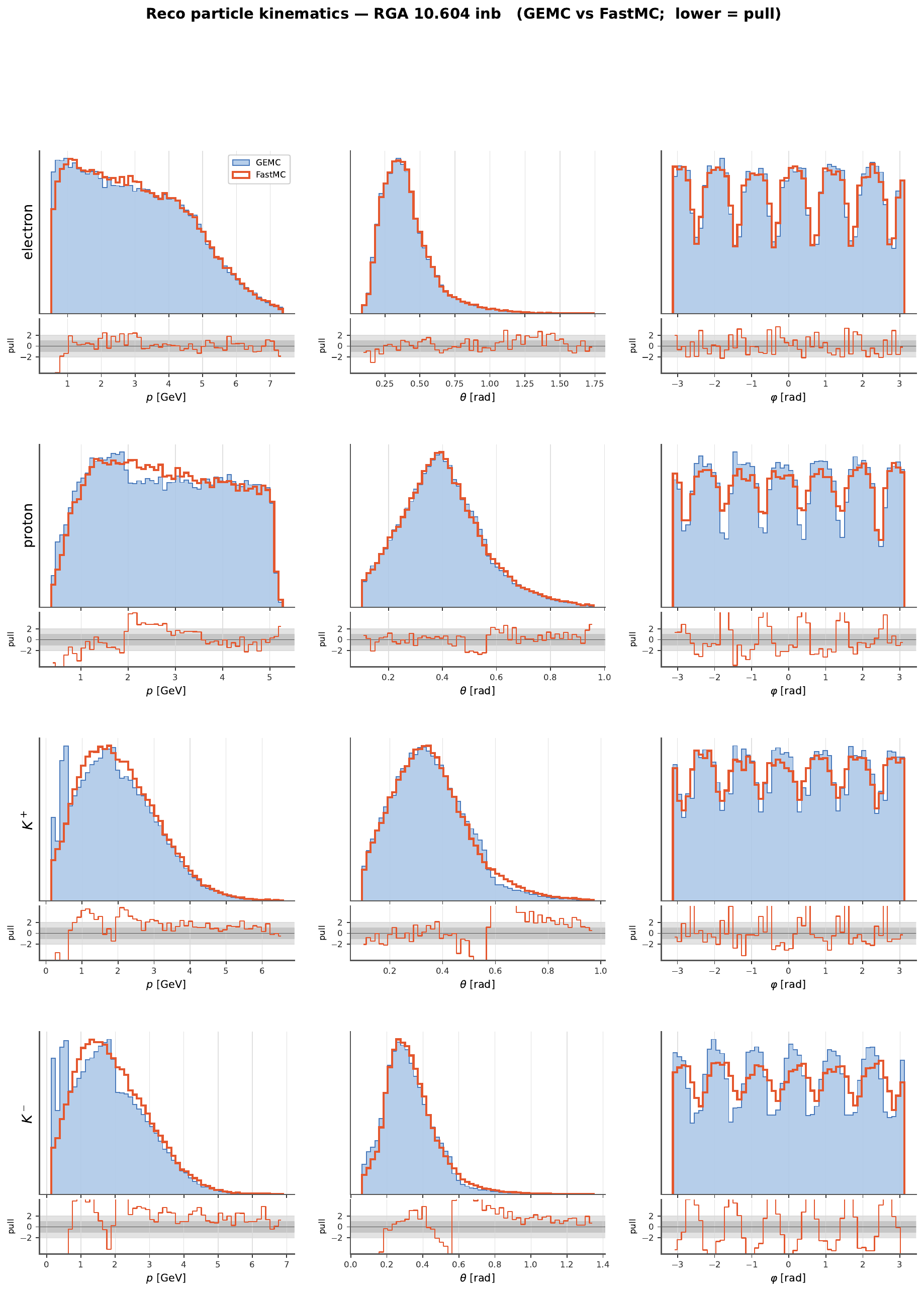}
\caption{Reconstructed particle kinematics $(p,\theta_{\rm lab},\phi_{\rm lab})$ for all four final-state
particles in the same configuration as Fig.~\ref{fig:fmcval}: GEMC against events passed through
the surrogate's resolution model, with the same pull convention.}
\label{fig:fmcvalkin}
\end{figure}

\emph{The GEMC-trained response surrogate.} The realistic response used in the transfer tests
(Secs.~\ref{sec:val-det}, \ref{sec:detamort}; Fig.~\ref{fig:fmcfull}) is trained directly on the
collaboration's GEMC productions for the CLAS12 run configurations that recorded the corresponding
data, covering all beam-energy and torus-polarity settings: $10.604$ and $10.594$~GeV in both torus
polarities, $10.2$~GeV inbending, and $6.535$ and $7.546$~GeV outbending, at analysis-level event
selection. It consists of two networks. An acceptance classifier returns the per-event
survival probability conditioned on 61 features: the seven physics variables
($Q^2$, $x_B$, $t$, $\varepsilon$, $\cos\theta$, $\varphi$, $\Phi$), the laboratory momenta
$(p,\theta_{\rm lab},\phi_{\rm lab})$ of all four final-state particles, the run configuration (beam
energy, torus polarity, beam current), and laboratory-azimuth Fourier harmonics of order $6$--$24$ for each
particle, which resolve the six-sector structure of the spectrometer. A ten-component
Gaussian-mixture density models the twelve momentum residuals ($p,\theta_{\rm lab},\phi_{\rm lab}$ per particle)
conditioned on the same laboratory kinematics, so acceptance \emph{and} resolution are taken
from GEMC event by event, not parameterized by hand. Folding through this surrogate rather than GEMC
itself is the computational choice noted in Sec.~\ref{sec:val-det}; nothing in the
extraction chain is aware of which response produced the events. The surrogate is trained on
$1.4\times10^{7}$ accepted GEMC events (overall accept rate $0.254$) across the seven
configurations and validated against held-out GEMC events in Figs.~\ref{fig:fmcval}
and~\ref{fig:fmcvalkin}. The predicted acceptance tracks GEMC across all seven kinematic
projections to $1.4\%$ on average, and the per-particle momentum-residual distributions match, in
every configuration. The acceptance classifier is also calibrated, its reliability curve lying on
the diagonal, so the probability it assigns is the probability GEMC realizes.

\subsection{Diffusion-model and training details}\label{app:diffusion}
\begin{table}[t]
\caption{Diffusion-model and training settings (the two processes are the standard DDPM
forward corruption and learned reverse denoising).}
\label{tab:diffusion}
\footnotesize
\begin{ruledtabular}
\begin{tabular}{@{}lp{0.56\columnwidth}@{}}
diffusion steps $T$ & 200 (linear $\beta_t\!:10^{-4}\!\to\!2\times10^{-2}$) \\
conditioning $x$ & 76 components $[\,69\ \text{moments}\,|\,3\ \text{rates}\,|\,3\ \varepsilon\,|\,P_b\,]$;
161 with a longitudinal target and 332 for the full polarized program (App.~\ref{app:polarized}) \\
network & 4 layers, SiLU, time embedding; width 320 (fixed detector), 384 with a 144-dim
response signature (detector-amortized) \\
optimizer & Adam, lr $2\times10^{-4}$, 400 epochs \\
training pairs $N_0$ & $4.8\times10^{5}$ (flat prior, scale aug.\ $U(0.03,1)$) \\
ensemble & $K=20$ independently initialized members \\
sampler & ancestral, $T$ steps; posterior draws per bin 400 \\
\end{tabular}
\end{ruledtabular}
\end{table}

\begin{figure*}[!ht]\centering
\includegraphics[width=\linewidth]{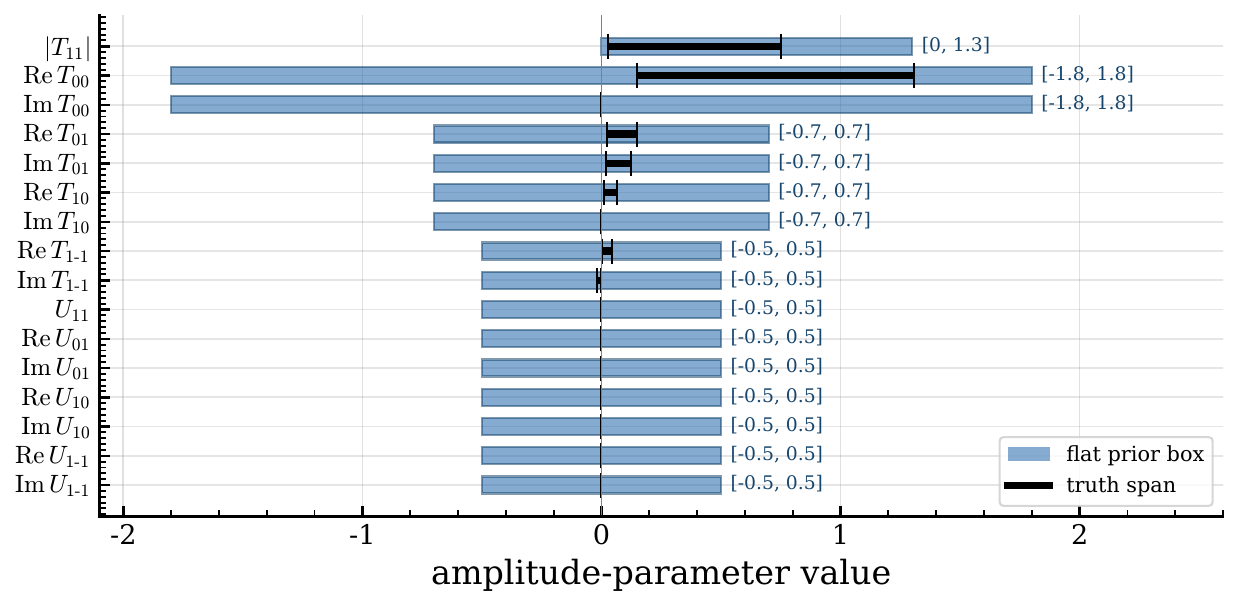}
\caption{The flat prior over the 16 amplitude parameters (blue boxes) from which the training set is
drawn, with the injected-truth span over the $(Q^2,|t|)$ grid (black bars) shown for reference. The
truth lies well inside the prior for every parameter, so the network is trained on a region that
comfortably contains the test cases.}
\label{fig:priors}
\end{figure*}
Variance-preserving schedule, $T=200$, $\beta_t$ linear $10^{-4}\!\to\!2\!\times\!10^{-2}$; noise
network with sinusoidal time embedding, a conditioner sub-network, and four SiLU layers (width
320 for the fixed-detector networks, 384 with the 144-dimensional response-signature input in the
detector-amortized variant);
Adam (lr $2\!\times\!10^{-4}$), $400$ epochs, $4.8\!\times\!10^5$ simulated pairs; 400 reverse-diffusion
draws per bin. Scale augmentation (amplitudes $\times\,U(0.03,1)$) covers the small-amplitude regime.
These are the unpolarized-tier settings; the polarized and detector-amortized trainings differ in
sample count, epochs, and learning rate, as recorded in the released configuration.
Code and configuration are released with the paper.

\subsection{Calibration}\label{app:cal}
\begin{figure*}[!t]\centering
\includegraphics[width=\linewidth]{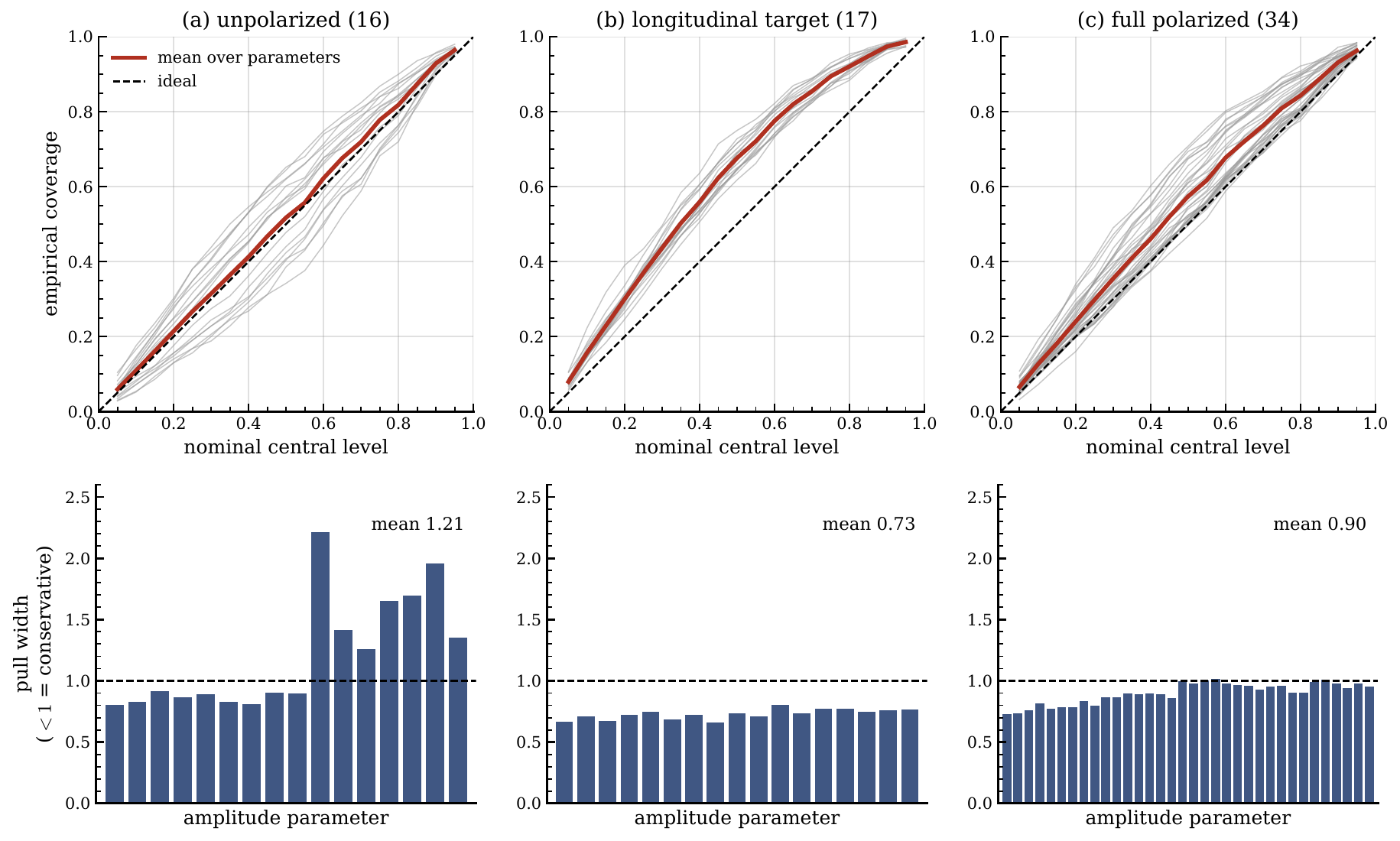}
\caption{\textbf{Simulation-based calibration of the total (deep-ensemble) uncertainty, all three
extraction tiers on one footing.} Top row: empirical vs.\ nominal central-interval coverage, one
faint curve per amplitude parameter and their mean in red, against the ideal diagonal. Bottom row:
per-parameter pull width, with unity marked. (a) unpolarized (16 parameters), (b) longitudinal
target (17), (c) the full polarized program (34); the polarized tiers with no sign alignment
applied, so the signed posterior is calibrated as measured. Coverage is rank-based and computed
identically for all three, so the panels are directly comparable. Every tier lies on or above the
diagonal ($0.72/0.96$, $0.85/0.99$, and $0.76/0.96$ at nominal $0.70/0.95$): the quoted bands
over-cover. The mean pull width is $1.21$, $0.73$ and $0.90$: the unpolarized tier is the
one carrying residual over-confidence, concentrated in the $U$ sector (App.~\ref{app:unc}).}
\label{fig:sbc}
\end{figure*}

A maximum-likelihood fit carries an asymptotic coverage guarantee through its parabolic errors; a
learned posterior $q_\psi(A\,|\,x)$ carries none, which is why the simulation-based calibration
(SBC) of Sec.~\ref{sec:val-sbc} is run before any error bar here is quoted. The pull and its width,
the diagnostic used throughout this appendix, are defined in App.~\ref{app:unc}.
The suite is run per trained ensemble, so the unpolarized, longitudinal-target, and full
polarized extractions each carry their own SBC and every closure figure is backed by the
calibration of its own ensemble (Fig.~\ref{fig:sbc}).
SBC of the total (ensemble) uncertainty used for the bands
above, over 500 prior draws per ensemble, gives uniform rank histograms (not shown) and, for
the unpolarized ensemble, empirical central-interval coverage $0.72/0.96$ at nominal
$0.70/0.95$ [Fig.~\ref{fig:sbc}(a)]. The central intervals therefore over-cover, so the quoted
uncertainty is a safe statement. The per-parameter pull widths are not uniform across the
amplitude set: $0.80$--$0.93$ throughout the natural-parity ($T$) sector, but $1.27$--$2.16$ in
the unnatural-parity ($U$) sector, for a mean of $1.21$. The $U$ amplitudes are the content that
unpolarized data constrain least, and the ensemble is correspondingly over-confident there; this
is the unpolarized counterpart of the weakly constrained flip combination identified in
App.~\ref{app:uml}, and it is why the longitudinal-target tier, which measures the $U$ sector
directly, is the one that certifies it. A single detector-amortized network is markedly over-confident there (mean pull width $2.2$),
so the deep ensemble supplies most of the calibration; App.~\ref{app:unc} makes the same
decomposition for the fixed-detector ensemble, where a single member is much closer to calibrated.
The polarized ensembles pass the same suite with no sign alignment anywhere, the signed
posterior calibrated as measured: mean pull std $0.73$ (longitudinal-target) and $0.90$
(full polarized program), with empirical central-interval coverage $0.85/0.99$ and $0.76/0.96$
at nominal $0.70/0.95$ [Fig.~\ref{fig:sbc}(b,c)]; like the unpolarized bands, both are
conservative.

\subsection{Acceptance correction}
\begin{figure*}[!ht]\centering
\includegraphics[width=\linewidth]{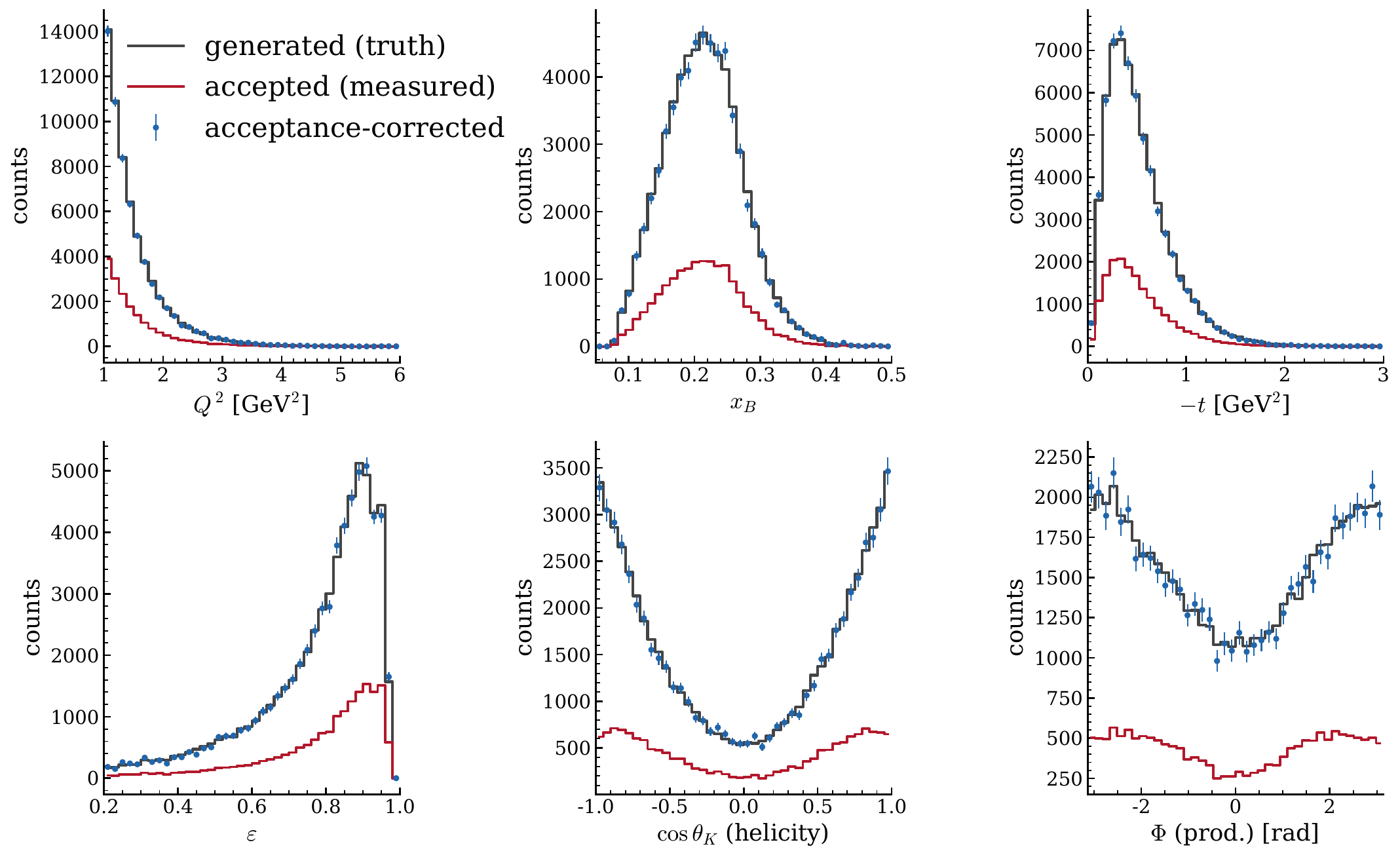}
\caption{Acceptance-correction validation: generated/truth (black), accepted/measured (red), and
acceptance-corrected ($\times1/\eta$, blue). The corrected curves recover the generated distributions
and remove the detector's six $\Phi$ sector gaps and the $\cos\theta$ flattening, confirming that
an incorrect $\eta$ would directly bias the angular shape and hence the SDMEs.}
\label{fig:acccorr}
\end{figure*}

We throw events following the full physics (production cross section $\times$ the non-flat angular
distribution $W$), pass them through the parameterized detector, and weight accepted events by $1/\eta$
(Fig.~\ref{fig:acccorr}). The accepted (measured) distributions are strongly distorted, with the six sector gaps in $\Phi$
and a flattened $\cos\theta$, while the $1/\eta$ correction recovers
the generated distributions within statistics. This shows that the acceptance folded into our
simulator and the explicit correction are mutually consistent, and that an incorrect $\eta$ would
directly bias the SDMEs and their kinematic dependence.

\section{Results and validation}
This appendix collects the extraction results and the tests that validate them: the unpolarized and longitudinally polarized tiers, the independent unbinned maximum-likelihood cross-check, the uncertainty quantification, and the closure loop from inferred amplitudes back to events.

\subsection{Unpolarized and longitudinal-target results}\label{app:modeAB}
\begin{figure*}[!t]\centering
\includegraphics[width=\linewidth,height=0.92\textheight,keepaspectratio]{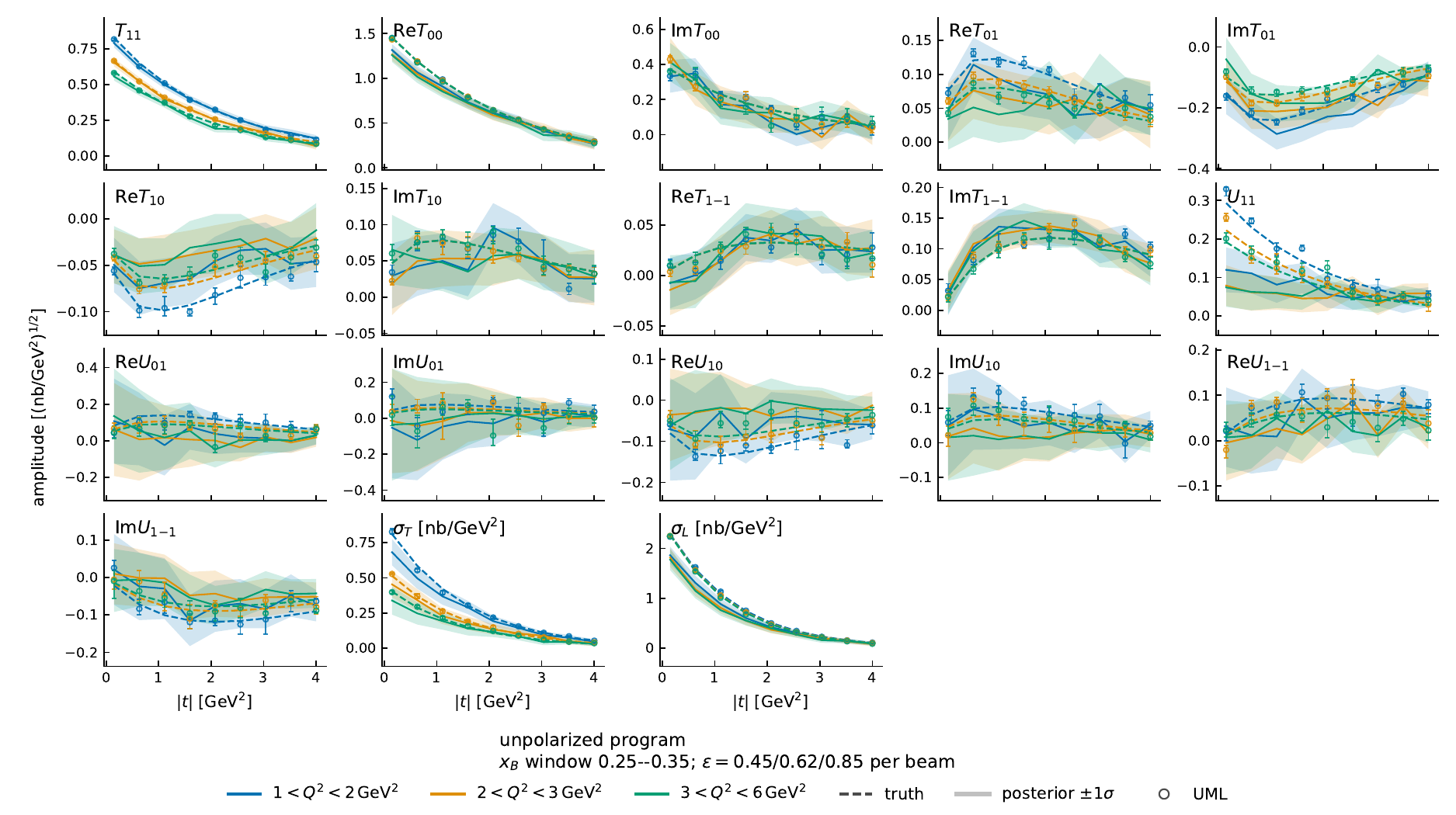}
\caption{\textbf{Unpolarized extraction ($\phi$).} The 16 nucleon-helicity-non-flip amplitude parameters
and the derived $\sT,\sL,R$ vs.\ $|t|$ for three $Q^2$ rings,
extracted (solid $\pm$ total-uncertainty band, App.~\ref{app:unc}) against the injected truth
(dashed); open circles show the independent unbinned maximum-likelihood fit with explicit
Rosenbluth separation.}
\label{fig:closure}
\end{figure*}

\begin{figure*}[!t]\centering
\includegraphics[width=\linewidth,height=0.92\textheight,keepaspectratio]{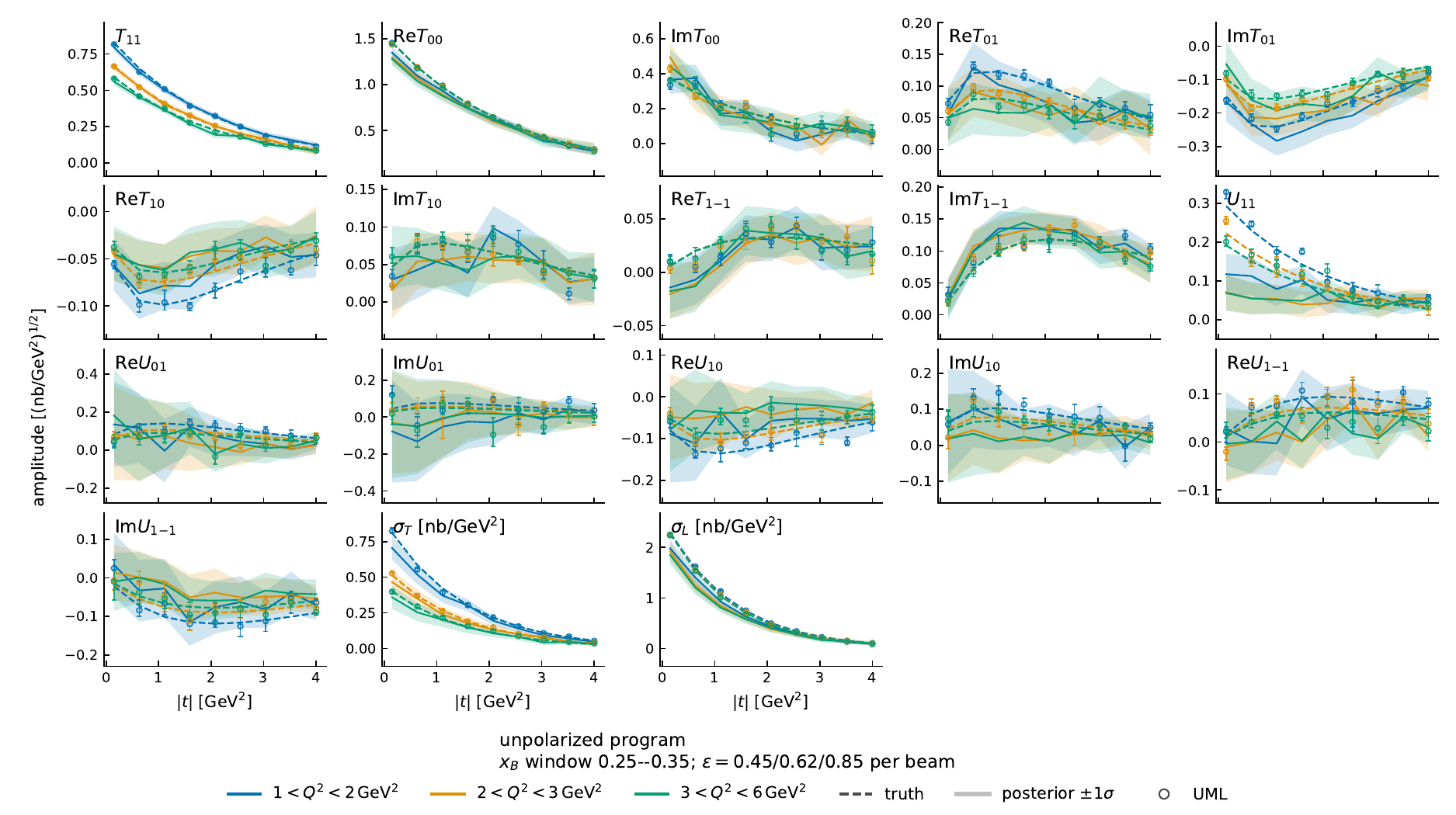}
\caption{\textbf{Unpolarized extraction: the $\rho^0$ full case.} The 16 amplitude parameters and derived
$\sT,\sL,R$ vs.\ $|t|$ for three $Q^2$ rings in $\rho^0\to\pi^+\pi^-$, with the
detector-amortized network: extracted (solid $\pm$ band) against the injected truth (dashed),
with the unbinned maximum-likelihood markers. The truth carries non-zero unnatural-parity
amplitudes ($U_{11},U_{01},U_{10},U_{1\text{-}1}$), which are recovered: the positive
counterpart of the $\phi$ null test (Fig.~\ref{fig:closure}).}
\label{fig:meson}
\end{figure*}

\emph{The unpolarized extraction.} Figure~\ref{fig:closure} shows the unpolarized 16-parameter extraction: the
amplitude parameters and derived $\sT,\sL,R$ vs.\ $|t|$ (to the full CLAS12 range
$|t|\le4\,$GeV$^2$) for three $Q^2$ rings track the injected truth across the plane, including
the steep $t$-slopes and the $Q^2$ ordering. The statement is per bin: the network operates
independently in each $(\Qsq,t)$ bin within the fixed $x_B$ window and never receives the bin's
own $(\Qsq,t)$ coordinates as inputs -- its only non-observable inputs are the run conditions
$\varepsilon$, $P_b$ and, in the polarized modes, $P$ (Sec.~\ref{sec:framework:sbi}) -- so the
recovered trends emerge point by point from the per-bin observables alone. The four unnatural-parity amplitudes are returned as well: for the
natural-parity-dominated $\phi$ truth they come out consistent with zero ($|\hat U|\lesssim0.02$,
an upper bound at the noise scale under the $U_{11}\ge0$ folding of App.~\ref{app:modulations}),
so the natural-parity restriction of the standard analysis emerges as a result, never an input; on a truth with non-zero unnatural amplitudes they are recovered
(the $\rho^0$ case, Fig.~\ref{fig:meson}). The large $\sL$ is reproduced (e.g.\ $2.42$ vs.\ truth $2.43$ at low
$|t|$); the bands are the total (statistical $\oplus$ deep-ensemble model) uncertainty of
App.~\ref{app:unc}.

\emph{The $\rho^0$ full case.} The same procedure re-run on $\rho^0\!\to\!\pi^+\pi^-$, with only the forward model re-specified and basis/architecture/calibration unchanged, closes at comparable quality (Fig.~\ref{fig:meson}). Here the injected truth carries sizable non-zero unnatural-parity amplitudes, which the extraction recovers: the same procedure that finds the unnatural sector consistent with zero for the natural-parity--dominated $\phi$ measures it when present, so the extraction does not depend on the meson.

\begin{figure*}[!t]\centering
\includegraphics[width=\linewidth]{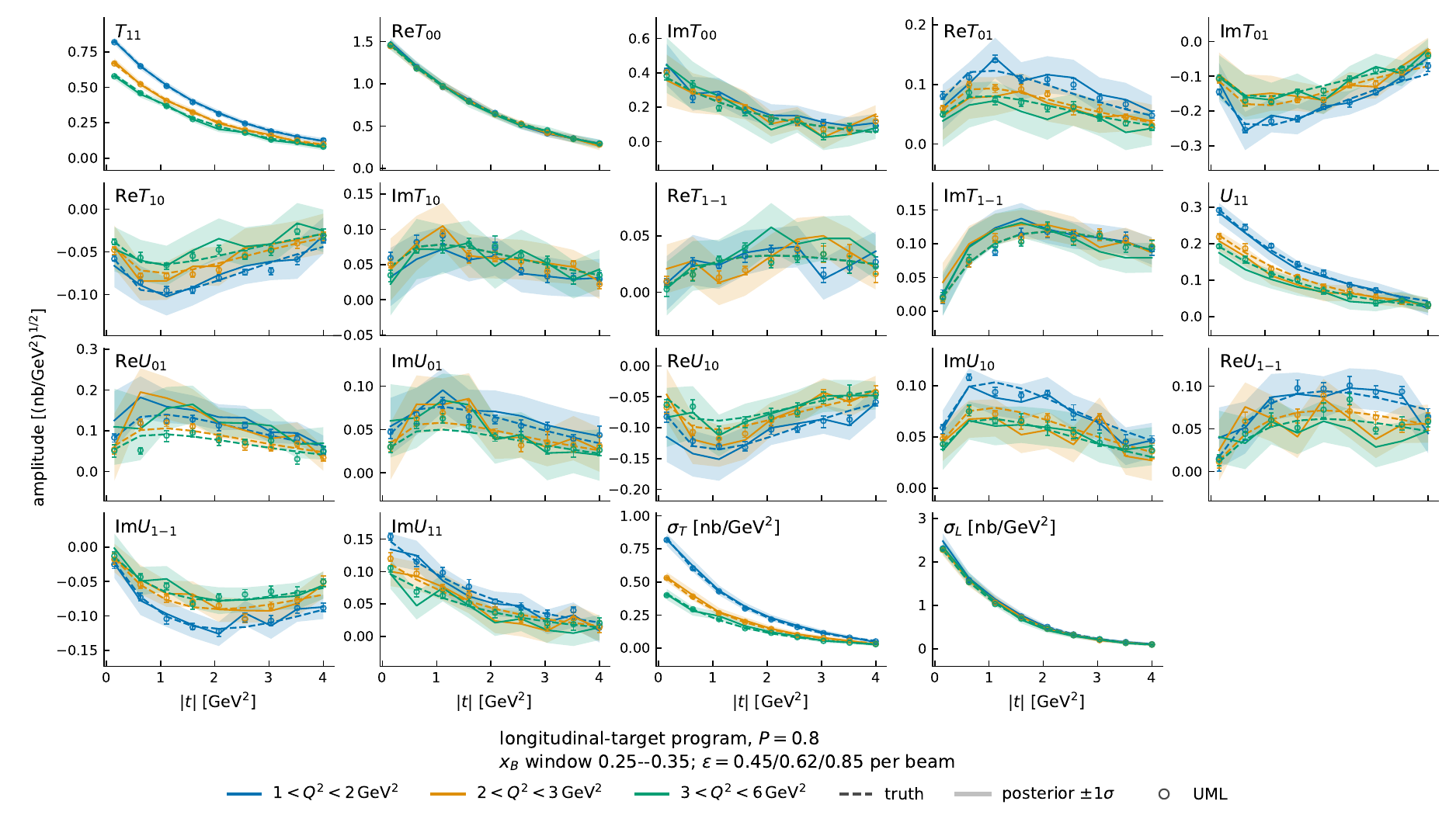}
\caption{\textbf{Longitudinal-target extraction, $\phi\to K^+K^-$ ($P=0.8$).} The 17 signed parameters (16 non-flip
$+\,\mathrm{Im}\,U_{11}$) and derived $\sT,\sL$ vs.\ $|t|$ for three $Q^2$ windows: signed
posterior band against the injected truth (dashed); the unbinned-ML fit of the same events
agrees (App.~\ref{app:uml}).}
\label{fig:polB}
\end{figure*}

\emph{The longitudinal-target extraction.} With a longitudinally polarized target the $U$-sector sign and phase become
observables (17 parameters; App.~\ref{app:polarized}). Figure~\ref{fig:polB} shows the signed
17-parameter closure vs.\ $|t|$: the unnatural amplitudes, including $\mathrm{Im}\,U_{11}$, are
recovered with their signs, with no folding anywhere (the posterior concentrates
$\approx97\%$ of its mass on the true sign branch), and the unbinned-ML fit of the same events
agrees (App.~\ref{app:uml}). This is the intermediate rung between the unpolarized extraction and the complete
polarized-target extraction of the main text (Fig.~\ref{fig:polC}).

\emph{Detector transfer of the polarized ensembles.} The polarized networks are amortized over a
detector family by the same construction as the unpolarized ensemble (Sec.~\ref{sec:detamort}),
though over a smaller family. For the full polarized extraction, applied with no retraining across a
13-detector family, the four held-out hard-edged six-sector CLAS12-like geometries are
reconstructed with $\sL$ closure RMS $0.038$--$0.132$ (median $0.101$) and the four held-out
smooth fields with $0.052$--$0.108$, against $0.085$ for the in-family nominal detector, the same
level the unpolarized ensemble reaches on the corresponding classes ($0.080$--$0.148$ and
$0.048$--$0.091$), so enlarging the inference from 16 to 34 parameters does not cost detector
generality. Only the extreme stress geometries degrade ($0.307$--$0.632$), further than in the
unpolarized case: with the flip sector included, a detector structurally far from the trained
family is the one regime where the transfer is not safe.

\begin{figure*}[!t]\centering
\includegraphics[width=\linewidth]{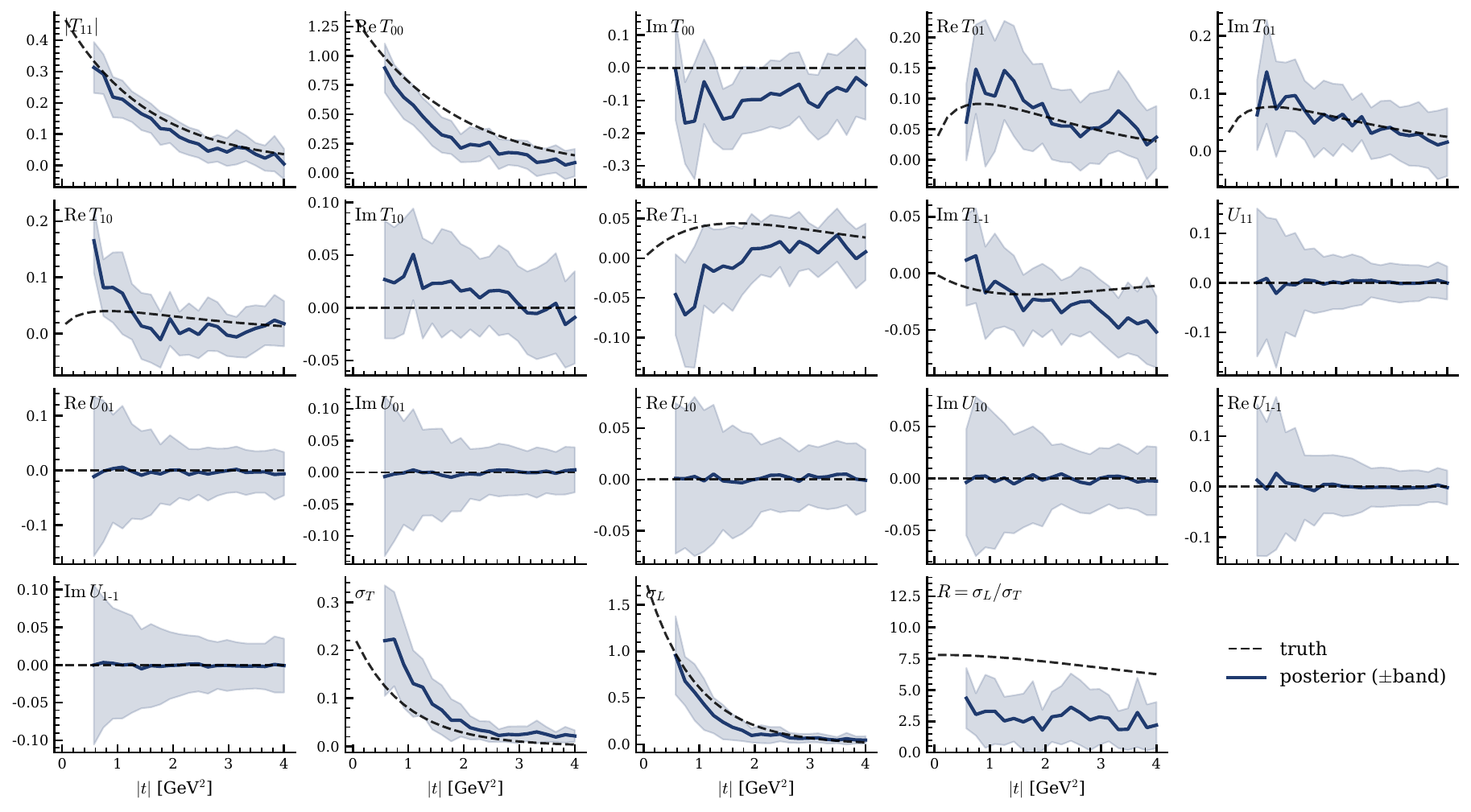}
\caption{\textbf{Complete unpolarized extraction through the GEMC-trained detector response.}
All 16 amplitude parameters vs.\ $|t|$, truth (dashed) against the extraction through the
GEMC-trained surrogate response that the ensemble never saw in training. $\sL$ closure RMS
$0.080$ over the 21 usable points, at the median level of the seven-dimensional detector family. Gaps at the three lowest
$|t|$ points mark where the spectrometer acceptance collapses and the per-point calibration fails
its statistics target at all three beam energies.}
\label{fig:fmcfull}
\end{figure*}

\emph{Full unpolarized closure through the GEMC-trained response.} Figure~\ref{fig:fmcfull}
extends the transfer test of Fig.~\ref{fig:detagnostic} from two summary quantities to the complete
unpolarized extraction: all 16 amplitude parameters versus $|t|$, with the events folded through the
GEMC-trained surrogate and the same never-retrained ensemble conditioned on its measured response.
The closure reaches $\sL$ RMS $0.080$ over the 21 usable $|t|$ points, the level of the median over
the full seven-dimensional detector family ($0.081$); this is a different and more demanding test
than the two-quantity transfer of Fig.~\ref{fig:detagnostic}, whose parameterized-ring family has
its own median ($0.095$), so the two sets of numbers are not directly comparable. The three lowest-$|t|$ points are shown as gaps: there the physical
acceptance of the simulated spectrometer collapses and the per-point calibration does not reach its
statistics target at all three beam energies (accepted counts $0/1/11$ against $\gtrsim2700$
everywhere else), a property of the instrument, not of the inference.

\begin{figure*}[!t]\centering
\includegraphics[width=\linewidth]{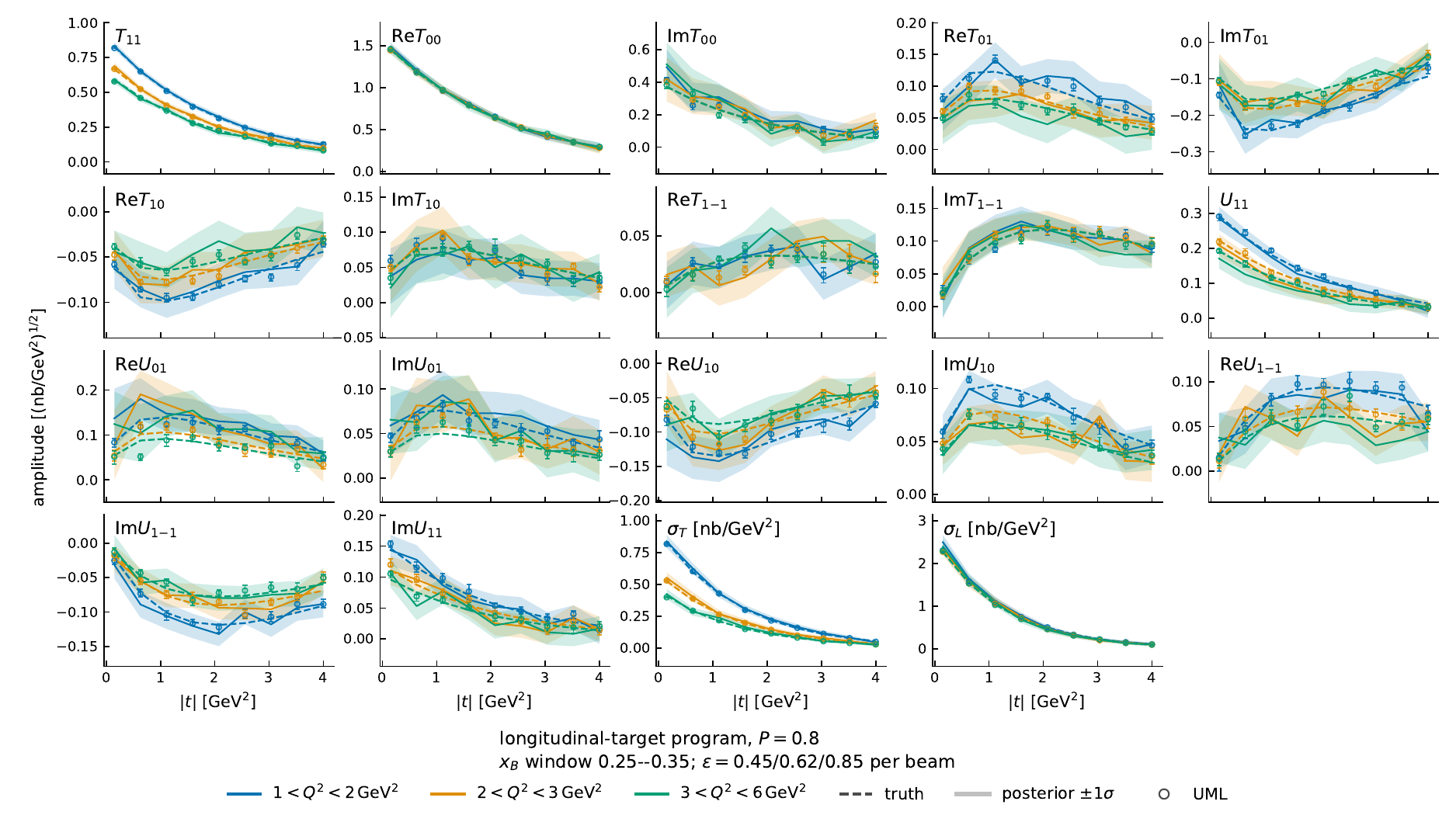}
\caption{\textbf{$\rho^0$: longitudinal-target extraction ($P=0.8$).} As
Fig.~\ref{fig:polB}, for $\rho^0\!\to\pi^+\pi^-$: the 17 signed parameters (16 non-flip
$+\,\mathrm{Im}\,U_{11}$) and derived $\sT,\sL$ vs.\ $|t|$ for three $Q^2$ windows,
signed posterior band against the injected truth (dashed).}
\label{fig:polrhoB}
\end{figure*}

\begin{figure*}[!t]\centering
\includegraphics[width=\linewidth,height=0.92\textheight,keepaspectratio]{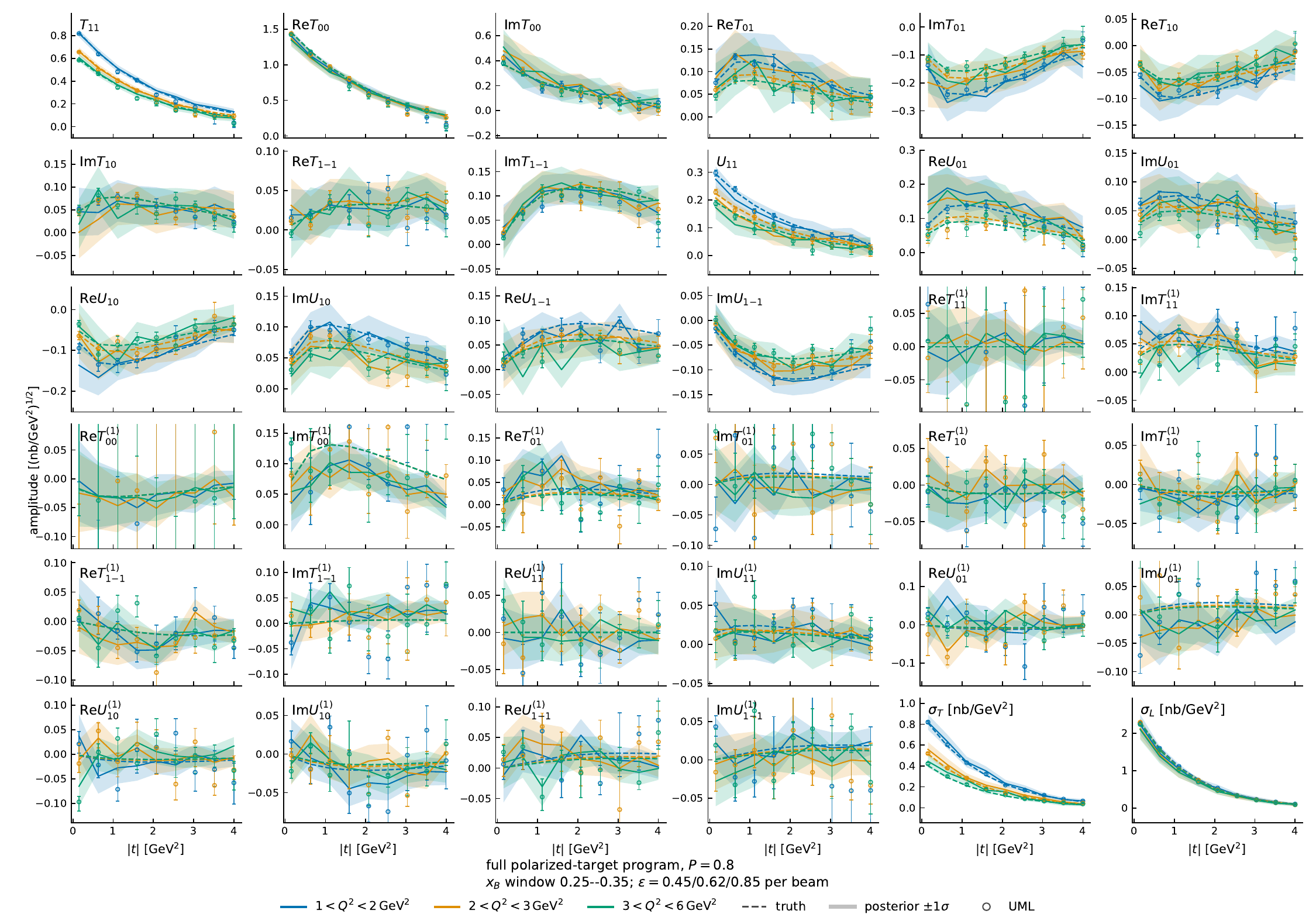}
\caption{\textbf{$\rho^0$: full polarized-target extraction.} As Fig.~\ref{fig:polC},
for $\rho^0$: the complete 34-parameter set and derived $\sT,\sL$ vs.\ $|t|$ for three
$Q^2$ windows, signed posterior (band) against the injected truth (dashed).}
\label{fig:polrho}
\end{figure*}

\emph{The $\rho^0$ polarized ladder.} The same longitudinal-target and full
polarized-target extractions for $\rho^0$ (identical recipe, $\rho^0$ truth set and
rates) close equally well, with closure pull RMS $0.75$ (longitudinal-target, $U$-sign
posterior fraction $0.97$) and $0.78$ (full program, flip-sign fraction $0.95$)
(Figs.~\ref{fig:polrhoB} and~\ref{fig:polrho}), completing for the $\rho^0$ the ladder whose
unpolarized rung is Fig.~\ref{fig:meson}.

\begin{figure*}[!t]\centering
\includegraphics[width=0.95\linewidth]{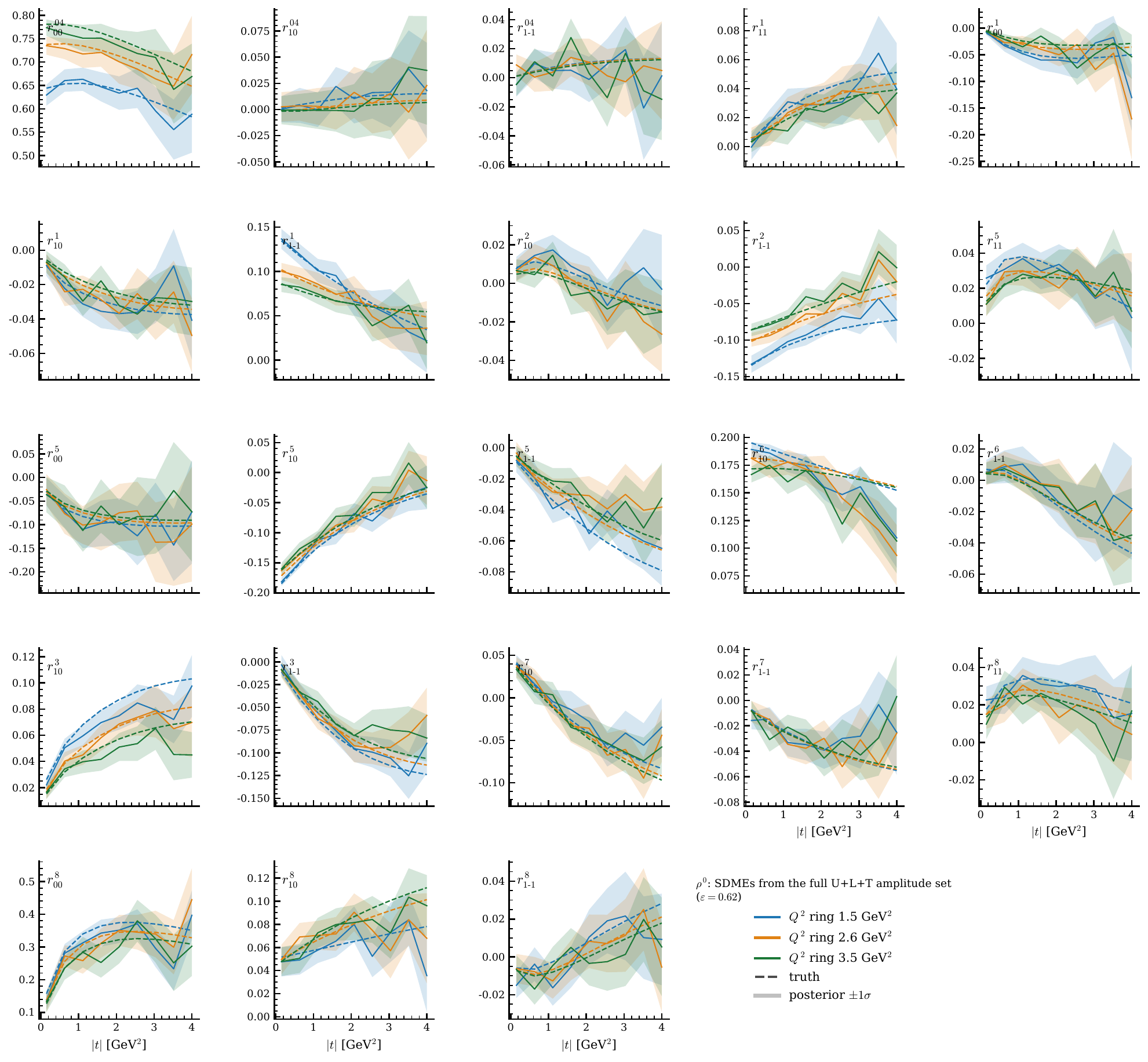}
\caption{\textbf{$\rho^0$: all 23 SDMEs as derived quantities} from the full
polarized-target (U+L+T) amplitude posterior, per posterior draw through
$r^{\alpha}_{\lambda\lambda'}=r(u(A))$ at the middle-energy $\varepsilon$: posterior mean
$\pm1\sigma$ (bands) against the injected truth (dashed) for three $Q^2$ rings.}
\label{fig:sdmerho}
\end{figure*}

\emph{SDMEs from the full amplitude set.} Because the U+L+T extraction determines the
complete signed amplitude set, every element of the spin-density matrix follows as a
derived quantity: each posterior draw is mapped through $u=u(A)$ and normalized to
$r^{\alpha}_{\lambda\lambda'}$, so the bands carry the full nonlinear error propagation
(no Gaussian approximation). Figure~\ref{fig:sdmerho} shows all 23
$r^{\alpha}_{\lambda\lambda'}$ for the $\rho^0$ versus $|t|$ in the three $Q^2$ rings at
the middle-energy $\varepsilon$: the posterior tracks the injected truth across every
element, including the unnatural-parity--sensitive ones. The amplitude-level extraction
absorbs the standard SDME analysis; it does not displace it.

\subsection{Independent unbinned maximum-likelihood cross-check}\label{app:uml}
As an independent check we fit the same simulated events by extended unbinned maximum
likelihood (unbinned ML) in the amplitude parameterization: the fit parameters are the 16 helicity
amplitudes $A$, with the intensity $W$ evaluated through the bilinears $u(A)$, so that
positivity of the spin-density matrix is automatic; the SDMEs follow as derived quantities
$r(u(A),\varepsilon)$ rather than being floated freely as in a linear SDME
fit~\citep{COMPASS:2022xig}. The fit is unbinned only in the decay
angles $(\cos\theta,\varphi,\Phi)$: like the diffusion extraction, it is performed per kinematic
cell, within the fixed $x_B$ window and $(\Qsq,|t|)$ bin, jointly over the three beam energies.
We minimize
\begin{widetext}
\begin{equation}
-\ln L(A)=\sum_{\varepsilon}\Big[-\sum_i\ln W(\Omega_i;A,\varepsilon)
+N_\varepsilon\ln Z_\varepsilon(A)\Big],\qquad
Z_\varepsilon(A)=\big\langle W(\Omega;A,\varepsilon)\,\eta(\Omega)\big\rangle_{\rm MC},
\end{equation}
\end{widetext}
jointly over the three beam energies, with the acceptance-weighted Monte Carlo normalization
$Z_\varepsilon$ (the angular factor $\eta(\Omega_i)$ on the data is amplitude-independent and drops
out of the minimization). The angular likelihood is scale-free: under $A\to cA$ both $W$ and
$Z_\varepsilon$ scale equally, so the fit determines the amplitude direction, hence all 23 SDMEs and
the relative phases, but not the overall normalization. That absolute scale ($\sT$ and $\sL$
separately) is exactly the information the diffusion model recovers jointly from the rate features (the
rate-based $L/T$ separation), which the angular fit alone does not constrain. The likelihood analysis reaches
it in a separate, explicit Rosenbluth step that fits the acceptance-corrected absolute yields
linearly in $\varepsilon$ across the three beam energies. The diffusion's rate-based separation
and this explicit one agree across the grid (Fig.~\ref{fig:closure}): the
per-bin $\sL$ pull between them has width $0.5$, and the two central values are equally accurate
against the truth (median $|\hat\sigma_L-\sL|/\sL$ of $5.1\%$ and $5.3\%$); the learned intervals
are the wider, conservative ones, quantified below. We therefore
compare all three, truth, diffusion, and unbinned ML, on the SDMEs, on $R$, and on the separated
$\sT,\sL$: at the scale-free SDME level the unbinned-ML fit, the diffusion posterior, and the truth
coincide, so the two methods are equivalent where the angular likelihood is well posed; the distinction
is that the diffusion additionally delivers the amplitudes and the absolute $\sT,\sL$ jointly (its
rate-based separation), where the likelihood reaches them only through the separate explicit Rosenbluth
step. Angular-fit uncertainties come from the numerical Hessian at the optimum, pseudo-inverted
to drop the flat scale direction; the Rosenbluth $\sT,\sL$ errors are the compound-Poisson yield
uncertainties propagated through the linear fit.

\begin{figure}[!ht]\centering
\includegraphics[width=\linewidth]{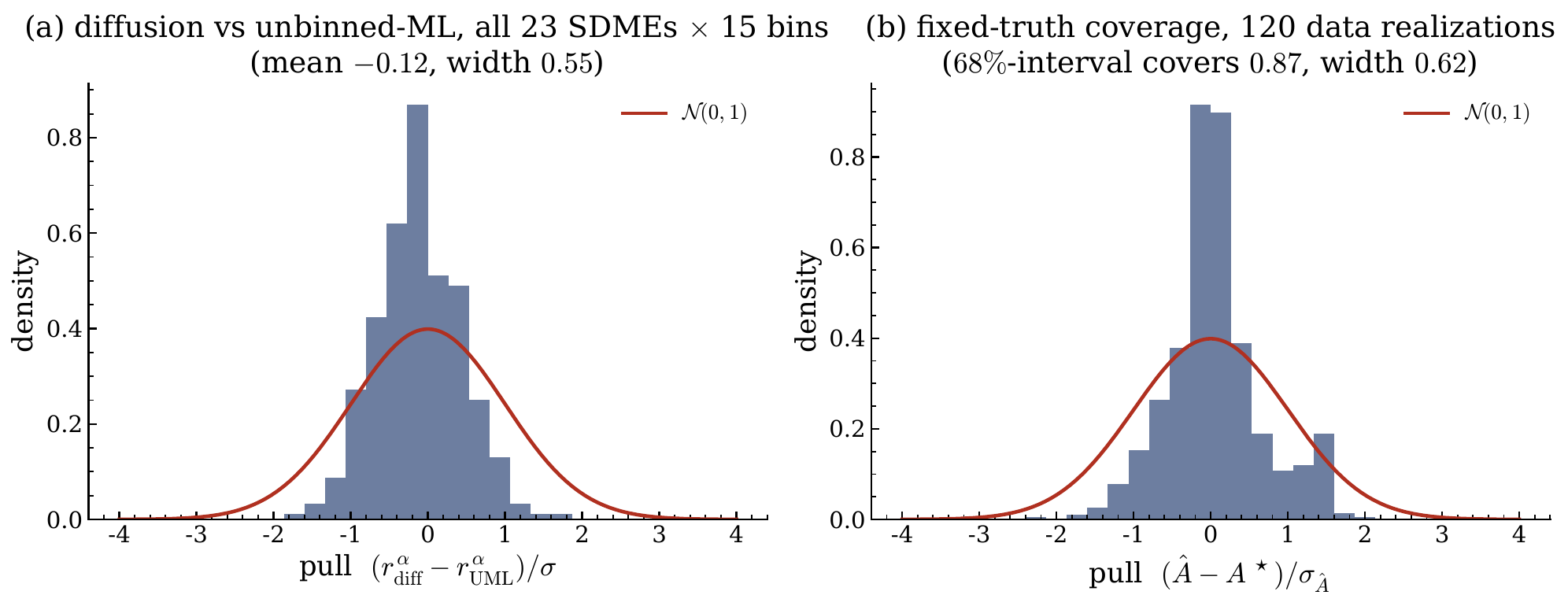}
\caption{(a) Diffusion vs.\ unbinned-ML agreement across the grid: the per-SDME pull over all
23 SDMEs and $15$ $(\Qsq,t)$ bins has mean $-0.12$, a small but statistically resolved offset, and
sub-unit width ($0.55$, consistent with the conservative per-method uncertainties).
(b) Fixed-truth (frequentist) coverage at one bin from $120$ independent data realizations:
$(\hat A-A^\star)/\sigma_{\hat A}$ is centered at zero with width $0.62$, and the $68\%$ interval covers
the truth $87\%$ of the time, conservative, consistent with the prior-averaged SBC.}
\label{fig:valextra}
\end{figure}

\emph{Agreement across the grid, and fixed-truth coverage.} The unbinned-ML comparison extends to the
whole grid (Fig.~\ref{fig:valextra}). Panel (a) histograms the
per-SDME pull $(r^\alpha_{\rm diff}-r^\alpha_{\rm ML})/\sqrt{\sigma_{\rm diff}^2+\sigma_{\rm ML}^2}$
over all 23 SDMEs and 15 $(\Qsq,t)$ bins: it has mean $-0.12$ and width
$0.55$, so the two methods agree everywhere to well within their combined errors. The offset is
small on that scale, though with $345$ entries it is statistically resolved. Panel (b) tests fixed-truth
(frequentist) coverage, complementary to the prior-averaged SBC: at one bin we repeat the experiment
over $120$ independent data realizations and histogram $(\hat A-A^\star)/\sigma_{\hat A}$; the $68\%$
credible interval covers the fixed truth $87\%$ of the time (binomial s.e.\ $\approx3\%$; pull width
$0.62$), confirming, now at a fixed point, that the reported ensemble bands err on the conservative side.

\emph{The statistical price of the compression.} The posterior conditions on the 76 compressed
features, not on the event sample itself, and the same grid measures what that compression costs
in single-bin precision. A raw width comparison is not meaningful, because the two methods do not
run at the same statistics: the likelihood fits a fixed $2500$ events per beam energy at every
point, while the feature statistics follow the acceptance-folded rate and fall by more than an
order of magnitude across the $|t|$ range. We therefore rescale the posterior widths to the
likelihood's event count ($\sigma\propto 1/\sqrt{N}$), remove the ensemble (model) share of the
posterior variance, measured per tier by the decomposition of App.~\ref{app:unc}, and deflate by
the measured per-amplitude SBC pull
widths ($0.80$--$0.93$ in the natural-parity sector, Fig.~\ref{fig:sbc}) so that both methods are
compared at nominal coverage. After these corrections the
posterior credible intervals are a median factor $2.3$ wider than the unbinned-likelihood limit
over the nine natural-parity amplitudes (interquartile range $2.0$--$2.9$) and $2.7$ over the 23
SDMEs ($2.2$--$3.6$): the compression costs a factor of five to seven in effective single-bin
statistics. The unnatural-parity block is excluded from this accounting because it is prior-shrunk
for an unpolarized target, so its interval width measures the prior, not the data. For the
separated cross sections, which the likelihood analysis reaches only through the explicit
Rosenbluth step at the same luminosity, the corrected width ratios are $1.0$ for $\sT$, $2.2$ for
$\sL$, and $1.3$ for $R$: the rate features carry $\sT$ at no statistical cost, and the premium is
confined to $\sL$. This factor is the price of conditioning on a fixed, amortized feature vector;
what it buys is what the likelihood chain does not provide: calibrated coverage, the handling of
the discrete ambiguities, the joint correlations across all parameters, and the train-once
amortization over bins and detectors.

\emph{The polarized flip sector: a genuine failure of per-bin maximum likelihood.} The agreement
above holds wherever the angular likelihood is well posed. It breaks down in one place, and the
breakdown is informative. In the full polarized program each bin carries 34 parameters, of which most of the
18 nucleon-helicity-flip parameters enter the rate suppressed by powers of $\sqrt{t'}$
(App.~\ref{app:polarized}); the whole block is only
weakly constrained by a single bin's $\sim\!2.4\times10^{4}$ events. There the per-bin maximum-likelihood
estimator does not merely lose precision, it fails outright, while the posterior does not. At a
representative bin ($\Qsq=2.6$~GeV$^2$, $|t|=0.63$~GeV$^2$), where the true flip parameters have
rms $0.033$, the maximum-likelihood estimate lands $0.060$ from the truth in rms and the posterior mean
$0.019$: the likelihood maximum is a worse estimator than simply reporting zero, and the posterior is
a better one.

We have verified that this is a property of the estimator and not of our implementation of it. The
optimizer is not at fault: $24$ randomly seeded global restarts and a start placed exactly at the truth
all converge to the same optimum, which genuinely sits at the boundary of the physically allowed flip
region. Gross likelihood misspecification is excluded: averaging the score over $16$ independent
realizations, the gradient at the truth shows no large systematic offset in any of the 34
parameters (rms $|z|=1.31$, max $3.23$; for 34 standard normals the expected rms is unity, so the
residual spread is mildly above expectation, but far below the scale of the flip-sector
discrepancy under discussion). The disparity is not a matter of unequal prior information:
granting the likelihood fit the same bounded parameter support the network is trained on removes an
otherwise accessible $U\!\to\!-U$ mirror solution and improves the non-flip block, yet leaves the flip
failure intact ($0.060$ against the posterior's $0.019$). Nor is it an artifact of the nominal statistics:
raising the luminosity thirtyfold, to $7.3\times10^{5}$ events in the bin, far beyond what any realistic
binning of the measurement affords, leaves the maximum-likelihood flip error flat at $0.081\to0.087$,
where $1/\sqrt{N}$ scaling would give $0.081\to0.015$. The estimator remains consistent in the strict
asymptotic limit, once the data localize the flip parameters well inside the physical envelope; the
point is that this limit lies far outside the regime the measurement occupies.
Finally, the posterior is not merely reproducing its prior. Its width \emph{decreases} with $|t|$ while
the prior envelope widens, reaching $0.13$--$0.20$ of the prior at large $|t|$; and over truths drawn
from the prior it tracks the flip parameters with regression slopes $0.52$--$0.81$ and a pull width
of $0.94$.

The mechanism is the distinction between the mode and the bulk of a distribution in many dimensions.
In the weakly constrained 18-dimensional flip sector, Poisson fluctuations displace the likelihood
\emph{maximum} to the edge of the allowed region, whereas the posterior \emph{mass} remains
concentrated near the truth. Maximizing therefore fails where marginalizing succeeds. This is
confined to the flip sector: for the 16 non-flip parameters the same likelihood fit is well
calibrated (pull widths $0.96$ and $1.13$), which is why we retain it as the cross-check throughout.

Two consequences follow for how the comparison is displayed. First, the uncertainties: where the fit
reaches the $\sqrt{t'}$ envelope the Hessian is not a valid error estimate, because classical maximum-likelihood
asymptotics do not hold on a parameter-space boundary~\citep{Chernoff:1954eli,selfliang1987}, and the
constrained optimum can carry a negative curvature direction that a pseudo-inverse silently discards,
assigning zero variance to the least-determined direction. The likelihood-fit uncertainties in
Figs.~\ref{fig:polB},~\ref{fig:polC} and~\ref{fig:polrho} are therefore percentile intervals from a Poisson
bootstrap ($30$ replicates, resampled within each setting--energy--helicity group, with the full
estimator re-run on every replicate), which needs no Hessian and stays valid under both the boundary
and the multi-modality; the intervals are asymmetric where the constraint makes them so. Second, the
scope of the posterior's success: of the 18 flip parameters, 17 are recovered with the fidelity quoted
above, while $\mathrm{Re}\,T^{(1)}_{00}$, the softest direction of the per-bin likelihood, is only
weakly determined and is visibly shrunk toward the prior. We therefore claim the flip sector as
measured up to that one combination, not in full.

\emph{Second-truth cross-check.} The diffusion-vs-likelihood agreement is not specific to one
truth or one detector: repeating the comparison on an independently chosen held-out truth with
distinct $t$-slopes, phases, a different L/T balance, and non-zero unnatural amplitudes,
folded through a different non-parametric field detector (same $K=20$ ensemble, no
retraining), gives per-SDME diffusion-vs-unbinned-ML pull width $0.71$ and learned-vs-explicit
Rosenbluth per-bin $\sL$ pull width $0.49$, both unbiased against the truth.

\subsection{Uncertainty quantification}\label{app:unc}
Every error we report is a posterior uncertainty. For each bin we draw $N_s=400$ samples from
the conditional diffusion model $q_\psi(A\,|\,x)$ and summarize them by their central credible
interval. Because these are samples from the full posterior, the intervals are non-Gaussian and may
be asymmetric where the geometry demands it, near the positivity boundary or for small
amplitudes, with no covariance-matrix or linearization assumption.

\emph{Statistical origin.} The posterior width is driven by the finite per-bin event count, which
enters the conditioning features in two ways: the angular moments $\langle f_k\rangle$ carry sampling
noise $\sim\!1/\sqrt{N}$, and the rate features $Y(\varepsilon)$ are Poisson with mean
$\propto(\sT+\varepsilon\sL)\times$acceptance (App.~\ref{app:impl}). The simulator reproduces both, so
the network is trained on feature noise matched to the data and the posterior width reflects the
statistics of each bin, with no fixed prescription imposed.

\begin{figure*}[!t]\centering
\includegraphics[width=\linewidth,height=0.76\textheight,keepaspectratio]{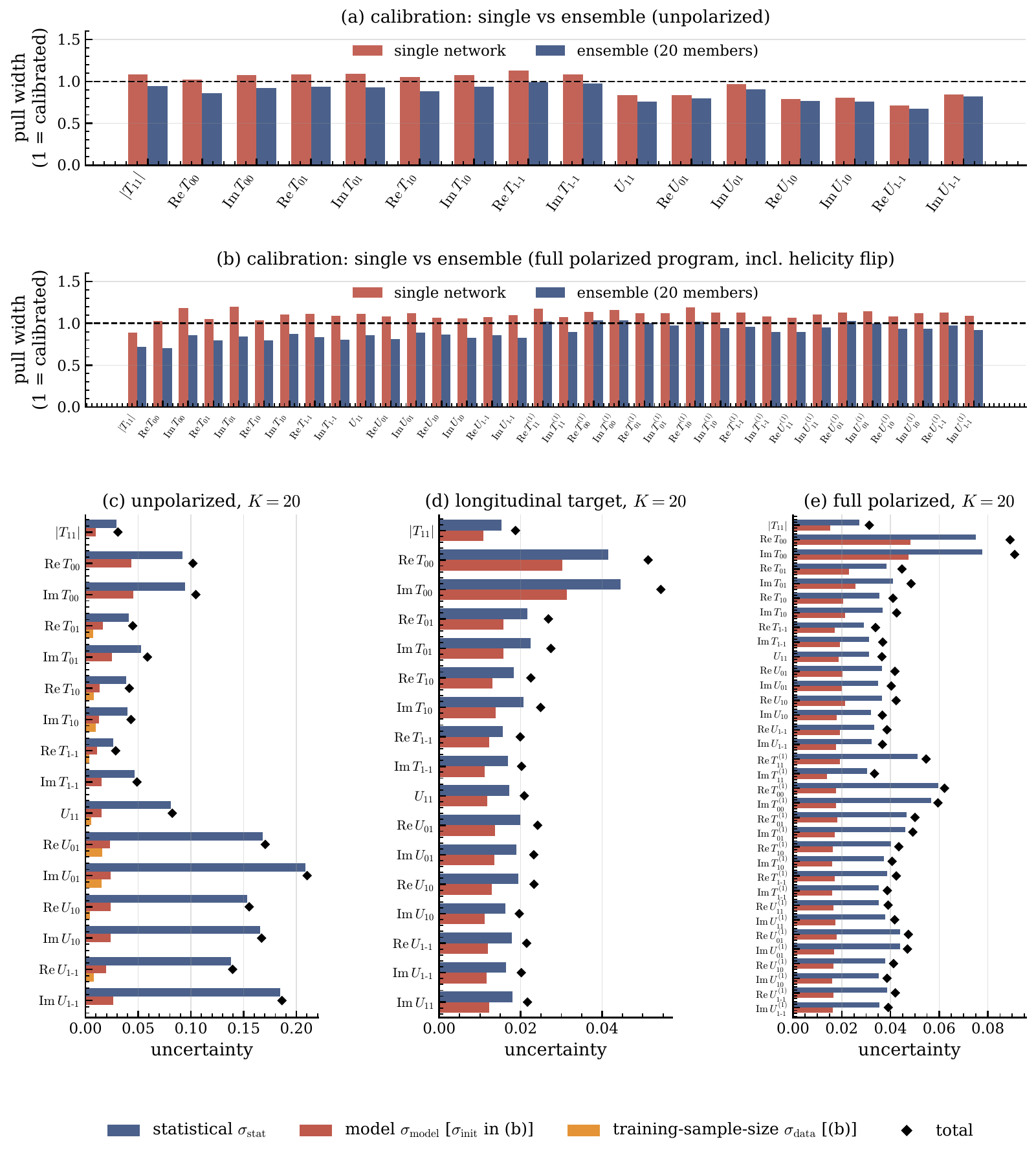}
\caption{Statistical versus model uncertainty on the extracted amplitudes, for all three
extraction modes ($500$ calibration draws each).
(a) Per-amplitude pull width of the unpolarized ensemble for a single network (red) and the
ensemble mixture (blue); unity is calibrated, $>\!1$ over-confident, $<\!1$ conservative, so the
calibrated uncertainty lies empirically between the two.
(b) The same comparison for the full polarized program: all 34 parameters, including the 18
nucleon-helicity-flip amplitudes absent from (a). (c--e) Per-amplitude
uncertainty budget from the law of total variance, for the unpolarized,
longitudinal-target and full polarized-target ensembles ($K=20$ each): the
statistical (aleatoric, blue) and model (epistemic, red) terms, with the quadrature
total (diamonds). In the unpolarized panel (c) alone the model term is further split into its
optimization/architecture floor $\sigma_{\rm init}$ (red, from a shared-data ensemble) and the
training-sample-size component $\sigma_{\rm data}$ (orange; direct two-ensemble measurement,
Fig.~\ref{fig:trainsize}). All three ensembles are trained at the same production sample size,
$N_0=4.8\times10^{5}$, the largest used in this work, so the training-sample-size term sits at the
same floor in every tier. It is measured only at the unpolarized tier, where it accounts for a
median $6\%$ of the model variance; since the model term is itself subdominant in the polarized
tiers (d),(e), $\sigma_{\rm data}$ is negligible in their totals and the model term is shown
undivided there.
The unpolarized $U$ sector is reported on the
$U_{11}\!\geq\!0$ branch of the exact $U\!\to\!-U$ degeneracy. Across panels the added
polarized information shrinks the statistical uncertainty of the $U$ sector by an order of
magnitude while the model term stays subdominant throughout.}
\label{fig:ampunc}
\end{figure*}

\emph{Calibration.} The diagnostic is the pull, $z=(\hat A-A^\star)/\sigma_{\hat A}$. Averaged over many simulated
truths, a correctly-sized error bar gives a pull with unit standard deviation (the ``pull width''):
the estimate sits about $1\sigma$ from truth. A pull width
$>1$ means the errors are understated, the $68\%$ interval catches the truth less than
$68\%$ of the time (over-confident); a pull width $<1$ means the errors are
overstated, the interval catches the truth more often than claimed (conservative).
Simulation-based calibration measures this coverage directly.
The single-network posterior (the statistical width alone) is mildly over-confident: its
per-parameter pull widths cluster near unity at mean $\approx1.07$, so a modest global inflation
restores nominal coverage [Fig.~\ref{fig:ampunc}(a), red]. The total (ensemble) uncertainty
actually used for the bands sits on the conservative side: its SBC (App.~\ref{app:cal},
Fig.~\ref{fig:sbc}) gives empirical central-interval coverage $0.72/0.96$ at nominal $0.70/0.95$,
i.e.\ it over-covers, with the residual per-parameter over-confidence confined to the $U$ sector.
Empirically the well-calibrated value lies between the two.

\emph{Statistical vs.\ model uncertainty.} Figure~\ref{fig:ampunc} separates the two, using
a deep ensemble~\citep{Lakshminarayanan:2017tnw} of $K=20$ networks trained independently (different initializations and training
draws). For each test case we draw posterior samples from every member and apply the law of total
variance: the mean within-member variance is the statistical (aleatoric) uncertainty
$\sigma_{\rm stat}^2$, and the variance of the member posterior means is an ensemble-based proxy for the
model (epistemic) uncertainty $\sigma_{\rm model}^2$ (with $K=20$ members a
conservative, necessarily noisy estimate), with $\sigma_{\rm tot}^2=\sigma_{\rm stat}^2+\sigma_{\rm model}^2$
[panel (c)]. The model term is not negligible: it is comparable to the statistical term, and
largest for the weakly-constrained $\mathrm{Im}\,T_{00}$ and $T_{01}$ components. Panels (d)
and (e) apply the same statistical/model split to the polarized ensembles: the added polarized
information shrinks the statistical term (most dramatically in the $U$ sector, by an order
of magnitude relative to the unpolarized case) while the model term remains subdominant,
so the uncertainty budget of the full 34-parameter extraction is statistics-dominated. All three tiers are
trained at the same production sample size $N_0=4.8\times10^{5}$, so the training-sample-size term
is at the same floor throughout; the further division of the model term into $\sigma_{\rm init}$
and $\sigma_{\rm data}$ below is therefore measured once, for the unpolarized ensemble, and is not
repeated for the polarized tiers, where the term it subdivides is already subdominant.

\begin{figure*}[!ht]\centering
\includegraphics[width=\linewidth]{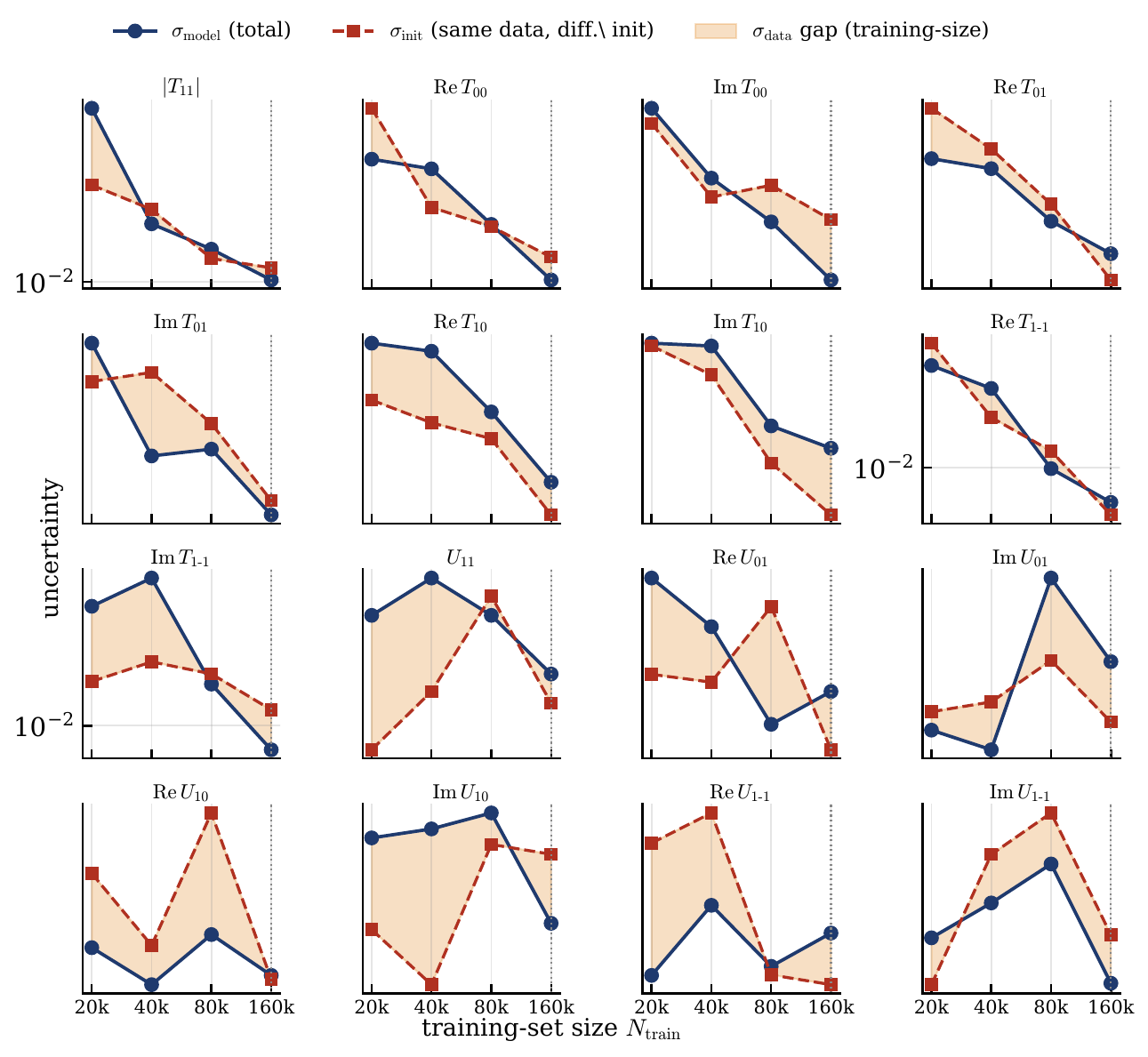}
\caption{Model (epistemic) uncertainty versus training-sample size, per amplitude. At each
$N_{\rm train}$, the total $\sigma_{\rm model}$ from a fresh-data ensemble (blue) and the
$\sigma_{\rm init}$ floor from a shared-data ensemble (same pairs, different init; red dashed); their gap
(shaded) is the training-sample-size part $\sigma_{\rm data}$ (dotted vertical: the reference size
$N_{\rm train}\!=\!160{,}000$ at which this two-ensemble sweep is anchored). $\sigma_{\rm model}$ falls with the training budget, but $\sigma_{\rm init}$ tracks
it closely: the model uncertainty is dominated by optimization/initialization noise at fixed data,
with only a small reducible training-sample-size gap (split used in Fig.~\ref{fig:ampunc}(c)).}
\label{fig:trainsize}
\end{figure*}

One might worry that this model term merely reflects the finite size of the training sample. We test this directly
with a pair of ensembles at the production size: a shared-data ensemble (identical training pairs,
different initializations) measures the optimization/architecture stochasticity $\sigma_{\rm init}$ that
remains with the data held fixed, while the nominal fresh-data ensemble (different data draws
and initializations) gives the total $\sigma_{\rm model}$; the part attributable to the finite
training sample is then $\sigma_{\rm data}=\sqrt{\sigma_{\rm model}^2-\sigma_{\rm init}^2}$. We find that
the shared-data ensemble reproduces most of the spread ($\sigma_{\rm init}\!\approx\!\sigma_{\rm model}$):
the model term is dominated by optimization/initialization noise at fixed data, and the pure
training-sample-size component is small (median $\sigma_{\rm data}^2/\sigma_{\rm model}^2\approx0.06$ at
the production size, and at the noise level of the measurement for several amplitudes).
Figure~\ref{fig:ampunc}(c) shows this split. This model term is therefore not primarily a
training-sample-size effect; it does fall with the training budget (Fig.~\ref{fig:trainsize}),
reducible by more training pairs and epochs, better optimization, or a more
expressive score network, and we retain it in the reported total as a conservative choice. Panel (a)
shows the
consequence for calibration, for the fixed-detector ensemble studied in this figure: a single
network is mildly over-confident (mean pull width $\approx1.07$),
whereas the full ensemble, whose mean is sharper and whose error budget includes the
model term, is essentially calibrated (mean pull width $\approx0.98$). The detector-amortized
ensemble is calibrated separately (App.~\ref{app:cal}) and carries a larger mean pull width
($1.21$), the excess confined to the $U$ sector; the two ensembles are distinct objects and their
pull widths are not directly comparable.

Because these two objects are easily confused, we state exactly what is reported. The single-network
conditional posterior $q_\psi(A\,|\,x)$ and the ensemble are not two competing candidates for the
physical posterior. Each member's conditional posterior represents the statistical uncertainty, the
irreducible spread induced by the finite event count entering the features; the dispersion
\emph{among} members represents the model uncertainty, which is a property of the learned
approximation, not of the data. The reported interval is constructed in one step from the two
together: posterior draws from all $K=20$ members are pooled into a single sample set, and the
quoted band is a quantile of that pooled set, so that by the law of total variance it carries
$\sigma_{\rm stat}^2+\sigma_{\rm model}^2$ without either term being fitted or added by hand. Every
band in this paper is that pooled-ensemble quantile. Its calibration is measured by the SBC of
App.~\ref{app:cal}, run separately for each trained ensemble so that every figure is backed by the
calibration of the ensemble that produced it, and it over-covers, which is the direction we accept
deliberately.
The recalibrated single-network alternative ($\sigma\!\to\!1.07\,\sigma$, which restores nominal
coverage on average) is reported here only to bracket the well-calibrated value from the other side;
it is not what the figures show.

\emph{Improving the calibration.} The single-network over-confidence can be reduced post-hoc by the
per-parameter recalibration $\sigma\!\to\!w\,\sigma$, and at the model level by more reverse-diffusion
steps, a finer noise schedule, more training pairs and epochs, and a more expressive score
network, all of which tighten the gap between $q_\psi$ and the true posterior and shrink the
ensemble spread.

\begin{figure*}[!ht]\centering
\includegraphics[width=\linewidth]{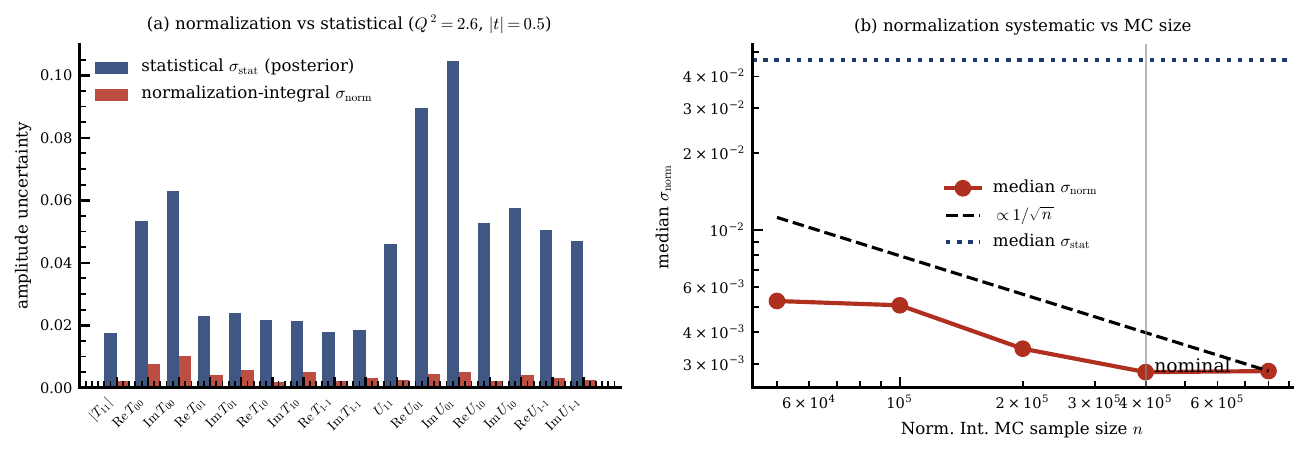}
\caption{Normalization-integral systematic. (a) Per-amplitude, the systematic from the finite-MC
acceptance normalization ($\sigma_{\rm norm}$, red) is everywhere below the statistical posterior width
($\sigma_{\rm stat}$, blue) at a representative bin. (b) $\sigma_{\rm norm}$ falls as
$1/\sqrt{N_{\rm cal}}$ with the calibration MC sample size $N_{\rm cal}$ (dotted: median
$\sigma_{\rm stat}$; vertical line: nominal $N_{\rm cal}$), so it is controllable and already sub-dominant.}
\label{fig:normcheck}
\end{figure*}

\emph{Normalization-integral systematic.} The conditioning moments
$\langle f_k\rangle=(Mu)/(b\!\cdot\!u)$ use acceptance-folded tensors $(b,M)$ estimated from a finite
flat-MC sample through $\eta$, i.e.\ the acceptance normalization integral is itself a Monte Carlo
estimate (the analogue of the $Z_\varepsilon$ normalization of the unbinned-likelihood fit) and carries
MC noise. Propagating that noise (recomputing $(b,M)$ over independent MC seeds and re-extracting) gives
a per-amplitude systematic $\sigma_{\rm norm}$ that is everywhere well below the statistical width
(median $\sigma_{\rm norm}/\sigma_{\rm stat}\approx0.3$ at the nominal $N_{\rm cal}=4\times10^5$ calibration sample)
and falls as $1/\sqrt{N_{\rm cal}}$ (Fig.~\ref{fig:normcheck}), so it is sub-dominant and reducible to negligibility
with a larger calibration sample at no methodological cost. It contributes $\lesssim5\%$ in quadrature to
the total and is not separately propagated in the reported bands.

\begin{figure*}[!t]\centering
\includegraphics[width=\linewidth]{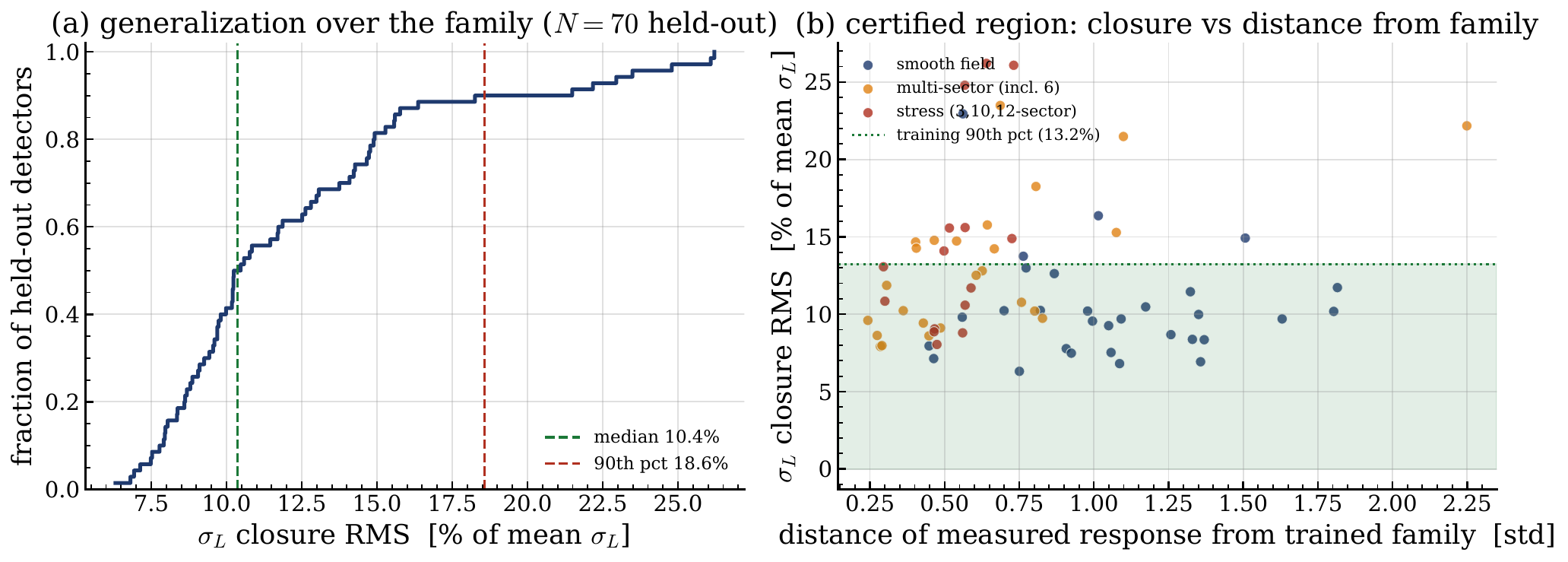}
\caption{Generalization confidence of the trained network over a large held-out set ($70$
detectors). (a) Cumulative distribution of the $\sL$ closure RMS (as \% of the mean $\sL$): the
median is $10.4\%$ and the $90$th percentile $18.6\%$, so $90\%$ of held-out detectors are recovered to
within ${\sim}19\%$. (b) The same closure RMS versus the distance of each detector's measured
response from the trained family (standardized units; training detectors lie near $1$): closure remains
bounded across the full range, including the structurally distant stress geometries; the shaded band is
the in-family ($90$th-percentile) reference. The certification is over the detector family
in simulation and does not include simulation-versus-data mismatch.}
\label{fig:genconf}
\end{figure*}

\emph{Generalization confidence over the detector family.} To quantify how the trained
network (Sec.~\ref{sec:detamort}) generalizes, we evaluate it on $70$ held-out detectors spanning the
trained family: smooth fields, multi-sector geometries (including six-sector), and extreme
$3,10,12$-sector stress geometries (Fig.~\ref{fig:genconf}). The $\sL$ closure RMS has median $10.4\%$ of the mean $\sL$ and
$90$th percentile $18.6\%$ [panel (a)], and stays bounded even for the most structurally distant
detectors [panel (b), closure versus the standardized distance of each detector's measured response from
the training set]. For any detector whose measured response lies in this family, $\sL$ is recovered to
within ${\sim}19\%$ at $90\%$ confidence, and a candidate detector is placed against the family by its
response distance. This statement concerns generalization across the detector family and across
unseen amplitudes and event samples, all within the trained forward model; it is independent of,
and does not bound, the simulation-versus-data systematic, which arises only when the simulated forward
model differs from the real apparatus and is estimated separately (below).

\emph{Propagation to derived quantities.} The SDMEs and the cross sections $\sT,\sL,R$ are not fit
separately. Each posterior amplitude sample $A^{(s)}$ (pooled across the $K=20$ ensemble members,
so the spread already carries the statistical $\oplus$ model uncertainty) is mapped through
$u=u(A)$ and the Diehl$\to$Schilling--Wolf relations to a sample of every observable, and the reported
band is the quantile of those samples; correlations among the amplitudes therefore propagate exactly
to the SDMEs and cross sections. This is the total band shown in Figs.~\ref{fig:polC} and~\ref{fig:closure}.
The ratio $R=\sL/\sT$ inherits a band that widens at large $|t|$ where
$\sT\to0$ and the ratio is intrinsically ill-conditioned.

\emph{Comparison fit.} The unbinned maximum-likelihood cross-check (App.~\ref{app:uml}) instead
reports the Gaussian (Cram\'er--Rao) uncertainty from the numerical Hessian at the optimum,
pseudo-inverted with a spectral cutoff that drops the flat directions (the overall-scale
direction, and for the per-point polarized fits the near-degenerate flip-sector modes), so the
quoted errors reflect the locally constrained subspace. The per-point polarized fits also
impose the $\sqrt{t'}$ threshold envelope on the flip block, the external input that the
observability ladder of App.~\ref{app:polarized} identifies as required for the per-bin
flip separation; without it the recoil-rotation direction degenerates at small $|t|$. The two
error estimates agree at the scale-free SDME level.

\emph{Systematics.} The uncertainties above are statistical. The leading systematic is
forward-model mismatch, an imperfect acceptance, resolution, or background model in the
simulator, which on real data is quantified by simulation-to-simulation transfer and by the
mutual consistency of the diffusion and likelihood SDMEs; the parameterized detector model used here
removes that source by construction.

\subsection{Closure loop: from inferred amplitudes back to events}\label{app:closureloop}
\begin{figure*}[!t]\centering
\includegraphics[width=\linewidth]{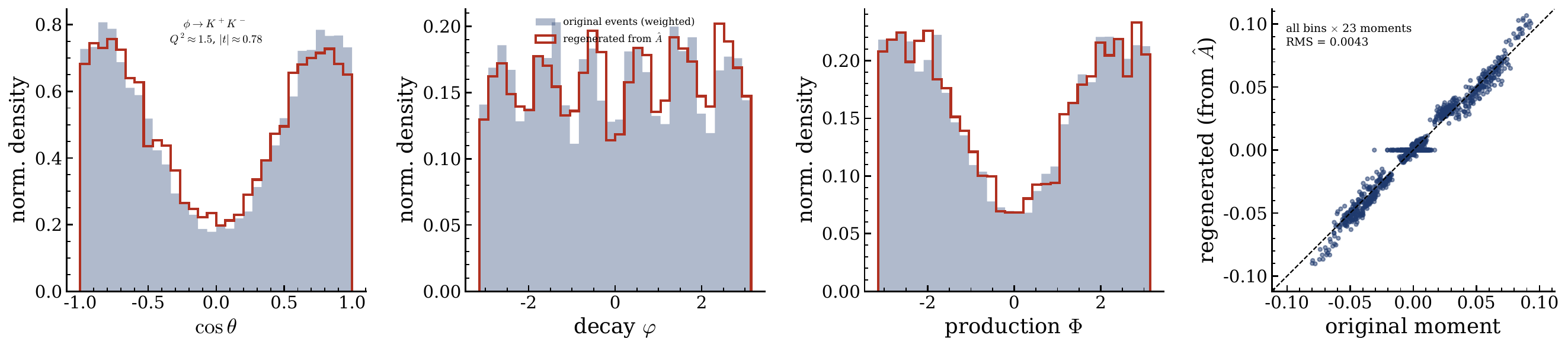}\\[2pt]
\includegraphics[width=\linewidth]{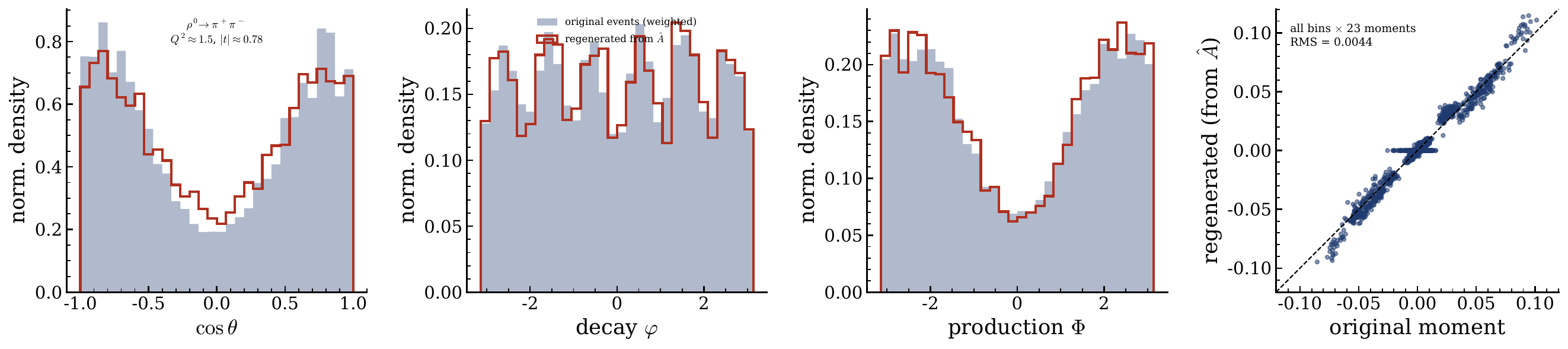}
\caption{\textbf{Closure loop} (inject $A$ $\to$ events $\to$ infer $\hat A$ $\to$ regenerate
$\to$ match), for $\phi\to K^+K^-$ (top) and $\rho^0\to\pi^+\pi^-$ (bottom). Left three
panels: original (weighted, filled) versus regenerated-from-$\hat A$ (line) decay angular
distributions in a representative bin. Right: regenerated versus original values of all
23 moments across every kinematic bin; the loop closes with RMS $0.0043$ ($\phi$) and
$0.0044$ ($\rho^0$).}
\label{fig:closureloop}
\end{figure*}

Every closure test so far compares the inferred amplitudes against the injected truth,
a comparison that is impossible on real data, where no truth exists. The closure loop is the
truth-free counterpart, run entirely in event space: amplitudes are injected, events are
generated and passed through the full chain, the amplitudes $\hat A$ are inferred per bin, and
then new events are generated from $\hat A$ through the same physics intensity and
overlaid on the originals. If the extraction has captured all the information the events
contain, the regenerated decay angular distributions and the 23 moments of every bin must
reproduce the originals; a bias in the inferred amplitudes, or a failure of the physical
parameterization to represent the injected state, would appear directly as a mismatch. The loop
therefore verifies that the inferred amplitudes regenerate the measured angular distributions,
which is the closure that remains available on experiment, where this test (infer, regenerate,
compare to data) is the only one that can be run. Its scope should be stated with it: the
posterior is conditioned on these same moments, so the loop establishes the internal consistency
of the inferred amplitudes with the features they were fitted to, not sufficiency against
information lying outside the moment basis; and because the regeneration uses a point estimate
$\hat A$, it tests the location of the posterior, not its width. Both channels close
(Fig.~\ref{fig:closureloop}): over all bins and all 23 moments the original-vs-regenerated
RMS is $0.0043$ for the $\phi$ and $0.0044$ for the $\rho^0$, with the angular distributions
overlapping within statistics.

\end{document}